
\input harvmac
\baselineskip=12pt

\def\frac#1#2{{#1\over#2}}

\def\half{\frac12}

\def\journal#1&#2(#3){\unskip, #1~\bf #2 \rm(19#3) }
\def\andjournal#1&#2(#3){\sl #1~\bf #2 \rm (19#3) }
\def\ie{{\it i.e.}}

\def\bra#1{\left\langle #1\right|}
\def\ket#1{\left| #1\right\rangle}

\def\exp{{\rm exp}}

\def\slash#1{\mathord{\mathpalette\c@ncel{#1}}}
\overfullrule=0pt
\def\steepslash{\c@ncel}
\def\frac#1#2{{#1\over #2}}

\def\hb{\hbar}
\def\p{\partial}
\def\inbar{\,\vrule height1.5ex width.4pt depth0pt}
\def\IQ{\relax\,\hbox{$\inbar\kern-.3em{\rm Q}$}}
\def\IB{\relax{\rm I\kern-.18em B}}
\def\IC{\relax\hbox{$\inbar\kern-.3em{\rm C}$}}
\def\IP{\relax{\rm I\kern-.18em P}}
\def\IR{\relax{\rm I\kern-.18em R}}
\def\ZZ{\relax\ifmmode\mathchoice
{\hbox{Z\kern-.4em Z}}{\hbox{Z\kern-.4em Z}}
{\lower.9pt\hbox{Z\kern-.4em Z}}
{\lower1.2pt\hbox{Z\kern-.4em Z}}\else{Z\kern-.4em Z}\fi}

\catcode`\@=12

\def\npb#1(#2)#3{{ Nucl. Phys. }{B#1} (#2) #3}
\def\plb#1(#2)#3{{ Phys. Lett. }{#1B} (#2) #3}
\def\pla#1(#2)#3{{ Phys. Lett. }{#1A} (#2) #3}
\def\prl#1(#2)#3{{ Phys. Rev. Lett. }{#1} (#2) #3}
\def\mpla#1(#2)#3{{ Mod. Phys. Lett. }{A#1} (#2) #3}
\def\ijmpa#1(#2)#3{{ Int. J. Mod. Phys. }{A#1} (#2) #3}
\def\cmp#1(#2)#3{{ Comm. Math. Phys. }{#1} (#2) #3}
\def\cqg#1(#2)#3{{ Class. Quantum Grav. }{#1} (#2) #3}
\def\jmp#1(#2)#3{{ J. Math. Phys. }{#1} (#2) #3}
\def\anp#1(#2)#3{{ Ann. Phys. }{#1} (#2) #3}

\def\o{\omega}

\def\t{\theta}

\def\cD{{\cal D}}
\def\half{{1\over 2}}

\def\{{\lbrace}
\def\}{\rbrace}
\def\({\lbrack}
\def\){\rbrack}

\def\tu{\theta_u}
\def\tv{\theta_v}
\def\tw{\theta_w}
\def\tx{\theta_x}
\def\ub{{\bar u}}
\def\vb{{\bar v}}
\def\wb{{\bar w}}

\def\tK{{\tilde K}}
\def\dis{{\rm dis}_1(X^*)}
\def\ni{\noindent}
\def\hb{\hfill\break}
\def\bra{\langle}
\def\ket{\rangle}
\def\pd{\partial}

\def\Z{{\bf Z}}

\def\da#1{{\pd \over \pd a_{#1}}}

\def\({\left(}
\def\){\right)}

\def\hb{\hfill\break}
\overfullrule=0pt


\def\o{\omega}

\def\t{\theta}
\def\half{{1\over 2}}

\def\{{\lbrace}
\def\}{\rbrace}
\def\Blb{\Bigl\{}
\def\Brb{\Bigr\}}
\def\({\left(}
\def\){\right)}
\def\tx{{\Theta_x}}

\def\ni{\noindent}
\def\hb{\hfill\break}
\def\bra{\langle}
\def\ket{\rangle}

\footline{\hss\tenrm--\folio\--\hss}
\rightline{HUTMP-93/0801}
\Title{LMU-TPW-93-22}
{\vbox{ \centerline{Mirror Symmetry, Mirror Map and Applications}
        \vskip2pt
        \centerline{to Calabi-Yau Hypersurfaces} } }

\centerline{S. Hosono$^*,$\footnote{$^{\dagger}$}
{ On leave of absence from Dept. of Math., Toyama Univ., Toyama 930,
  Japan  (address after Sep.1, 1993) }
, A. Klemm$^{**}$\footnote{$^{\diamond}$}{Address after Oct.1, 1993
Dept. of Math., Harvard Univ., e-mail: klemm@string.harvard.edu;
after April 1, 1994 CERN,
Theory Division},
S. Theisen$^{**}$ and S.-T. Yau$^*$}
\bigskip
\centerline{$^*$Department of Mathematics, Harvard University}
\centerline{ Cambridge, MA 02138, U.S.A. }
\medskip\centerline{$^{**}$Sektion Physik der Universit\"at M\"unchen}
\centerline{Theresienstra\ss e 37, D - 80333 M\"unchen, FRG}
\vskip .5in

\centerline{\bf Abstract:}
\ni Mirror Symmetry, Picard-Fuchs equations and
instanton corrected Yukawa couplings are discussed within the framework
of toric geometry. It allows to establish mirror symmetry of Calabi-Yau
spaces for which the mirror manifold had been unavailable in previous
constructions. Mirror maps and Yukawa couplings are explicitly
given for several examples with two and three moduli.
\Date{August 1993}

\vfill\eject

\newsec{Introduction}

Mirror symmetry
\ref\mirror{For reviews, see {\it Essays on Mirror Manifolds}
     (ed. S.-T. Yau), Int. Press, Hong Kong, 1992}
started from the trivial observation
\ref\dixon{L. Dixon, in {\it Superstrings, Unified Theories and
     Cosmology 1987}, G. Furlan et al, eds.,
     World Scientific 1988}
\ref\lvw{W. Lerche, C. Vafa and N. Warner, \npb324(1989)427}
that the relative
sign of the two $U(1)$-charges of $(2,2)$
super-conformal field theories is simply a
matter of convention. Geometrically,
however, if one interprets certain symmetric
$(2,2)$ superconformal theories as  string compactifications
on Calabi-Yau spaces, the implications are far from
trivial and imply identical string propagation
on topologically distinct manifolds for
which the cohomology groups
$H^{p,q}$ and $H^{q,3-p}$, $p,q=1,\ldots, 3$
are interchanged.

Within the classes of Calabi-Yau spaces that have been
investigated by physicists, namely complete intersections
in projective spaces
\ref\cdls{P. Candelas, A. M. Dale, C. A. L\"utken, R. Schimmrigk :
             {\sl Complete Intersection Calabi-Yau Manifolds I, II },
             \npb298(1988)493, \npb306(1988)113},
toroidal orbifolds
\ref\orb{L. Dixon, J. A. Harvey, C. Vafa and E. Witten :
             Nucl. Phys. B261 (1985) 678 and Nucl. Phys. B274 (1986) 285;
             D. G. Markushevich, M. A. Olshanetsky and A. M. Perelomov,
             Commun. Math. Phys. 111 (1987) 247;
             J. Erler, A. Klemm, Commun. Math. Phys. 153 (1993) 579}
and hypersurfaces or complete intersections in products
of weighted projective spaces
\ref\ks{A. Klemm and R. Schimmrigk, {\sl Landau-Ginzburg String Vacua},
             CERN-TH-6459/92, to appear in Nucl. Phys. B;
             M. Kreuzer and H. Skarke, Nucl. Phys. B388 (1993) 113},
one does indeed find approximate mirror symmetry, at least on the
level of Hodge numbers, which get interchanged
by the mirror transformation: $h^{p,q}\leftrightarrow h^{q,3-p}$.
Most of the known candidates for mirror pairs of Calabi-Yau manifolds
are hypersurfaces or complete intersections in products of weighted
projective spaces and are related to string vacua described by
$N=2$ superconformal limits of Landau-Ginzburg models
\ref\lwv{C. Vafa and N. Warner, \plb218(1989)51;
         C. Vafa, Mod. Phys. Lett. A 4 (1989) 1169},
         \lvw,\ks .
For subclasses of these manifolds one can find discrete
symmetries such that the desingularized quotient
with respect to them yields a mirror configuration; see ref.
               \ref\roan{S.-S. Roan,
               {\sl Int. Jour. Math.}, Vol. {\bf 2},
               No. 4 (1991) 439;
               {\sl Topological properties of Calabi-Yau
               mirror manifolds}, preprint 1992}
               and for a somewhat more general construction ref.
               \ref\bh{P. Berglund
               and T. H\"ubsch, Nucl. Phys. 393B (1993) 377}.
Likewise the corresponding superconformal
field theory exhibits in subclasses symmetries
\ref\gp{B. Greene and M. Plesser,\npb338(1990)15}, which can be used
to construct the mirror SCFT by orbifoldization.
The Landau-Ginzburg models in the sense of ref. \lwv~
have been classified in \ks.
It turns out that the spectra in this class do not exhibit
perfect mirror symmetry, even after including quotients
\ref\kstwo{M. Kreuzer and H. Skarke, Nucl. Phys. B405 (1993) 305;
           M. Kreuzer, Phys. Lett. 314B (1993) 31}.
A gauged generalization of Landau-Ginzburg models was proposed in
\ref\wlg{E. Witten, Nucl. Phys. B403 (1993) 159}.
The associated Calabi-Yau spaces are realized as hypersurfaces or
complete intersections in more general toric varieties or Grassmanians.

A particularly appealing construction of Calabi-Yau manifolds,
within the framework of toric geometry, was given by Batyrev in
\ref\batyrev{V. Batyrev, {\sl Dual Polyhedra and the Mirror Symmetry
      for Calabi-Yau Hypersurfaces in Toric Varieties}, Univ. Essen
      Preprint (1992), to appear in  Journal of Alg. Geom.
      and Duke Math. Journal, Vol. 69, No 2, (1993) 349}.
It gives hypersurfaces in Gorenstein toric varieties and
unlike previous constructions it is manifestly mirror symmetric.
This is the approach we will take in this paper. We will
show that mirror partners, which are missing in the conventional
Landau-Ginzburg approach \ks, even when
including the quotients \kstwo \bh \roan,
can be constructed systematically as hypersurfaces
in these generalized Gorenstein toric varieties.

A much less trivial implication of mirror symmetry  than the
existence of Calabi-Yau spaces with flipped Hodge numbers,
is the isomorphism between the cohomology ring of the $(2,1)$-forms with
its dependence on the complex structure moduli and the quantum
corrected cohomology ring of the $(1,1)$-forms with its dependence
on the complexified K\"ahler structure parameters.
The most convincing evidence for this part of the mirror
conjecture is the successful prediction of the numbers
of certain rational curves for the quintic in
\ref\cdgp{P. Candelas, X. De la Ossa, P. Green
     and L. Parkes, \npb359(1991)21}
and other manifolds with $h^{1,1}=1$ in
\ref\morrison{D. Morrison, {\it Picard-Fuchs Equations and
     Mirror Maps for Hypersurfaces}, in ref.\mirror}
\ref\ktone{A. Klemm and S. Theisen, \npb389(1993)153}
\ref\font{A. Font, \npb391(1993)358}
\ref\libtei{A. Libgober and J. Teitelbaum, {\it Duke Math. Jour.,
     Int. Math. Res. Notices} 1 (1993)29}
\ref\ktthree{A. Klemm and S. Theisen, {\it Mirror Maps and
     Instanton sums for Complete Intersections in Weighted
     Projective space}, preprint LMU-TPW 93-08},
which test mirror symmetry, at least locally in moduli space
in the vicinity of the point of maximal unipotent monodromy.

Further evidence for mirror symmetry at one loop
in string expansion was provided by the successful prediction
\ref\bcov{M. Bershadsky, S. Cecotti, H. Ooguri and C. Vafa,
          Nucl. Phys. B405 (1993) 279}
of the number of elliptic curves for the manifolds discussed
in \ktone\font.

{}From a mathematical point of view mirror symmetry
is so far not well understood. Some of the problems
have been summarized in
\ref\manin{Yu. I. Manin, {\it Report on the Mirror Conjecture},
     Talk at the Mathematische Arbeitstagung, Bonn 1993}.
The question of mirror symmetry for rigid manifolds
($h^{2,1}=0$), which is again obvious from the
conformal field theory point of view, has been discussed
in \ref\cdp{P. Candelas, E. Derrick and L. Parkes,
     Nucl. Phys. B407 (1993) 115}.

Aside from the mathematicians' interest in the subject,
mirror symmetry has turned out to be an indispensable
tool for e.g. the computation of Yukawa-couplings for
strings on Calabi-Yau spaces. This is a problem of prime
physical interest, so let us briefly review some aspects.
We will restrict ourselves to strings on
Calabi-Yau spaces corresponding to symmetric $(2,2)$
conformal field theories, since they are on the one hand,
due to their higher symmetry,
easier to treat than e.g. the more general $(2,0)$
compactifications, and on the other hand general
enough to allow for potentially phenomenologically
interesting models.

The Yukawa couplings between mass-less matter fields,
in the following characterized by their $E_6$
representation, fall into four classes, symbolically
written as $\bra27^3\ket$, $\bra\overline{27}^3\ket$,
$\bra27\cdot\overline{27}\cdot1\ket$ and $\bra 1^3\ket$.
Here $27$ and $\overline{27}$
refer to the charged matter fields which accompagnie,
via the (right-moving)
extended world-sheet superconformal symmetry, the
complex structure and K\"ahler structure moduli,
respectively. The singlets are neutral matter fields
related to $End({\cal T}_X)$.
Unlike the singlets, the charged matter fields can be naturally
identified as physical states in two topological
field theories, which can be associated to every  $N=2$
superconformal theory by twisting, as described in
\ref\topolone{T. Eguchi and S.K. Yang, \mpla5(1990)1693;
R. Dijkgraaf, H. Verlinde and E. Verlinde, \npb352(1991)59;
E. Witten : {\sl Mirror Manifolds And Topological Field
                       Theory}, in
{\it Essays on Mirror Manifolds} (ed. S.-T. Yau), Int. Press, Hong Kong,
1992 120-158}.
Here we will be concerned only with the
couplings in these topological subsectors.

The $\bra 27^3\ket$ Yukawas depend solely on the
complex structure moduli and do not receive contributions
{}from sigma model and string loops; in particular,
the tree level results are not corrected by world-sheet
instanton corrections. In contrast to this, the
$\bra\overline{27}^3\ket$'s
are functions of the parameters of the possible deformations
of the K\"ahler class only and do receive perturbative and
non-perturbative corrections
\ref\digr{J. Distler and B. Greene, \npb309(1988)295}.
This makes their direct computation,
which involves a world-sheet instanton sum, virtually impossible,
except for the case of $\ZZ_n$ orbifolds
\ref\dfms{L. Dixon, D. Friedan, E. Martinec and S, Shenker,
          \npb282(1987)13;
           S. Hamidi and C. Vafa,\npb279(1987)465;
           J. Lauer, J. Mas and H. P. Nilles  \npb351 (1991) 353;
           S. Stieberger, D. Jungnickel, J. Lauer and
           M. Spalinski, Mod. Phys. Lett. A7 (1992) 3859;
           J. Erler, D. Jungnickel, M. Spalinski and
           S. Stieberger, \npb397(1993)379
}.
These difficulties can be circumvented by taking advantage
of mirror symmetry. This was first
demonstrated for the quintic threefold in \cdgp
and subsequently
applied to other models with one K\"ahler modulus in refs.
\morrison\ktone\ktthree.
The idea is the following: in order to compute the
$\bra\overline{27}^3\ket$ Yukawa couplings on the CY manifold $X$,
one computes the $\bra27^3\ket$ couplings on its mirror
$X^*$ and then returns to $X$ via the mirror map
which relates the elements
$b_i^{1,1}(X)\in H^1(X^*,{\cal T}_{X^*}^*)
\sim H^{1,1}_{\bar\partial}(X^*)$
to the $b_i^{2,1}\in H^1(X,{\cal T}_X)\sim H^{2,1}_{\bar\partial}(X)$
and their corresponding deformation parameters
$t_i^*$ and $t_i$ ($i=1,\dots h^{1,1}(X)=h^{2,1}(X^*)$).

In the Landau-Ginzburg models one can
straightforwardly compute ratios of $\bra 27^3\ket$ Yukawa
couplings by reducing
all operators of charge three, via the equations of motion,
to one of them. This fixes the Yukawa couplings however only
up to a moduli dependent normalization.
Information about the Yukawa couplings can also be obtained
{}from the fact that the moduli space of the  $N=2$ theory has a natural
flat connection
\ref\LSW{B. Blok and A. Varchenko, Int. J. Mod. Phys. A7 (1992) 1467;
         W. Lerche, D.J. Smit and N. Warner, Nucl. Phys. B372 (1992) 87}
\ref\CF{A.C. Cadavid and S. Ferrara, Phys. Lett. 267B (1991) 193;
        S. Ferrara and J. Louis, Phys. Lett. 278B (1992) 240;
        A. Ceresole, R. D'Auria, S. Ferrara, W. Lerche
        and J. Louis, Int. J. Mod. Phys. A8 (1993) 79}
\ref\KST{ A. Klemm, S.  Theisen and M. G. Schmidt,
        Int. J. Mod. Phys. A7 (1992) 6215}.
The route we will follow, which was first used in \cdgp,
is especially adequate for models with an interpretation
as Calabi-Yau spaces.

In this procedure, the Picard-Fuchs equations, i.e. the
differential equations satisfied by the periods of the holomorphic
three form as a function of the complex
structure moduli, play a prominent role. They allow for the
computation of the $\bra27^3\ket$ Yukawa couplings
and furthermore, the mirror map can be constructed from their
solutions. This has been abstracted from the results of
\cdgp~in \morrison~and further applied in refs.\libtei\ktthree.
In this paper we develop a way of getting
the Picard-Fuchs equations for a class of models with
more than one modulus. This construction uses some results
{}from toric geometry, which are especially helpful to give
a general prescription for the mirror map.

The mirror map also defines the so-called special coordinates
on the K\"ahler structure moduli space. In these
coordinates the $\bra\overline{27}^3\ket$
Yukawa couplings on $X$ are simply the third
derivatives with respect to the moduli
of a prepotential from which the
K\"ahler potential can also be derived.
Whereas the left-moving
$N=2$ superconformal symmetry of (2,2) compactifications
is necessary for having
$N=1$ space-time supersymmetry, it is the additional
right-moving symmetry which is responsible for the
special structure
\ref\dkl{L. Dixon, V. Kaplunovsky and J. Louis,
         Nucl. Phys. B329 (1990) 27;
         S. Cecotti, S. Ferrara and L. Girardello,
         Int. J. Mod. Phys. A4 (1989) 2465}.

The paper is organized as follows. In Section 2 we describe
those aspects of toric geometry which are relevant for us
and give some illustrative examples of mirror pairs.
We also state the rules for computing topological couplings
using toric data.
In Section 3 we discuss the Picard-Fuchs
equations for hypersurfaces in weighted projective space
and show how to set them up.
Section 4 contains applications to two and three
moduli models.
We compute the Yukawa couplings and discuss
the structure of the solutions of the Picard-Fuchs equations.
In Section five we show how to find the appropriate variables
to describe the large complex structure limit and the mirror map.
In the last section we interpret our
results for the Yukawa couplings as the instanton corrected
topological coupling. We conclude with some observations
and comments.

\newsec{Toric Geometry: Mirror Pairs and Topological Couplings}

In this section we will describe the aspects of the
geometry of hypersurface (complete intersection) Calabi-Yau spaces,
which we need later to facilitate the derivation of the
Picard-Fuchs equation, and to define the mirror map on the level
of Yukawa couplings.
These types of Calabi-Yau spaces arise naturally
{}from the Landau-Ginzburg approach to two dimensional
$N=2$ superconformal theories \lvw\wlg.
The hypersurfaces with
$ADE$ invariants are related to tensor products of
minimal $N=2$ superconformal field theories.

Some important geometrical properties of these manifolds
are however easier accessible in the framework of toric geometry
\roan\batyrev.
We therefore want to give in the first part of this section a
description of Calabi-Yau hypersurfaces
in terms of their toric data.
We summarize the construction of mirror pairs of Calabi-Yau
manifolds given in \batyrev~and describe the map between
the divisors related to $(1,1)$-forms
and the monomials corresponding to the variation of the
complex structure and hence to the $(2,1)$-forms.
In the second part of this section we give the toric
data for manifolds with few K\"ahler moduli which we will
further discuss in later sections.
In section (2.3) we use the toric description to construct
the mirrors which were missing in \ks\kstwo.
In section (2.4) we summarize results for the topological triple
couplings of complete intersection manifolds
using toric geometry.
As they are the large radius limit of the $\bra\overline{27}^3\ket$
Yukawa couplings, we will need this information for the mirror
map.

\subsec{The families of Calabi-Yau threefolds}

Consider a (complete intersection) Calabi-Yau variety $X$ in
a weighted projective space $\IP^n(\vec w)=\IP^n(w_1,\cdots,w_{n+1})$
defined as the zero locus of transversal quasihomogeneous polynomials
$W_i \;(i=1,\cdots,m)$ of degree deg$(W_i)=d_i$ satisfying
$\sum_{i=1}^m d_i=\sum_{j=1}^{n+1} w_j$;
\eqn\cicy{
X=X_{d_1,\ldots,d_m}(\vec w)=
\{[z_1,\cdots,z_n]\in \IP^n(\vec w)\;\vert\;
W_i(z_1,\cdots,z_{n+1})=0 \;\; (i=1,\cdots,m) \;\} \;\; .
}
Due to the action $z_i\to\lambda^{w_i}z_i,\,\lambda\in\IC^*$,
whose orbits define points of $\IP^n(\vec w)$, the
weighted projective space has singular strata
${\cal H}_S=\IP^n(\vec w)\cap \{z_i=0\,\,\forall i\in
\{1,\ldots, n+1\}\setminus S\}$ if the subset
$\{\omega_i\}_{i\in S}$ of
the weights has a non-trivial common factor $N_S$.
We consider only well-formed hypersurfaces where
$X$ is called well-formed
if  $\IP^n(\vec w)$ is well-formed, i.e. if the weights of any set of
$n$ projective coordinates are relative prime
and if $X$ contains no codimension $m+1$
singular strata of $\IP^n(\vec w)$.
In fact, every projective space is isomorphic to a well formed
projective space and furthermore, one can show, using the explicit
criteria for transversality given in \ref\dk{D. Dais and A. Klemm:
{\sl On the Invariants of Minimal Desingularizations of
$3$-dimensional Calabi-Yau Weighted Complete
Intersection Varieties}, in preparation}, that transversality together
with $\sum_{i=1}^m d_i=\sum_{j=1}^{n+1} w_j$ already
implies well-formedness for $X_{d_1,\dots,d_m}$.

Hence the possible singular sets on $X$ are either
points or curves. For singular points these singularities are locally
of type $\IC^3/\ZZ_{N_S}$
while
the normal bundle of a singular curve has locally a
$\IC^2/\ZZ_{N_S}$ singularity. Both types of singularities
and their resolution can be described by methods of toric geometry.
The objects which we will be concerned with are families
of Calabi-Yau manifolds describable
in toric geometry, as explicated below.

To describe the toric variety $\IP_\Delta$, let us consider an
$n$-dimensional
convex integral polyhedron $\Delta\subset \IR^n$ containing the origin
$\nu_0=(0,\cdots, 0)$. An integral polyhedron is a polyhedron
whose vertices
are integral, and is called {\it reflexive} if its dual defined by
\eqn\dual{
\Delta^*=\{\;(x_1, \cdots, x_n) \;\vert\;
\sum_{i=1}^n x_i y_i \geq -1 \;{\rm for \; all \; }
(y_1, \cdots, y_n)\in \Delta \;\}
}
is again an integral polyhedron. Note if $\Delta$ is reflexive,
then $\Delta^*$
is also reflexive since $(\Delta^*)^*=\Delta$.
We associate to $\Delta$ a complete rational fan $\Sigma(\Delta)$
as follows:
For every $l$-dimensional face $\Theta_l\subset\Delta$ we define a
$n$-dimensional cone $\sigma(\Theta_l)$ by $\sigma(\Theta_l)
:=\{ \lambda(p'-p)
 \; \vert \; \lambda\in\IR_+ , p\in\Delta , p'\in \Theta_l\; \}$.
$\Sigma(\Delta)$ is then given as the collection of $(n-l)$-dimensional
dual cones $\sigma^*(\Theta_l) \; (l=0,\cdots, n)$ for all faces of
$\Delta$. The
toric variety $\IP_\Delta$ is the toric variety associated to the fan
$\Sigma(\Delta)$, i.e. $\IP_\Delta:=\IP_{\Sigma(\Delta)}$ (see
\ref\oda{ T. Oda, {\sl Convex Bodies and Algebraic Geometry,
              (An Introduction to the Theory of Toric Varieties}),
              {\sl Ergebnisse der Mathematik und ihrer Grenzgebiete},
              3.Folge, Bd. {\bf 15}, Springer-Verlag (1988);
              V.I. Danilov, Russian Math. Surveys, 33 (1978) 97}
for detailed constructions).

Denote by $\nu_i \; (i=0,\cdots,s)$ the integral points in $\Delta$ and
consider an affine space $\IC^{s+1}$ with coordinates
$(a_0, \cdots,a_{s})$.
We will consider the zero locus $Z_f$ of the Laurent polynomial
\eqn\laurentpolynomial{
f(a,X) = \sum_{i=0}^s a_i X^{\nu_i},\quad f(a,X)
\in \IC[X_1^{\pm 1}, \cdots, X_{n}^{\pm 1}]}
in the algebraic torus $(\IC^*)^n \subset \IP_\Delta$,
and its closure $\bar
Z_f$ in $\IP_\Delta$.
Here we have used the
convention $X^\mu:= X_1^{\mu_1}\ldots X_{n}^{\mu_{n}}$.

$f:=f_\Delta$ and $Z_f$ are called $\Delta$-regular if
for all $l=1,\dots n$
the $f_{\Theta_l}$ and $X_i{\p\over\p X_i}f_{\Theta_l},\,
\forall i=1,\dots n$ do not vanish simultaneously in
$(\IC^*)^n$. This is equivalent to the transversality condition
for the quasi-homogeneous polynomials $W_i$.
When we vary the parameters $a_i$ under the condition of
$\Delta$-regularity, we will have a family of
Calabi-Yau varieties.

The ambient space $\IP_{\Delta}$
and so $\bar Z_{f}$ are in general singular.
$\Delta$-regularity ensures that the only singularities of $\bar Z_{f}$
are the ones inherited from the ambient space.
$\bar Z_{f}$ can be resolved to a Calabi-Yau manifold $\hat Z_f$
iff $\IP_{\Delta}$
has only Gorenstein singularities, which is the case
iff $\Delta$ is reflexive \batyrev.

The families of  the Calabi-Yau manifolds $\hat Z_f$ will be denoted
by ${\cal F}(\Delta)$. The above definitions proceeds in exactly
symmetric way for the dual polyhedron $\Delta^*$ with its integral points
$\nu_i^*\;(i=0,\cdots,s^*)$.

In ref. \batyrev~ Batyrev observed that a pair of reflexive polyhedra
$(\Delta, \Delta^*)$ naturally gives us a pair of mirror
Calabi-Yau families $({\cal F}(\Delta),{\cal F}(\Delta^*))$ as
the following identities ($n\geq 4$) on the Hodge numbers
($(n-1)$ is the dimension of the Calabi-Yau space) hold
\eqn\hodgenumbers{
\eqalign{h^{1,1}(\hat Z_{f,\Delta})&=h^{n-2,1}(\hat Z_{f,\Delta^*})\cr
&=l(\Delta^*)-(n+1)-\sum_{{\rm codim}\Theta^*=1}
l^\prime(\Theta^*)
+\sum_{{\rm codim}\Theta^*=2 }
l^\prime(\Theta^*)l^\prime(\Theta)\cr
h^{1,1}(\hat Z_{f,\Delta^*})&=h^{n-2,1}(\hat Z_{f,\Delta})\cr
&=l(\Delta\,)-(n+1)-\sum_{{\rm codim}\Theta=1}l^\prime(\Theta)
+\sum_{{\rm codim}\Theta=2}l^\prime(\Theta)
l^\prime(\Theta^*).}
}
Here $l(\Theta)$ and $l^\prime(\Theta)$ are the number of integral
points on a face $\Theta$ of $\Delta$ and
in its interior, respectively
(and similarly for $\Theta^*$ and $\Delta^*$).
An $l$-dimensional face $\Theta$ can be represented by specifying
its vertices
${\rm v}_{i_1},\cdots,{\rm v}_{i_{k}}$. Then the dual face defined by
$\Theta^*=\{ x\in \Delta^*\; \vert \; (x,{\rm v}_{i_1})= \cdots
= (x,{\rm
v}_{i_k})=-1 \}$ is a $(n-l-1)$-dimensional face of $\Delta^*$.
By construction
$(\Theta^*)^*=\Theta$, and we thus have a
natural duality
pairing between $l$-dimensional faces of $\Delta$ and
$(n-l-1)$-dimensional
faces of $\Delta^*$.  The last sum in each of the
two equations in (2.4) is over pairs of dual faces.
Their contribution cannot be associated with a monomial
in the Laurent polynomial. In the language of
Landau-Ginzburg theories, if appropriate, they correspond to
contributions from twisted sectors. We will denote by
$\tilde h^{2,1}$ and $\tilde h^{1,1}$ the expressions
\hodgenumbers~ without the last terms.

Three dimensional Calabi-Yau hypersurfaces in $\IP^4(\vec w)$
were classified in \ks.
A sufficient criterion for the possibility to
associate to such a space a reflexive polyhedron is
that $\IP^n(\vec w)$ is Gorenstein,
which is the case if  ${\rm lcm}[w_1,\ldots,w_{n+1}]$
divides the degree $d$
\ref\belrtametti{M. Beltrametti and L. Robbiano,
Expo. Math. 4 (1986) 11.}.
In this case we can define a simplicial, reflexive polyhedron
$\Delta(\vec w)$ in terms of the weights, s.t. $\IP_\Delta(\vec w) \simeq
 \IP(\vec w)$.
This associated $n$-dimensional integral convex polyhedron
is the convex hull of the integral vectors $\mu$ of the exponents
of all quasihomogeneous monomials $z^\mu$ of degree $d$,
shifted by $(-1,\ldots,-1)$:
\eqn\polyhedron{
\Delta(\vec w):=
\{(x_1,\ldots, x_{n+1}) \in
\IR^{n+1}|\sum_{i=1}^{n+1} w_i x_i=0,x_i\geq-1\}.}
Note that this implies that the origin is the only
point in the interior of $\Delta$.

If the quasihomogeneous polynomial $W$ is Fermat, i.e. if it consists of
monomials $z_i^{d/w_i}\,\,(i=1,\cdots,5)$,
$\IP^4(\vec w)$ is clearly Gorenstein,
and $(\Delta,\Delta^*)$ are thus simplicial.
If furthermore at least one weight
is one (say $w_5=1$) we may choose
$e_1=(1, 0, 0, 0, -w_1)$,  $e_2 = (0, 1, 0, 0,-w_2)$,
$e_3=(0, 0, 1, 0, -w_3)$ and $e_4 = (0, 0, 0, 1,-w_4)$
as generators for $\Lambda$, the lattice induced
{}from the $\ZZ^{n+1}$ cubic
lattice on the hyperplane $H=\{(x_1,\ldots, x_{n+1}) \in
\IR^{n+1}|\sum_{i=1}^{n+1} w_i x_i=0\}$.
For this type of models we then always obtain as
vertices of $\Delta(\vec w)$
\eqn\verticesD{
\eqalign{
&\nu_1=(d/w_1-1, -1, -1, -1)  \; ,\; \cr
&\nu_4=(-1, -1, -1, d/w_4)   \; ,\;  \cr }
\eqalign{
&\nu_2=(-1, d/w_2-1, -1, -1)  \; ,\; \cr
&\nu_5=(-1,-1,-1,-1)                 \;  \cr}
\eqalign{
&\nu_3=(-1, -1, d/w_3-1, -1)  ,\cr
& \cr}
}
and for the vertices of the dual simplex $\Delta^*(w)$ one finds
\eqn\vertices{
\eqalign{
&\nu^*_1=(1,0,0,0)  \; ,\;
 \nu^*_2=(0,1,0,0)  \; ,\;
 \nu^*_3=(0,0,1,0)  \; ,\;
 \nu^*_4=(0,0,0,1)  \cr
&\nu^*_5=(-w_1,-w_2,-w_3,-w_4)}.
}
We can now describe the monomial-divisor mirror
map \ref\topologychange{D.Aspinwall, B.Greene and D.Morrison,
Phys.Lett. 303B (1993) 249} for these models.
Some evidence for the existence of such a map was given
by the computations in
\ref\alr{P. Aspinwall, A. L\"utken and G. Ross,
\plb241(1990)373; P. Aspinwall and A. L\"utken,
{\it A New Geometry from Superstring Theory}, in
\mirror.}. The subject was further developed in \roan\batyrev.
The toric variety $\IP_{\Delta^*(w)}$ can be
identified with
\eqn\Hfive{
\eqalign{
\IP_{\Delta^*(w)} &\equiv {\bf H}^5(\vec w) \cr
&=\{[U_0,U_1,U_2,U_3,U_4,U_5]\in\IP^6
|\prod_{i=1}^5 U_i^{w_{i}} = U^d_0 \},
}}
where the variables $X_i$ in eq. \laurentpolynomial~ are related to the
$U_i$ by
\eqn\toricident{
[1,X_1,X_2,X_3,X_4,{1\over \prod_{i=1}^4 X_i^{w_{i}}}] =
[1,{U_1 \over U_0},{U_2 \over U_0},{U_3 \over U_0},
   {U_4 \over U_0},{U_5 \over U_0}].}
Let us consider the etale mapping
$\phi:\IP^4(\vec w)\rightarrow {\bf H}^5(\vec w)$ given by
\eqn\etalemap{
[z_1,z_2,z_3,z_4,z_5] \mapsto [z_1z_2z_3z_4z_5,
z_1^{d/w_1},z_2^{d/w_2},z_3^{d/w_3},z_4^{d/w_4},z_5^{d/w_5}].}

In toric geometry, this etale mapping replaces the orbifold
construction for the mirror manifolds described in \gp.
Furthermore, the integral
points in $\Delta^*(\vec w)$ are mapped to
monomials of the homogeneous
coordinates of $\IP^4(\vec w)$ by
\eqn\monomialdivisor{
\mu=(\mu_1,\mu_2,\mu_3,\mu_4)\mapsto\phi^*(X^{\mu} U_0)=
{\left(\prod_{i=1}^5 z_i\right)^{(1+\sum_{i=1}^4 \mu_i)}
\over \prod_{i=1}^4 z_i^{\mu_{i} d/w_i }}.}
Since in toric geometry the integral points inside $\Delta^*(\vec w)$
describe the exceptional divisors which are introduced
in the process of the resolution of
the toric variety $\IP_{\Delta(\vec w)}$, the map \monomialdivisor~
is called the monomial-divisor map.

\subsec{Models with few moduli}
We are interested in studying systems with few K\"ahler moduli.
For Fermat hypersurfaces in $\IP^4(\vec w)$ we find five two moduli
systems\foot{In addition, five non-Fermat examples can be found in \ks.}.
In table 1 we display these models, their Hodge numbers,
the points on faces of dimensions one and two of $\Delta^*$ and the
face $\Theta^*$ these points lie on, specified by its
vertices.
Points lying on a one-dimensional edge
correspond to exceptional divisors over singular curves
whereas the points lying in the interior of two-dimensional
faces correspond to exceptional divisors over singular points
(cf. sec. 2.3 below).
There is also always one
point in the interior, $\nu_0^*=(0,0,0,0)$, corresponding to the
canonical divisor of $\IP^4(\vec w)$ restricted to $\hat X$.
We also give the exceptional divisor $E$ and the $G$-invariant
monomial $Y$ related to it via the monomial-divisor mirror map.
Here $G$ is the
group which, by orbifoldization, leads to the mirror configuration.
Its generators $g^{(k)}=(g^{(k)}_1,\ldots,g^{(k)}_{n+1})$ act by
\eqn\gaction{
g: z_i\mapsto\exp\left(2\pi i g_i{w_i\over d}\right)z_i
}
on the homogeneous coordinates of $X_d(\vec w)$. Note that
this action has always to be understood modulo the
equivalence relation $z_i\sim \lambda^{w_i} z_i$. For Fermat
hypersurfaces $G$ consists of all $g^{(k)}$  with
$\sum_{i=1}^5 g^{(k)}_i w_i/d=1$.
The generators of $G$ are
also displayed in the table. Here we have suppressed
$g^{(0)}=(1,1,1,1,1)$, which is present in all cases.
The first four models of the table have a singular $\ZZ_2$ curve $C$
and the exceptional divisor is a ruled surface which is locally
$C\times\IP^1$. The last example has a $\ZZ_3$ singular point
blown up into a $\IP^2$.
$$\vbox{
{\tenpoint{
{\offinterlineskip\tabskip=0pt
\ni{\bf Table 1:} Hypersurfaces in $\IP^4(\vec w)$ with $h^{1,1}=2$
\halign{\strut\vrule#&~#~&\vrule#&\hfil$#$~&\vrule#&\hfil$#$~&\vrule#
&\hfil$#$~&\vrule#&\hfil$#$~&\vrule#&\hfil $#$~&\vrule#\cr
\noalign{\hrule}
& &&X_{8}(2,2,2,1,1)
&&X_{12}(6,2,2,1,1)
&&X_{12}(4,3,2,2,1)
&&X_{14}(7,2,2,2,1)
&&X_{18}(9,6,1,1,1)& \cr
\noalign{\hrule}
&$h^{1,1}$
&& 2 (0)
&& 2 (0)
&& 2 (0)
&& 2 (0)
&& 2 (0) & \cr
&$h^{2,1}$
&& 86(2)
&& 128(2)
&& 74(4)
&& 122(15)
&& 272(0)  & \cr
&$\nu^*_6$
&& (-1,-1,-1,0)
&& (-3,-1,-1,0)
&& (-2,-1,-1,-1)
&& (-3,-1,-1,-1)
&& (-3,-2,0,0) & \cr
&$\Theta^*$
&& (1,2,3)
&& (1,2,3)
&& (1,3,4)
&& (2,3,4)
&& (1,2) & \cr
&$E$
&& C\times \IP^1
&& C\times \IP^1
&& C\times \IP^1
&& C\times \IP^1
&& \IP^2 & \cr
&$Y$
&& z_4^4 z_5^4
&& z_4^6 z_5^6
&& z_2^2 z_5^6
&& z_1   z_5^7
&& z_3^6 z_4^6 x_5^6  & \cr
\noalign{\hrule}
&$G$
&& (0,0,0,7,1)
&& (0,0,0,11,1)
&& (0,0,0,5,2)
&& (0,0,0,6,2)
&& (0,0,0,17,1)  & \cr
&
&& (0,0,3,0,2)
&& (0,0,5,0,2)
&& (0,0,5,0,2)
&& (0,0,6,0,2)
&& (0,0,17,0,1)  & \cr
&
&& (0,3,0,0,2)
&& (0,5,0,0,2)
&&
&&
&&& \cr
\noalign{\hrule}}}}}}
$$
The Hodge numbers are in accordance with the formulas
for the invariants of twisted Landau-Ginzburg models \lvw~ or the
counting of chiral primary fields in the $A$-series $N=2$ superconformal
minimal tensor product models\foot{The first model in table 1
corresponds to a tensor product of five minimal $N=2$
superconformal $A$-models at levels $(2,2,2,6,6)$.
If one replaces the two level 6 $A$-models by level 6
$D$-models, the spectrum and the couplings of the chiral states
does not change. Geometrically the latter model corresponds
to a complete intersection of $p_1=\sum_{i=1}^5 z_i^4$
and $p_2=z_4 z_6^2+z_5 z_7^2$ in $\IP^4\times\IP^1$. It would be
interesting to see how these two geometrical constructions
are related.}.
Contributions which come from
the last terms in \hodgenumbers~correspond to twisted vacua
in the CFT or Landau-Ginzburg approach. Their contribution
to $h^{1,1},h^{2,1}$ is indicated in parantheses;
e.g. in the $X_{14}(7,2,2,2,1)$ model we
have $l^\prime(\Theta(2,3,4))\cdot l^\prime(\Theta^*(1,5))=1\cdot 15$
states from the twisted sector.
Similarly, for the five three moduli the data are collected in
table 2.
$$\vbox{
{\tenpoint{
{\offinterlineskip\tabskip=0pt
\ni{\bf Table 2:} Hypersurfaces in $\IP^4(\vec w)$ with $h^{1,1}=3$
\halign{\strut\vrule#&~#~&\vrule#&\hfil$#$~&\vrule#&\hfil$#$~&\vrule#
&\hfil$#$~&\vrule#&\hfil$#$~&\vrule#&\hfil$#$~&\vrule$#$\cr
\noalign{\hrule}
& &&X_{12}(6,3,1,1,1)
&&X_{12}(3,3,3,2,1)
&&X_{15}(5,3,3,3,1)
&&X_{18}(9,3,3,2,1)
&&X_{24}(12,8,2,1,1)& \cr
\noalign{\hrule}
&$h^{1,1}$
&& 3 (1)
&& 3 (0)
&& 3 (0)
&& 3 (0)
&& 3 (0) & \cr
&$h^{2,1}$
&& 165(0)
&& 69(6)
&& 75(12)
&& 99(4)
&& 243(0)  & \cr
&$\nu^*_6$
&& (-2,-1,0,0)
&& (-1,-1,-1,0)
&& (-1,-1,-1,-1)
&& (-3,-1,-1,0)
&& (-3,-2,0,0) & \cr
&$\nu^*_7$
&& {\rm twisted\,sector}
&& (-2,-2,-2,-1)
&& (-3,-2,-2,-2)
&& (-6,-2,-2,-1)
&& (-6,-4,-1,0) & \cr
&$\Theta^*$
&& (1,2)
&& (1,2,3)
&& (2,3,4)
&& (1,2,3)
&& (1,2,3),(1,2) & \cr
&$E$
&& \IP^2,\IP^2
&& C\times(\IP^1\wedge \IP^1)
&& C\times(\IP^1\wedge \IP^1)
&& C\times(\IP^1\wedge \IP^1)
&& C\times\IP^1,\Sigma_2 & \cr
&$Y$
&& z_3^4 z_4^4 z_5^4, -
&& z_4^4 z_5^4, z_4^2 z_5^8
&& z_1^2 z_5^5, z_1 z_5^{10}
&& z_4^6 z_6^6, z_4^3 z_5^{12}
&& z_3^6 z_4^6 z_5^6, z_4^{12} z_5^{12} &\cr
\noalign{\hrule}
&$G$
&&(0,0,0,11,1)
&&(0,0,0,5,2)
&&(0,0,0,4,3)
&&(0,0,5,0,3)
&&(0,0,0,23,1)&\cr
&
&& (0,0,11,0,1)
&& (0,0,3,0,3)
&& (0,0,4,0,3)
&& (0,5,0,0,3)
&& (0,0,11,0,2)&\cr
&
&&
&& (0,3,0,0,3)
&&
&&
&&&\cr
\noalign{\hrule}}}}}}
$$
The first model in table 2
has two singular $\ZZ_3$ points which are each blown up
to a $\IP^2$. The second through the fourth models have singular
$\ZZ_3$ curves for which the exceptional divisor is a ruled
surface which is locally the product of the curve $C$ and
a Hirzebruch-Jung {\it Sph\"arenbaum}. The last model has a singular
$\ZZ_2$ curve with an exceptional $\ZZ_4$ point which is blown up
to a Hirzebruch surface $\Sigma_2$.

Finally we list a class of models whose K\"ahler moduli stem from
non-singular ambient spaces, the product of ordinary projective spaces.
The simplest model in this class is the bi-cubic
in $\IP^2\times \IP^2$ whose defining equation is
\eqn\bicubic{
(z_1^3+z_2^3+z_3^3)w_1w_2w_3 +
z_1z_2z_3(w_1^3+w_2^3+w_3^3) =0 \;\; ,
}
where $z_1,z_2,z_3$ and $w_1,w_2,w_3$ are homogeneous
coordinates for each $\IP^2$, respectively.
We write the family of this type as
$X_{(3|3)}({1,1,1}|{1,1,1})$. In table 3 we list all Calabi-Yau
hypersurfaces of this type, together with their Hodge numbers.
$$\vbox{
{\tenpoint{
{\offinterlineskip\tabskip=0pt
\ni {\bf Table 3:} Hypersurfaces in products of projective spaces\hb
\halign{\strut\vrule#&~#~&\vrule#&\hfil$#$~&\vrule#&\hfil$#$~&\vrule#
&\hfil$#$~&\vrule#&\hfil$#$~&\vrule#\cr
\noalign{\hrule}
&
&&X_{(3|3)}(_{1,1,1}|_{1,1,1})
&&X_{(4|2)}(_{1,1,1,1}|_{1,1})
&&X_{(3|2|2)}(_{1,1,1}|_{1,1}|_{1,1})
&&X_{(2|2|2|2)}(_{1,1}|_{1,1}|_{1,1}|_{1,1})
    & \cr
\noalign{\hrule}
&$h^{1,1}$
&& 2
&& 2
&& 3
&& 4
  & \cr
&$h^{2,1}$
&& 83
&& 86
&& 75
&& 68& \cr
\noalign{\hrule}}}}}}
$$
The polyhedra associated to these models are the direct product of the
polyhedra which describes each projective space, e.g., for the bi-cubic
model it is given by $\Delta(1,1)\times\Delta(1,1) \in \IR^4$.

We will see in section five that these kinds of non-singular Calabi-Yau
manifolds will provide good examples for which one can compare
the instanton expansions with calculations in algebraic geometry
\ref\BatyrevStraten{V. Batyrev and D. van Straten, {\it Generalized
Hypergeometric functions and Rational Curves on Calabi-Yau Complete
Intersections in Toric Varieties}, Essen preprint, 1993.}.

Related few moduli models can be obtained by passing to
products of weighted projective spaces, such as e.g.
$X_{(4|3)}(2,1,1|1,1,1)$ with $h^{2,1}=75$ and
$h^{1,1}=3$.
For details about complete intersections in products of ordinary
projective spaces we refer to ref.\cdls.

\subsec{\it Reflexive Polyhedra for Calabi-Yau hypersurfaces
in non-Gorenstein $\IP^n(\vec w)$ }

Let us now consider examples of Calabi-Yau hypersurfaces in $\IP(\vec w)$
for which the ambient space is non-Gorenstein. We will show that
$\Delta(\vec w)$ defined in \polyhedron~is reflexive also for
these spaces.
We claim an isomorphism between $\hat X(\vec w)$ and
$\hat Z_{f_{\Delta(\vec w)}}$,
indicated by the fact that the Newton polyhedra of the constraints
are isomorphic and the Hodge numbers coincide.
Passing to $\hat Z_{f_{\Delta^*(\vec w)}}$ we obtain
a mirror configuration. The relation between $ X(\vec w)$ and
$Z_{f,\Delta(\vec w)}$ is that the latter is a partial resolution,
namely of the non-Gorenstein singularities, of the former.

The manifold which we treat as an example, appears in the
classification of ref. \ks. Its mirror manifold can however not be
constructed using the methods of \bh~ nor as an abelian
orbifold w.r.t. symmetries of the polynomials of the
models in \ks. We consider the
following hypersurface in $\IP^4(\vec w)$
\eqn\nonfermat{
z_{1}^{25}+z_{2}^{8}z_{1}+z_{3}^{3}z_{5}+
z_{4}^{3}z_2+z_{5}^{3}z_1+z_5^2 z_2^3=0\in\IP^4(3,9,17,22,24).
}
One can choose the
generators of $\Lambda$ as
$e_1 = (-8, 0, 0, 0, 1)$,  $e_2 = (-17, 0, 3, 0,0)$,
$e_3=(-13, 0, 1, 1, 0)$ and $e_4=(-3, 1, 0, 0, 0)$ . In this basis
the 10 vertices of $\Delta(\vec w)$, which has $33$ integral points, are
$$
\eqalign{
&\nu_1=(-1, -1, 2, -1 )  \; ,\;
 \nu_2=(-1, -1, 2, 0 )  \; ,\;
 \nu_3=(-1, 0, -1, -1 )  \; ,\;
 \nu_4=(-1, 0, -1, 7 )  \cr
&\nu_5=(-1, 0, 0, 3 )  \; ,\;
 \nu_6=(-1, 1, -1, -1 ) \; ,\;
 \nu_7=(-1, 1, -1, 1 ) \; , \;
 \nu_8=(0, 1, -1, -1 )  \cr
&\nu_9=(1, 0, -1, 2 )  \; ,\;
 \nu_{10}=(2, 0, -1, -1 ) \cr
}\,.
$$
The dual polyhedron $\Delta^*(\vec w)$ with $44$ integral points
has the following 12 vertices
$$\eqalign{
&\nu^*_1=(-9, -18, -14, -3   )  \; ,\;
 \nu^*_2=(-8, -17, -13, -3   )  \; ,\;
 \nu^*_3=(-5, -11, -8, -2   )  \cr
& \nu^*_4=(-5, -10, -8, -2   )  \; ,\;
 \nu^*_5=( -3, -6, -5, 0  )  \; ,\;
 \nu^*_6=(-2, -7, -5, -1   ) \; ,\;
 \nu^*_7=(-2, -6, -4, -1   ) \cr
& \nu^*_8=(0, -3, -2, 0   ) \; ,\;
 \nu^*_9=(0, 0, 0, 1   )  \; ,\;
 \nu^*_{10}=(0, 0, 1, 0   ) \; ,\;
 \nu^*_{11}=(0, 3, 1, 0   ) \; , \;
 \nu^*_{12}=(1, 0, 0, 0   )  \cr
}$$
In table 4 we list
the numbers $l(\Theta)$ of lattice points on the faces of
dimension $0,\ldots, 4$.
$$\vbox{
{\tenpoint{
{\offinterlineskip\tabskip=0pt
\ni{\bf Table 4:} Toric data for hypersurface in
$\IP^4(3,9,17,22,24)$\hb
\halign{\strut\vrule#&~$#$~&\vrule#&~$#$~&\vrule#&~$#$~&\vrule#
&~$#$~&\vrule#&~$#$~&\vrule#&~$#$~&\vrule#\cr
\noalign{\hrule}
&&\multispan{3}$\hfil\Delta(\vec w)$\hfil
&&&&\multispan{3}$\hfil\Delta^*(\vec w)$\hfil&& \cr
\noalign{\hrule}
&{\rm dim}\Theta&&l(\Theta)&&{\Theta}
&&{\rm dim}\Theta^*&&l(\Theta^*)&&{\Theta^*}&\cr
\noalign{\hrule}
&4&&1&& &&4&&1&& &\cr
\noalign{\hrule}
&0&&10&& &&3&&4&& &\cr
\noalign{\hrule}
&1&&1&&(8, 10, 12)&&2&&0&&(6,7)&\cr
& &&2&&(9, 10, 11)&& &&0&&(3,10)&\cr
& &&7&&(10, 11, 12)&& &&0&&(3,4)&\cr
& &&0&&(5,8,11,12)&& &&3&&(1,2)&\cr
& &&0&&(1,2,5,6,8)&& &&3&&(2,8)&\cr
& &&0&&(1,5,11)&& &&1&&(2,10)&\cr
& &&0&&(1,5,9,10)&& &&3&&(8,10)&\cr
& &&0&&(1,10,11)&& &&1&&(9,10)&\cr
\noalign{\hrule}
&2&&7&&(10,11)&&1&&2&&(3,4,9,10)&\cr
& &&1&&(9,12)&& &&0&&(1,3,6)&\cr
& &&0&&(5,8)&& &&2&&(1,2,8)&\cr
& &&0&&(5,11)&& &&2&&(1,2,10)&\cr
& &&0&&(4,12)&& &&1&&(2,4,5)&\cr
& &&0&&(2,6)&& &&1&&(2,5,8)&\cr
& &&0&&(1,5)&& &&2&&(2,8,10)&\cr
& &&0&&(1,11)&& &&2&&(2,9,10)&\cr
& &&0&&(8,10)&& &&2&&(6,7,8)&\cr
& &&0&&(1,10)&& &&2&&(8,9,10)&\cr
\noalign{\hrule}
&3&&4&& &&0&&12&& &\cr
\noalign{\hrule}}}}}}
$$
\ni
For dim$\Theta=1$ and 2 we also indicate on which edges the points lie
and specify the corresponding two-dimensional dual faces of
$\Delta^*$.
Applying now eq. \hodgenumbers~we obtain
$h^{1,1}(\hat Z_{f_\Delta})=h^{2,1}(\hat Z_{f_{\Delta^*}})=35$
and
$h^{2,1}(\hat Z_{f_{\Delta}})=h^{1,1}(\hat Z_{f_{\Delta^*}})=38$.
As $\IP_{\Delta^*}$ is Gorenstein while $\IP(\vec w)$ is not, we
see a difference in the structure of the singularities,
i.e. not all exceptional divisors which correspond to curve
and point singularities on $X_d(\vec w)$ in $\IP(\vec w)$
are represented by points on faces of dimension one and two
in $\Delta^*$. The mirror of the manifold \nonfermat~ is
the hypersurface $\hat Z_{f_{\Delta^*}}$ in $\IP_{\Delta^*}$.

We have looked at a large number (several thousand)
of  models which appear in
the lists of refs.\ks\kstwo~including especially those for
which no mirrors could be found, even
after considering all abelian orbifolds\foot{We thank M. Kreuzer
for providing a
list of these manifolds.}, and verified that
they always lead to reflexive polyhedra
and that thus the corresponding $\IP_{\Delta^*}$ is
Gorenstein. This in particular entails that one can explicitly
construct all mirrors for these manifolds as hypersurfaces
in $\IP_{\Delta^*}$.
A general combinatorial proof that quasi-smoothness and vanishing
first Chern class of $X_d(\vec w)$ are equivalent to
reflexivity of $\Delta(\vec w)$, will be published elsewhere.
It has however been shown in ref.
\ref\TY{T. Yonemura, Tohoku Math. J. 42 (1990) 351}
that a reflexive polyhedron in three dimensions can be
associated to every $K_3$ hypersurface in $\IP^3(\vec w)$.

\subsec{\it Topological triple couplings}

We now want to give a recipe of how to compute topological triple
couplings
or intersection numbers of divisor classes
on the CY three-fold $\hat X$, which is the global minimal
desingularization $\pi:\hat X \rightarrow X$ of
$X=X_{d_1,\ldots,d_m}(\vec w)$ defined in ~\cicy~.
Proofs can be found in \oda,
   \ref\roancoup{S.-S. Roan:
   {\sl Topological Couplings of Calabi-Yau Orbifolds},
   {\sl Max-Planck-Institut Series}, No. 22 (1992),
   to appear in {\sl J. of Group Theory in Physics}}
and \dk. A related application to orbifolds
of tori is discussed in
\ref\MOP{D. Markushevich, M. Olshanetsky and A. Perelomov,
      Commun. Math. Phys.  111 (1987) 247}.
If ${\cal H}_S$ is a singular stratum of $\IP^n(\vec w)$, we denote
by $M\subset \{1,\ldots,m\}$
the subset which consists of the indices of
those defining polynomials $W_j$ which
do not vanish identically on ${\cal H}_S$. The singular
sets ${\cal S}_S$ on $X$ can be described as
$X_{\{d^\prime_j\}_{j\in M}} (\{w^\prime_i\}_{i\in S})$ (the relation
between $w'_i,d'_j$ and $w_i,d_j$ is explained below).
Their dimension is $|S|-|M|-1$ and, as mentioned before, only points
and curves occur. ${\cal S}_S$ is a weighted projective
space ($|M|=0$), a hypersurface ($|M|=1$) or a complete
intersection ($|M|>1$) in a weighted projective space.

For singular points we distinguish between isolated points and
exceptional points; the latter are singular points on singular
curves or the points of intersection of singular curves where
the order of the isotropy group $I$
of the exceptional points is higher than that of the
curve.

For the singular sets we get, through the process of blowing up,
exceptional divisors which are K\"ahler.
We use the following notation:
$D_i$ and $E_j$ denote the exceptional divisors on $\hat X$
coming from the resolution of the singular curves
and points, respectively.
$J$ is the divisor on $\hat X$ associated to the generating
element of ${\rm Pic}(X)$, cf.
\ref\dolgachev{I. Dolgachev, {\sl Weighted Projective Varieties.}
                In: {\sl Group Actions and Vector Fields},
                Proc. Polish - North Amer. Sem. Vancouver (1981)
                LNM Vol. {\bf 965}, Springer-Verlag (1982) 34}.

Each irreducible exceptional divisor provides,
by Poincar\'e duality, a harmonic $(1,1)$ form, which
we will denote by $h_J,h_E$ and $h_D$.
$h^{1,1}(\hat X)$
is $\#$ exceptional divisors $+1$.
The topological triple couplings are then given as
e.g. $E_i\cdot D_j\cdot J=\int_{\hat X}
h_{E_i}\wedge h_{D_j}\wedge h_J$.

In toric geometry the topological data
of singular points are represented by a three-dimensional
lattice and a simplicial cone defined by three lattice vectors
{}from which, however, the lattice points within the cone cannot
all be reached as linear combinations with positive integer
coefficients. For Abelian singularities of type $\IC^3/\ZZ_{N_S}$
the local desingularization process consists of adding further
generators such that this becomes possible. This corresponds to a
subdivision of the cone into a fan. The endpoints of the vectors
generating the fan all lie on a plane,
called the trace $\Delta_{}$ of the
fan. This is a consequence of the fact that the isotropy
group of singular points is a subgroup of $SU(3)$, necessary for
having a trivial canonical bundle on $\hat X$.
The exceptional divisors are thus in 1-1 correspondence with
lattice points in $\Delta_{Tr}$, whose location is given by
$$
{\cal P}=\left\lbrace\sum_{i=1}^3\vec e_i {n_i\over N_S}\left|
(n_1,n_2,n_3)\in\ZZ^3,
\pmatrix{e^{2\pi i {n_1\over N_S}}&&\cr&e^{2\pi i {n_2\over N_S}}
&\cr&&e^{2\pi i {n_3\over N_S} }\cr}\right.
\in I,\sum_{i=1}^3 n_i=N_S,n_i\geq 0\right\rbrace
$$
Here elements of $I$ describe the action of the isotropy
group on the coordinates of the normal bundle of the singular
point and $\vec e_1,\vec e_2,\vec e_3$ span an equilateral triangle
{}from its center.

For an isolated singular point there are only points
in the interior of the
triangle, whereas for an exceptional singular point there are also
points on its edges, corresponding to the exceptional divisors
that arise from resolving the curves on which the point lies
\foot{Not all exceptional divisors have a
toric description, only $\tilde h^{1,1}$ of them do.
The remaining ones cannot be trated by the methods
outlined here.}.
If an
exceptional point is the intersection of two or three curves,
there will be points on two or three sides of the triangle.
For points on
curves with $A_{N_S - 1}$-type $\IC^2/\ZZ_{N_S}$ singularity, there
are $N_S-1$ points on a side of the triangle.
The possible\foot{Not all triangulations lead
to a projective algebraic desingularization,
see \dk~for local criteria.}
triangulations of $\Delta_{Tr}$ with its points in the interior,
on the edges and
its three vertices, correspond to the different
desingularisations on which some intersection numbers
will depend. The number of triangles into which the trace
is subdivided is equal to $N_S$, the order of the
isotropy group.
Let us now discuss the various possibilities in turn.

\vskip0.3cm
\noindent
{\bf (A)}:
$$
J^3={\prod_{j=1}^m d_j\over\prod_{i=1}^{n+1} w_i}n_0^3\,
\qquad (n-m=3\,{\rm for\,\,threefolds})
$$
where
$n_0$ is the least common multiple of the orders $N_S$ of the
isotropy groups of all singular points, e.g.
for a manifold given by a single constraint of
Fermat type, this is the least common multiple of the
common factors of all possible pairs of weights.

\vskip0.3cm
\noindent
{\bf (B)}:
The action of the isotropy group on the fibers of the normal bundle to
curves with an $A_{N_S-1}$ singularity
is generated by $g={\rm diag}(\alpha,\alpha^{N_S-1})$ where
$\alpha=e^{2\pi i/(N_S)}$.
Resolving these singular curves adds
$N_S-1$ exceptional divisors which are $\IP^1$ bundles over
the curves $C$.
For the intersection numbers one finds \dk \hb
\item{$(a)$:}
$$
D_i\cdot D_j\cdot D_k=0\,\qquad{\rm for}\,\,i\neq j\neq k\neq i
$$
\item{$(b)$:}
$$\eqalign{
D_{j-1}^2\cdot D_j&=\psi(\sigma(j-N_S+1);\vec w^\prime;
\vec d^\prime)-\half\chi_C\cr
D_{j}^2\cdot D_{j-1}&=\psi(\sigma(N_S-j);\vec w^\prime;
\vec d^\prime)-\half\chi_C\cr
D_i^2\cdot D_j&=0\quad{\rm for}\quad|i-j|>1}
$$
Here $\chi_C$ is the Euler number of the singular curve
$X_{\{d^\prime_j\}_{j\in M}} (\{w^\prime_i\}_{i\in S})$, embedded in a
well-formed weighted projective space, i.e.
$w_i^\prime=w_i/m_i$ and $d_j^\prime=d_j/m$ where
$m={\rm lcm}(\{c_j\}_{j\in S})$,
$m_i={\rm lcm}(\{c_j\}_{j\in S\setminus \{i\}})$
and $c_i={\rm gcd}(\{w_j\}_{j\in S\setminus \{i\}})$.
Since ${\rm gcd}(w_i,c_i)=N_S$, there exist, for all $n\in\ZZ$, two
integers $a_i(n)$ and $b_i(n)$, such that $N_S n=a_i(n) w_i+b_i(n)c_i$
with $0\le a_i(n) < c_i/N_S$.
We then define
$$
\sigma(n)={N_S n-\sum_{i\in S} a(n)_i w_i
\over m}
$$
The function $\psi(n;\vec w^\prime;\vec d^\prime)$ is defined to be
$$
\psi(n;\vec w^\prime;\vec d^\prime)=\phi(n;\vec w^\prime;\vec d^\prime)
-\phi(\sum d^\prime_i-\sum w^\prime_j-n;\vec w^\prime;\vec d^\prime)
$$
where
$$
\phi(n;\vec w;\vec d)={1\over n!}{d^n\over dx^n}\left.{\prod(1-x^{d_i})
\over
\prod(1-x^{w_i})}\right|_{x=0}
$$
with $\phi(0;\vec w;\vec d)=1$ and $\phi(n;\vec w;\vec d)=0$
for $n<0$.
\item{$(c)$:}
$$
D_i^3=
\cases{4\chi_C\,\,&for $C$ without exceptional points\cr
4\chi_C-\sum_{j=1}^r(l^i_j-1) \,\,
             &for $C$ with exceptional points.\cr}
$$
As for the second contribution for curves with exceptional
points, we recall that each
exceptional divisor $D_i$ over $C$ corresponds to a point $P_{ij}$
on the side of the triangle belonging to
the $j$-th exceptional point over $C$. Now $r$ is the total number
of exceptional points over $C$ and $l^i_j$ are the number of
links between the point $P_{ij}$ and other points of the
$j$-th triangle which do not lie on the same side.
\hb
\item{$(d)$:}
$$
J^2\cdot D=0
$$
\item{$(e)$:}
$$
J\cdot D_j^2=-{2\over N_S}\left
(\psi(\sigma(n_0);\vec w^\prime;
\vec d^\prime)-\half\chi_C\right)
$$
\item{$(f)$:}
$$
D_i\cdot D_j\cdot J=\cases{{1\over N_S}\left
(\psi(\sigma(n_0);\vec w^\prime;
\vec d^\prime)-\half\chi_C\right)\,{\rm for}\,|i-j|=1 \cr
0\quad{\rm otherwise}\cr}
$$

\vskip0.3cm
\noindent
{\bf (C)} For the intersection of the divisors resulting from the
resolution of singular points, one obtains \oda~ \hb
\item{$(a)$:}
$$
E_i^3=12-\xi_i
$$
where $\xi_i$ is the number of triangles which have the point
$v_i$ corresponding to $E_i$ as a vertex.
\item{$(b)$:} $E_i^2\cdot E_j\neq0$ iff the points $v_i,v_j$ belong
to a common 2-simplex. If $u$ and $u^\prime$ are the two unique
additional points such that $\langle v_i,v_j,u\rangle$ and
$\langle v_i,v_j,u^\prime\rangle$ are 2-simplices, then
we have the relation
$$
(E_i^2\cdot E_j)v_i+(E_i\cdot E_j^2)v_j+u+u^\prime=0
$$
{}from where we can determine the intersection numbers.\hb
\item{$(c)$:} $E_i\cdot E_j\cdot E_k=1$ ($i\neq j\neq k\neq i$)
if $\langle v_i,v_j,v_k\rangle$
is a two-simplex; these couplings vanish otherwise.\hb
\item{$(d)$:}
$$
J^2\cdot E_i=J\cdot E_i^2=J E_i E_j=0
$$

\vskip0.3cm
\noindent
{\bf (D)}: What is left are $(a)$
the intersections between $E_i$ and $D_j$ and
the intersection of divisors over different but intersecting curves.
These cases are again easily described in
terms of the toric diagram and do in fact follow from
$({\bf C}(b))$, where the points $v_i,v_j$ may now also
lie on the sides of the triangle, in which case they represent
exceptional divisors over the curve. And
$(b)$ $E D J = 0$.

Let us finally discuss some examples:
Consider the two-moduli model $X_8(2,2,2,1,1)$.
The singular set consists of one singular $A_1$ curve
$C=X_4(1,1,1)$ which is already well-formed,
i.e. $\sigma(n)=n$. Its
isotropy group is $\ZZ_2$, and $\chi_C=-4$.
Also, $n_0=2$ and one easily computes
$\psi(2;1,1,1;4)=6$. We can then collect all triple intersections,
using an obvious notation, in the form $K^0=8J^3-8JD^2-16D^3$.

For the hypersurface $X_{24}(12,8,2,1,1)$ the singular sets
are an $A_1$ curve $C=X_{12}(6,4,1)\simeq X_6(3,2,1)$
with an exceptional $\ZZ_4$ point
$P=X_6(3,2)\simeq X_1(1,1)$.
Here $n_0=4$ and applying (${\bf A}$) gives $J^3$=8.
The points in $\Delta_{Tr}$ are $v=(1,0,0),u=(0,1,0),u'=(0,0,1),
v_E=({1\over 2},{1\over 4},{1\over 4}),v_D=(0,{1\over 2},{1\over 2})$,
i.e.
three corners, one internal point and one point
on the edge, the latter corresponds to the exceptional divisor $D$ of the
resolution of the $A_1$ singular curve. $\chi(C)=0$ and by (${\bf B}(c))$
we have $D^3 = 0$. Furthermore, $\sigma(4)=2$ and $\psi(2;3,2,1;6)=2$.
The unique triangulation of $\Delta_{Tr}$ consists of
four triangles with common point $v_E$.
Applying (${\bf B}(b)$),
($\bf C(a,b)$) and ($\bf D$) we finally obtain
$K^0=8\, J^3-2\, D^2 J- 2\,D^2 E + 8\, E^3$.

Let us summarize the intersection numbers for the two and
three moduli models. For the models with two moduli we find
\eqn\intersectionII{\eqalign{
X_8(2,2,2,1,1)\,    :& \quad K^0= 8\,J^3-8\,J D^2-16\,D^3\cr
X_{12}(6,2,2,1,1)\, :& \quad K^0= 4\,J^3-4\,J D^2-8\,D^3\cr
X_{12}(4,3,2,2,1)\, :& \quad K^0= 2\,J^3-6\,J D^2-24\,D^3\cr
X_{14}(7,2,2,2,1)\, :& \quad K^0= 2\,J^3-14\,J D^2-112\,D^3\cr
X_{18}(9,6,1,1,1)\, :& \quad K^0= 9\,J^3+9\,E^3\cr
}}
The toplogical coupling for the models with three moduli are
\eqn\intersectionIII{\eqalign{
X_{12}(6,3,1,1,1)\,:&\quad K^0= 18\,J^3+ 9\,E_1^3+9\,E_2^3\cr
X_{12}(3,3,3,2,1)\,:&\quad K^0= 6\,J^3-8\,J(D_1^2+D_2^2)+4\,J D_1 D_2 +
                       4\,D_2^2 D_1-16\,(D_1^3+D_2^3)\cr
X_{15}(5,3,3,3,1)\,:&\quad K^0= 3\,J^3-10\,J(D_1^2+D_2^2)+
                     5\,J D_1 D_2+5\,D_2^2 D_1-40\,(D_1^3+D_2^3)\cr
X_{18}(9,3,3,2,1)\,:&\quad K^0= 3\,J^3-4\,J(D_1^2+D_2^2)+2\,J D_1 D_2+
                     2\,D_2^2 D_1-8\,(D_1^3+D_2^3)\cr
X_{24}(12,8,2,1,1)\,:&\quad K^0=8\,J^3-2\,D^2 J-2\,D^2 E+8\,E^3
}}

The intersection numbers for
hypersurfaces in products of ordinary projective
spaces can be readily calculated following \cdls.
One finds
\eqn\prodcoup{\eqalign{
X_{(3|3)}(1,1,1|1,1,1)\,:&\quad  K^0= 3\,J_1^2 J_2 + 3\,J_1 J_2^2\cr
X_{(2|4)}(1,1|1,1,1,1)\,:&\quad  K^0= 2\,J_2^3+4\, J_1 J_2^2\cr
X_{(2|2|3)}(1,1|1,1|1,1,1)\,:&\quad   K^0= 2\,J_1 J_3^2+2\,J_2 J_3^2
                                    +3 J_1 J_2 J_3\cr
X_{(2|2|2|2)}(1,1|1,1|1,1|1,1)\,:&\quad  K^0= 2\sum_{i\neq j\neq k\neq i}
                                     J_i J_j J_k}
}

\newsec{Picard-Fuchs Differential Equations for Hypersurfaces}

Consider the holomorphic three form $\Omega(\psi)$ of a Calabi-Yau
three-fold $X$ as a function of the complex structure moduli
$\psi_i,\,i=1,\dots,2(h^{2,1}+1)$. Its derivatives
w.r.t. the moduli are elements of $H^3(X)$, which is finite
dimensional. This means that there must be linear combinations
of derivatives of the holomorphic three form which are exact.
Upon integration over an element of $H_3(X)$ this leads to linear
differential equations for the periods of $\Omega$,
the Picard-Fuchs (PF) equations.
Candelas, De la Ossa, Green and Parkes showed in \cdgp~
how the solutions of the PF
equation, together with their monodromy properties, allow for
the computation of the $\bra 27^3\ket$
Yukawa couplings, the K\"ahler potential for the complex structure
moduli space and also for an explicit construction of the
mirror map.

The discussion in \cdgp was limited to models with
one complex structure modulus only. Here we want to discuss
the PF equations for the case of several moduli. We start
with a review of a method to set up the Picard-Fuchs equations
due to Dwork, Griffiths and Katz. In the second part
of this section we show how one may use the toric data
of a Calabi-Yau hypersurface to construct the PF equations.

\subsec{Dwork-Griffiths-Katz Reduction Method}

As shown in ref.
\ref\griffiths{P. Griffiths, Ann. of Math. 90 (1969) 460}
\morrison,
the periods $\Pi_i(\psi)$ of the holomorphic three form $\Omega(\psi)$
can be written as
\eqn\periods{\Pi_i(\psi)=\int_{\gamma_1}\cdots\int_{\gamma_m}
             \int_{\Gamma_i}{\omega\over W_1(\psi)\cdots W_m(\psi)}\,,
             \quad i=1,\dots , 2(h^{2,1}+1)\,.}
Here
\eqn\measure{
\omega=\sum_{i=1}^{n+1}(-1)^i w_i z_i dz_1\wedge\dots\wedge
\widehat{dz_i}\wedge\dots\wedge dz_{n+1}\,;}
$\Gamma_i$ is an element of $H_3(X,\ZZ)$ and
$\gamma_j$ a small
curve around $W_j=0$ in the $n$-dimensional embedding space.
The observation that
${\partial\over\partial z_i}\left({f(z)\over
W_1^{p_1}\cdots W_m^{p_m}}\right)\omega$ is exact
if $f(z)$ is homogeneous with degree such that
the whole expression has degree zero,
leads to the partial integration rule, valid under
the integral ($\p_i={\p\over\p z_i}$):
\eqn\partialint{{f\p_i W_j
                \over W_1^{p_1}\cdots W_m^{p_m}}
                ={1\over p_j-1}{W_j \p_i f
                \over W_1^{p_1}\cdots W_m^{p_m}}
                -\sum_{k\neq j}{p_k\over p_j-1}
                {W_j\over W_k}
                {f\p_i W_k
                \over W_1^{p_1}\cdots W_m^{p_m}}}

In practice one chooses a basis $\{Q_k(z)\}$
for the local ring ${\cal R}$ with multi-degree $(d_1,\dots,d_m)$.
For hypersurfaces ${\cal R}=\IC [z_1,\dots,z_{n+1}]/(\partial_i W)$.
One then takes derivatives of the period w.r.t. the moduli until one
produces an integrand of the form $g(z)\over W_1^{p_1}\dots W_m^{p_m}$
such that $g(z)$ is not one of the $Q_i(z)$. One then expresses
$g(z)=\sum_{i=1}^{n+1}f_i(z,\psi)\partial_i W(z,\psi)$
and uses \partialint.
For complete intersections ${\cal R}=(\IC[z_1,\dots,z_{n+1}])^m
/(\sum_i(\p_i W_1,\dots,\p_i W_m)
+\sum_j W_j(\IC[z_1,\dots,z_{n+1}])^m)$
and for the basis elements of the ring one can choose
vector monomials, i.e. $m$-component vectors whose
only non-vanishing component is a monomial
\ref\Aleksandrov{A. Aleksandrov, Math. USSR Sbornik,
                 45 (1983) 1}.

The generalization to complete intersections
in products of projective spaces is straightforward
\cdls : one simply replaces the measure
$\omega$ by $\prod_r\omega_r$, with $\omega_r$
given by eq. \measure~ for each factor in the direct product
of projective spaces.

Note that the PF differential equations contain only those
complex structure moduli for which there exists
a monomial perturbation in the defining polynomials
(there are $\tilde h^{2,1}$ of them). This will also
be true for the method described in the following
subsections.

Above method of deriving the PF differential equations has been used
in \morrison\font\ktone for one modulus hypersurfaces and
in \ktthree~for one modulus complete intersections.
It applies in the form given above only to
complete intersections in products of projective spaces and
not for manifolds embedded in more general toric
varieties.
Applied to models with several moduli it becomes
rather complicated. However, one can extract the general
structure of the PF differential equations by inspecting
the structure of the local ring ${\cal R}$.

To see this
let us restrict our arguments to the case in which the mirror
manifold $X^*$ of a Calabi-Yau three fold $X$ can be obtained
by the orbifoldisation by a finite abelian group $G$
\gp, and consider the period integrals on the mirror manifold $X^*$.
In this case the local ring ${\cal R}^G$ for the mirror $X^*$
consists of the $G$-invariant elements of ${\cal R}=\IC[z_1,\cdots,
z_5]/(\pd_iW)$. We fix a basis of the ring ${\cal R}^G$ as
$\{\varphi_0 ;
\varphi_1\ldots,\varphi_{\tilde h^{2,1}};
\varphi_{{\tilde h^{2,1}}+1}\ldots,\varphi_{2\, {\tilde h^{2,1}}+1};
\varphi_{2\,{\tilde h^{2,1}} + 2}\}$
where the elements are grouped
according to their degrees $(0;d;2\,d;3\,d)$. The elements with
degree $d$ correspond to the perturbations which are parametrized
by the complex structure moduli $\psi_i$ in the untwisted sector.
(It will turn out that a choice for the monomials
$\varphi_i \; (i=1,\cdots, \tilde h^{2,1})$ which is determined
by the toric data of $\Delta^*$
by the monomial-divisor map \monomialdivisor
is a natural basis to study the mirror map.)
Then the period matrix $(\Pi_i^{\;j})$ defined by
$\Pi_i^{\;j}=k!\int_{\Gamma_i}{\varphi_j \over W^{k+1}}\o \;\;
(k={1\over d}{\rm degree}(\varphi_j)\,)$ satisfies the first
order system, called  Gauss-Manin system
\eqn\gaussmanin{
\pd_{\psi_k}\Pi = M^{(k)}(\psi) \Pi \quad (k=1,\cdots,\tilde h^{2,1}).
}
Here $M^{(k)}(\psi)$ are $(2\tilde h^{2,1}+2)\times (2\tilde h^{2,1}+2)$
matrices parametrized by $\psi_i$. This system is defined completely by
the local ring ${\cal R}^G$.
Our PF differential equations are a minimal set of (higher order)
differential equations which is equivalent to the Gauss-Manin system.

Now let us note that
the local ring ${\cal R}^G$ can be expressed as
\eqn\localring{
{\cal R}^G\cong \IC[\varphi_1,\cdots,\varphi_{\tilde h^{2,1}}]/{\cal I}
\quad .}
Here the ideal ${\cal I}$ is generated by the algebraic
relations of the form
$P(\varphi_1,\ldots,\varphi_{\tilde h^{2,1}})
\equiv 0 \; ({\rm mod} \; \pd_i W)$, \ie
\eqn\prepf{
P(\varphi_1,\ldots,\varphi_{\tilde h^{2,1}}) = \sum_{i=1}^5
Q_i(z_1,\ldots,z_5)\partial_{i}W ,
}
where $P$ and $Q_i$ are polynomials in the
$\varphi_i$ and $z_i$ respectively
whose coefficients are polynomials of the moduli parameters.
The relations \prepf~ can be readily translated into PF differential
operators for the period $\Pi_i(\psi)\equiv\Pi_i^{\;0}(\psi)$
by replacing monomials
$\varphi_1^{n_1} ,\ldots,\varphi_r^{n_r}$ by differential operators
$\partial_1^{n_1},\ldots,\partial_r^{n_r}$
and reducing successively the terms of type $Q_i\partial_i W$
by using \partialint.
Multiplication by $\varphi_i$ at the level
of the ring \localring~just translates to derivatives
with respect to the complex structure moduli
at the level of the PF differential equations.
Therefore the requirement that the relations \prepf~from which the PF
differential equations are derived generate ${\cal I}$ constitutes a
neccesary and sufficient
condition  that the PF differential equations are equivalent
to the Gauss-Manin system.

By simple analysis one now sees how many PF differential equations and
of which order one obtains.
For one modulus cases the ring will be of the form
$\{1,\varphi,\varphi^2,\varphi^3\}$ and
the truncation at degree $4\,d$ is done by an algebraic relation
$\varphi^4=\sum_i^5 Q_i \partial_i W$
leading to a fourth order PF differential equation.
For two moduli cases there will always be  one relation of degree $2\,d$
which truncates the three possible products $\varphi_i \varphi_j$
at level $2\,d$ to two dimensions.
This relation multiplied by $\varphi_1$,
$\varphi_2$ gives two, necessarily independent, relations at degree
$3\,d$. Hence there must be always one further relation of
degree $3\,d$.
Also, for Fermat hypersurfaces, the relations at degree
$3\,d$ always generate five independent
relations at degree $4\, d$ so that the
ring is trivial at this degree.
The full information about the period is therefore
contained in one second and one third order differential operator.

For higher dimensional moduli spaces the order of
the full set of differential equations depends on the details of the
ring \localring.
For example, in the case of the $X_{24}(12,8,2,1,1)$
model the three relations of the type \prepf,
generating the ideal at degree
$2\,d$, generate in fact the whole ideal.
Applying \partialint~ yields immediately the
three second order differential operators, given in Appendix A.

For the model $X_{12}(3,3,3,2,1)$ the
three relations at degree $2\,d$ only
yield seven independent relations at degree $3\,d$. Hence the system
has to be supplemented by two relations at order $3\,d$
in order to generate
$\cal I$. The system of Picard-Fuchs equations will therefore contain
three second and two third order equations, compare Appendix A.

For our purpose of constructiong mirror map, we need to find the point
where the monodromy of solutions for the PF differential equations
becomes maximally unipotent and the local solutions around this point
as well as the concrete form of the PF differential equations.
We will find that the toric data encoded in $\Delta^*$ provides us
all necessary information for this purpose.

\vskip0.3cm

\subsec{ Generalized hypergeometric equations and
PF differential equations}

We will now describe an equivalent but often
more efficient way to obtain
the FP differential equations satisfied by the period integral on
the mirror manifold $X^*$ of $X$.
We will mainly discuss, again, the case where
the mirror $X^*$ can be
obtained by orbifoldisation by a finite abelian group $G$\gp.
We will brievely comment on the general case at the end.
The following arguments for toric varieties are largely due
to Batyrev \batyrev.

As summarized in sect.2, in toric geometry the mirror
manifold $X^*$ is described by the toric data encoded in
the reflexive polyhedron $\Delta^*$. In this language the
period integrals are written as
\eqn\toricperiod{
\Pi_i(a)=\int_{\gamma_i}{1\over f(a,X)}\prod_{j=1}^n{dX_j \over X_j}\;\;,
}
with $\gamma_i\in H_n((\IC^*)^n\setminus Z_f)$.
The Laurant polynomial $f$ is given by
\eqn\f{
f(a,X)=\sum_{i=0}^{s^*}a_i X^{\nu_i^*}\quad ,
}
where $\nu_i^*$'s $(i=0,\cdots,s^*)$ are integral points in
$\Delta^*$ which do not lie in the interior of
codimension one faces of $\Delta^*$.

Now let us introduce the generalized hypergeometric system
of Gel'fand, Kapranov and Zelevinsky
\ref\GKZ{I.M.Gel'fand, A.V.Zelevinsky and M.M.Kapranov,
         Func. Anal. Appl. 28 (1989) 12 and
         Adv. Math. 84 (1990) 255}
which is defined for each
configuration of a given set of integral points
$A=\{v_0,\cdots,v_p\}$ in $\IR^n$. We consider the embedding
of these points in the plane with distance one from the
origin of $\IR^{n+1}$ by $\bar v_i=(1, v_i)$ and denote
$\bar A=\{\bar v_0,\cdots,\bar v_p\}$. We assume that the
integral vectors $\bar v_0,\cdots,\bar v_p$ span $\ZZ^{n+1}$.
Since we have $p+1$ integral points in $\IR^{n+1}$, there are
linear dependences described by the lattice
\eqn\latticeL{
L=\{(l_0, \cdots , l_p)\in \ZZ^{p+1} \vert
\sum_{i=0}^p l_i \bar v_i =0 \} \quad .
}
Obviously $\sum l_i=0$.
Considering the affine complex space $\IC^{p+1}$ with
coordinates $(a_0,\cdots,a_p)$, we define the
homogeneous differential operator
\eqn\Dl{
{\cal D}_l=\prod_{l_i>0}\(\da{i}\)^{l_i}
-\prod_{l_i<0}\(\da{i}\)^{-l_i}  \quad ,
}
for each element $l$ of $L$. In addition, we define
differential operators
\eqn\Z{
{\cal Z}_j=\sum_{i=0}^n \bar v_{i,j} a_i\da{i} - \beta_j
}
with $\beta \in \IR^{n+1}$ and $\bar v_{i,j}$ representing
$j$-th component of the vector $\bar v_i \in \IR^{n+1}$.
One can show \GKZ~ that
the operators \Dl~and \Z~define a consistent system
of differential equations
\eqn\ghyper{
{\cal D}_l\Phi(a)=0 \;\; (l\in L) \quad , \quad  {\cal Z}_j\Phi(a)=0 \;\;
(j=0,\cdots,n) }
which is called as $A$-hypergeometric system with exponent $\beta$.

In \batyrev~Batyrev remarked that for a reflexive polyhedron $\Delta^*$,
the period integral \toricperiod~ satisfies the $A$-hypergeometric
system with exponent $\beta=(-1,0, \cdots ,0)$ and
$A$ being the set of the integral points in $\Delta^*$
which do not lie in the interior of
faces of codimension one. Following Batyrev,
we will refer to this system as $\Delta^*$-hypergeometric system.

In general, the $\Delta^*$-hypergeometric system does not suffice
to derive the Picard-Fuchs differential equations.
It turns out that in general
we need to extend the system by supplementing further
differential operators. This depends heavily on the toric data of
$\Delta^*$.
However the system \ghyper~is quite useful because
$(i)$ for some models,
the $\Delta^*$-hypergeometric system provides the
Picard-Fuchs differential equations directly and $(ii)$ even
if this is not the case, this system gives finite
dimensional solution space in which the solution space
of the Picard-Fuchs differential equations is a subspace.
On the other hand we should be very careful when applying
the general results for the $A$-hypergeometric system
in \GKZ~to our $\Delta^*$-hypergeometric system because
the latter is not generic in that it is
(semi-non)resonant (see \GKZ~for details) and
the monodromy group is no longer irreducible.
This is reflected in the simple example below by the fact
that the fifth order operator we start with
factorizes, leaving a fourth order operator which
is precisely the PF differential operator for that case.

In order to obtain an idea of the $\Delta^*$-hypergeometric
system, let us study the case of the quintic hypersurface
in $\IP^4$. In this case, the integral points of the
reflexive polyhedron $\Delta^*$ are given by \vertices~
and the corresponding vertices
$\bar \nu_i^*=(1,\nu_i^*)\in \IR^5$ become
\eqn\quintici{
\eqalign{
&\bar\nu_0^*=(1,0,0,0,0)  \;\;,\;\;
 \bar\nu_1^*=(1,1,0,0,0)  \;\;,\;\;
 \bar\nu_2^*=(1,0,1,0,0)  \;\;,  \cr
&\bar\nu_3^*=(1,0,0,1,0)  \;\;,\;\;
 \bar\nu_4^*=(1,0,0,0,1)  \;\;,\;\;
 \bar\nu_5^*=(1,-1,-1,-1,-1)  \;\; . \cr }
}
As an integral base of the lattice $L$, which is one
dimensional in this case, we can choose
$l^{(1)}=(-5,1,1,1,1,1)$, i.e. $L=\ZZ\, l^{(1)}$.
The system \ghyper~ then becomes
\eqnn\lambdaone
\eqnn\lambdatwo
$$
\eqalignno{
& \Blb \sum_{i=0}^{5}a_i\da{i}+1 \Brb \Pi_i(a)=0  \quad ,
&\lambdaone \cr
& \( a_i\da{i}-a_5\da{5} \)\Pi_i(a) =0  \;\; (i=1,\cdots ,4) \;,
&\lambdatwo \cr
}
$$
together with
\eqn\quinticii{
\Blb \da{1}\da{2}\da{3}\da{4}\da{5} - \(\da{0}\)^5
\Brb\Pi_{i}(a)=0 \quad ,
}
for ${\cal D}_l$ with $l=l^{(1)}$.
If we translate the period integral \toricperiod~to the
more familiar expression
\eqn\quinticiii{
\Pi_i(a)=\int_\gamma\int_{\Gamma_i}
{\o \over a_1z_1^{5}+a_2z_2^{5}+a_3z_3^{5}+a_4z_4^{5}
+a_5z_5^{5} + a_0z_1z_2z_3z_4z_5}  \quad ,
}
utilizing the correspondence described by the monomial-divisor
map \monomialdivisor, we see that
\quinticii~originates from the trivial
relation in the integrand
$z_1^5\cdots z_5^5-(z_1z_2z_3z_4z_5)^5\equiv 0$.
The two equations (constraints)
\lambdaone~and \lambdatwo~ can be understood as the
infinitesimal form of
\eqn\quinticiv{
\eqalign{
& \Pi_i(\lambda^5 a_0, \cdots ,\lambda^5 a_5)
= \lambda^{-5} \Pi_i(a_0,\cdots, a_5)   \quad ,\cr
& \Pi_i(a_0, \cdot\cdot, \lambda_i^5 a_i, \cdot\cdot , \lambda_i^{-5}a_5)
 =\Pi_i(a_0, \cdots ,a_5) \quad (i=1, \cdots ,4) \quad,\cr
}}
with $\lambda, \lambda_i \in \IC^*$, which are verified
by a change of integration variables.
The PF differential equation can be extracted from the
$\Delta^*$-hypergeometric system by making the Ansatz
\eqn\quinticansatz{
\Pi_{i}(a)={1\over a_0}\tilde\Pi_{i}({a_1a_2a_3a_4a_5\over a_0^5})\quad,
}
which solves \lambdaone~and \lambdatwo.
Then equation \quinticiii~becomes
\eqn\quinticPF{
\tx\Blb\tx^4+x(5\tx+4)(5\tx+3)(5\tx+2)
(5\tx+1)\Brb\tilde\Pi_{i}(x)=0\quad ,
}
where $x={a_1a_2a_3a_4a_5\over a_0^5}$ and $\tx=x{d\over dx}$.
Since the factored operator $\tx$ has only constants as solutions,
we can remove this factor by introducing a constant.
However, the asymptotic behavior of the period $\tilde\Pi_{i}(a)$,
for $a_0\rightarrow \infty$ with the other $a_i$'s fixed,
tells us that this constant must be zero (cf. section 4).
We then obtain the generalized hypergeometric equation of fourth
order in \cdgp.

As the simplest example shows, the differential operators
${\cal D}_l \; (l\in L)$ represents the algebraic relations
among the $G$-invariant monomials
$\varphi_0,\cdots, \varphi_{s^*}$
which are the image of the integral points $\nu_i^*$ under
the monomial-divisor map \monomialdivisor. We will see
that if these monomials generate the $G$-invariant polynomial
ring $\IC[z_1,\cdots,z_5]^G$ then the independent algebraic
relations at lowest non-trivial degree
result in the Picard-Fuchs
differential equations, after factorization similar
to the example above.

\subsec{Extension of the $\Delta^*$-hypergeometric system}

Consider $G$-invariant monomials $\varphi_0,\cdots,\varphi_{s^*}$
which correspond to the integral points in $\Delta^*(\vec w)$
not lying in the interior of codimension one faces.
Then the orbifoldizationan of the zero locus of the
quasi-homogeneous polynomial
\eqn\W{
W(z,a)=\sum_{i=0}^{s^*} a_i \varphi_i(z)
}
describes Calabi-Yau hypersurface $X^*=X/G$.
The period integral \toricperiod~ in the toric
language is then translated to the form \partialint~ as
\eqn\period{
\Pi_i(a)=\int_\gamma\int_{\Gamma_i}{\o \over W(a)} \quad,
}
with $\Gamma_i \in H_n(X,\ZZ)$. As elucidated on the
example of the quintic hypersurface, the
differential operator ${\cal D}_l \; (l\in L)$ stems
{}from the algebraic relations satisfied by $\varphi_i$'s.
On the other hand the operators ${\cal Z}_i (i=0,\cdots,n)$
represent the constraints which reduce the apparent redundancy
in the description of the complex
structure deformation of $X^*$ \W~ that arise from
introducing parameters
$a_i$ for all $i\,\in\{0,\cdots,s^*\}$.
Apart from the problem of solving these constraints
by defining suitable variables, which will be
discussed later, the main idea of the $\Delta^*$-hypergeometric
system lies in the fact that
we can find algebraic relations among
the $G$-invariant monomials $\varphi_0,\cdots,\varphi_{s^*}$
which result in the PF differential equations.

In order to verify this for the models specified in the tables
of section two, we classify them into three types.
Type I: there are no integral points inside codimension one
faces of $\Delta^*(\vec w)$
and the $G$-invariant monomials
$\varphi_0,\cdots,\varphi_{s^*}$ generate the ring
$\IC[z_1,\cdots,z_{n+1}]^G$. Type II: there are $m>0$
integral points in the interior of
codimension one faces of
$\Delta^*(\vec w)$; if we include the corresponding
monomials $\varphi_{s^*+1}, \cdots,\varphi_{s^*+m}$ then
$\varphi_0,\cdots,\varphi_{s^*+m}$ are $G$-invariant
and together generate the ring
$\IC[z_1,\cdots,z_{n+1}]^G$. Type III: there are $m\geq0$
integral points inside codimension one faces of
$\Delta^*(\vec w)$ but we also need to consider $m'>0$ $G$-invariant
monomials $\tau_1,\cdots,\tau_{m'}$ of degree greater than
$d$ together with the degree $d$ $G$-invariant monomials
$\varphi_{1},\cdots,\varphi_{s^*+m}$
to generate the ring $\IC[z_1,\cdots,z_{n+1}]^G$.
According to this classification, the models
of type I are
\eqn\typeI{
\eqalign{
& X_{8}(2,2,2,1,1) \;,\;
X_{(3|3)}(1,1,1|1,1,1) \;,\;
X_{(2|4)}(1,1|1,1,1,1) \cr
& X_{(2|2|3)}(1,1|1,1|1,1,1) \;,\;
X_{(2|2|2|2)}(1,1|1,1|1,1|1,1)\,;
}}
for type II we have
\eqn\typeII{
X_{12}(6,2,2,1,1) \;,\;
X_{14}(7,2,2,2,1) \;,\;
X_{18}(9,6,1,1,1) \;,\;
X_{12}(6,3,1,1,1) \;,\;
X_{24}(12,8,2,1,1)
}
and finally for type III
\eqn\typeII{
X_{12}(4,3,2,2,1) \;,\;
X_{12}(3,3,3,2,1) \;,\;
X_{15}(5,3,3,3,1) \;,\;
X_{18}(9,3,3,2,1) \;.
}
For the models of each type we can now find the algebraic
relations which result in the PF differential
equations otherwise obtained through the reduction
method reviewed above. More precisely, for models
of type I, there are elements $l\in L$ for which the
operators ${\cal D}_l$ produce the PF differential
operators after solving the constraints and some
factorization as we have observed in the example.
For models of type II and III, in general,
not all of the PF differential operators follow
{}from ${\cal D}_l$ with some $l\in L$.
We miss the algebraic relations which involve
$\varphi_{s^*\!+\!1},\dots,\varphi_{s^*\!+\!m}$ and
$\tau_1,\dots,\tau_{m'}$.
We develop below a formal procedure for handling
the models of type II and then demonstrate the recipe applicable
to the most general case, type III, by treating
an example.

To formulate the recipe for the models of type II,
let us recall \batyrev~
that the integral points in the interior of codimension
one faces of $\Delta^*(\vec w)$ are related to the
automorphism group $G_{\Delta^*}$
of $\IP^n_{\Delta^*(\vec w)}$ by the formula
\eqn\autodim{
{\rm dim}G_{\Delta^*}=n+\sum_{{\rm codim}\Theta^*=1}l'(\Theta^*)
\quad.
}
The first term takes into account  the $n$-dimensional torus
action which exists canonically for toric varieties while
the second term indicates additional symmetries, which
can be written in infinitesimal form as
\eqn\auto{
z_i'=z_i + \sum_{k=1}^{m}\epsilon_{k} b_i^{(k)}(z)
\quad  (i=1,\cdots,5) \; ,
}
where $m={\rm dim}G_{\Delta^*}-n$.
In order to take advantage of these additional symmetries
we extend the quasi-homogeneous potential \W~to
\eqn\Wextend{
W(z,a)=\sum_{i=0}^{s^*+m} a_i \varphi_i(z) \quad.
}
We can then utilize the symmetries \auto~ to derive the relations
\eqn\autoward{
\int_\gamma\int_{\Gamma_i} {\o\over W(z,a)^2}
\sum_{i=1}^5 \sum_{k=1}^m \epsilon_{k} b_i^{(k)}(z)
{\pd W \over \pd z_i} = 0
\quad ,
}
using $\int_{\Gamma_i'}\o'=\int_{\Gamma_i}\o$ for
the automorphism \auto.
Since $b_i^{(k)}(z)$ has the same degree as $z_i$,
the term in the integrand can be written as a linear
combination of the degree $d$ $G$-invariant monomials.
All degree $d$ $G$-invariants can be obtained
by differentiating $\Pi_i(a)$ (c.f. eq.\period)
with $W(z,a)$ given by \Wextend.
Therefore we obtain independent differential operators
of the form
\eqn\autdiff{
{\cal Z}_k'=\sum_{i,j=0}^m C^{(k)}_{ij}a_i\da{{j}}\quad
}
for each $\epsilon_{k}\;(k=1, \cdots, m)$. In this way we
arrive at the linear system which extends the
$\Delta^*$-hypergeometric system to
\eqn\extended{
{\cal D}_l\Phi(a)=0 \;\; (l\in L'), \;\;
{\cal Z}_j\Phi(a)=0 \;\; (j=0,\cdots,n),\;\;
{\cal Z}'_k\Phi(a)=0 \;\;(k=1, \cdots, m) \;\; ,
}
where $L'$ is now the lattice of relations between all integral points
$\bar\nu_0^*,\dots,\bar\nu_{s^*+m}^*$ in $\bar\Delta^*(\vec w)$
(cf. eq.\latticeL).

\vskip0.3cm

As a simple but non-trivial example, let us consider the model
$X_{14}(7,2,2,2,1)$ with defining polynomial
\eqn\exampleIW{
W=a_1 z_1^2 + a_2 z_2^7 + a_3 z_3^7 + a_4 z_4^7 + a_5 z_5^{14}
  + a_0 \varphi_0
  + a_6 \varphi_6 + a_7 \varphi_7 + a_8 \varphi_8  \; ,
}
where $\varphi_0=z_1z_2z_3z_4z_5 , \varphi_6=z_1z_5^7,
\varphi_7=z_2z_3z_4z_5^8$ and
$\varphi_8=z_2^2z_3^2z_4^2z_5^2$. The latter two
correspond to integral points inside faces of $\Delta^*(\vec w)$
of codimension one. This leads to two additional symmetries \auto
which are easily recognized as
\eqn\exampleIauto{
z_1'=z_1 + \epsilon_1 \, z_2z_3z_4z_5 + \epsilon_2 \, z_5^7 \quad , \quad
z_i'=z_i \qquad (i=2,\cdots, 5) .
}
{}From eq.\autoward~we then get the extended system with two
additional linear operators
\eqn\exampleIops{
\eqalign{
{\cal Z}_1'=2a_1 \da{0} + a_0\da{8} + a_6 \da{7}  \quad , \cr
{\cal Z}_2'=2a_1 \da{6} + a_0\da{7} + a_6 \da{5}  \quad . \cr
}
}
The algebraic relations
$\varphi_6^2-z_1^2z_5^{14}=0$ and
$\varphi_0\varphi_8^3-z_2^7z_3^7z_4^7\varphi_6=0$
then lead to PF differential equations of second
order and, after factorizing a trivial first order operator,
of third order, respectively. To get the third order
equation we use the relations \exampleIops~to express
$\da{8}$ in terms of derivatives
with respect to $a_0,a_5$ and $a_6$ and set
$a_7=a_8=0$.

\vskip0.3cm

Let us now show how the models of type III are treated.
In this most general case we will have to go beyond
linear systems such as \extended. For illustrative
purposes we will treat the model $X_{12}(4,3,2,2,1)$ as an example.
We start with the perturbed potential
\eqn\PFIIIxii{
W=a_1z_1^3+a_2z_2^4+a_3z_3^6+a_4z_4^6+a_5z_5^{12}
+a_0\varphi_0+a_6\varphi_6 \quad ,
}
where the $G$-invariant monomials $\varphi_0=z_1z_2z_3z_4z_5$ and
$\varphi_6=z_2^2z_5^6$ correspond
to the origin and an integral point on a one-dimensional face
of $\Delta^*$, respectively.
$\Delta^*$ for this model has no
integral points in the interior of faces of codimension one.
However the operators ${\cal D}_l$ in the
$\Delta^*(\vec w)$-hypergeometric system miss
the algebraic relations among the generators of
$\IC[z_1,\cdots,z_5]^G$, because it turns out that
we need to incorporate the charge two invariants,
$\tau_1=z_1z_3^4z_4^4z_5^4, \; \tau_2=z_2z_3^3z_4^3z_5^9,\;
\tau_3=z_1^2z_3^2z_4^2z_5^8$ and $\tau_4=z_2^3z_3^3z_4^3z_5^3$
into the generators of the invariants $\IC[z_1,\cdots,z_5]^G$.
Though the algebraic relations which produce the PF
differential operators are not unique, we may choose to consider
the relations $\varphi_6^2-z_2^4z_5^{12}=0$ and
$\varphi_0^2\tau_1-z_1^3z_3^6z_4^6\varphi_6=0$.
The former relation directly gives us a differential equation
\eqn\typeIIIsecond{
\Blb \(\da{6}\)^2-\da{2}\da{5} \Brb\Pi_i(a)=0 \quad .
}
In contrast to this, we need to define
\eqn\typeIIIchargetwo{
\Pi'_i(a)=2\int_\gamma\int_{\Gamma_i}{\tau_1\over W^3}d\mu \quad ,
}
in order to express the latter algebraic relation as
\eqn\typeIIIalg{
\(\da{0}\)^2\Pi'_i(a)-\da{1}\da{3}\da{4}\da{6}\Pi_i(a)=0 \quad .
}
On the other hand, since up to total derivatives with
respect to the coordinates $z_i$ we have the relation
$a_0^3\tau_1=12a_0a_1a_2\varphi_0^2-24a_0a_2a_6\varphi_0\varphi_6
-12a_1 a_6^2\varphi_0 z_5^{12}$ we obtain
\eqn\typeIIIreduction{
\Pi_i'(a)=\Blb {12a_1a_2\over a_0^2}\(\da{0}\)^2
-{24a_1a_2a_6 \over a_0^3}\da{0}\da{6}
-{12a_1a_6^2 \over a_0^3}\da{0}\da{5} \Brb \Pi_i(a) \;.
}
If we now combine \typeIIIchargetwo~
and \typeIIIreduction, we find
a fourth order differential operator which annihilates
$\Pi_i(a)$. We again find the fourth order operator to factorize,
leading finally to a third order differential operator.

We want to close this subsection with a comment. The
discussion presented here was restricted to Fermat
hypersurfaces for which the mirror $X^*$ can be obtained
{}from $X$ as an orbifold, i.e. $X^*=\widehat{X/G}$.
With the exception of the analysis of the type III
models however, all the information that was used in the
derivation of the extended hypergeometric system and
of the PF differential equations, is directly
contained in $\Delta^*$. We can thus base our discussion
also on the expression \toricperiod~rather than \period.
The generalization for hypersurfaces in products of
projective spaces and to complete intersections is also
straightforward.

\vskip0.3cm

\subsec{ Application to Hypersurfaces with Two and Three Moduli}

In this subsection we will show how the general discussion
above applies to the models with few moduli that we
have listed in section 2.

Let us first go to a new gauge and define
\eqn\newgauge{
\Pi_i(a)={1\over a_0}\tilde \Pi_i(a) \quad .
}
The linear operators \Z~ then read
\eqn\tldeZ{
\tilde {\cal Z}_j=\sum_{i=0}^{s^*} \bar \nu^*_{i,j}a_i \da{i} \quad .
}
One then notices easily that
the constraints ${\cal Z}_j\tilde\Pi_i(a)=0$ are solved if
$\tilde\Pi_i$ depends on the variables $a_i$ through the
combination $a^l := a_0^{l_0}\cdots a_{s^*}^{l_{s^*}}$
for arbitrary $l\in L$. We therefore introduce variables
\eqn\xk{
x_k=(-1)^{l^{(k)}_0} a^{l^{(k)}}
:=(-1)^{l^{(k)}_0} a_0^{l^{(k)}_0}\cdots a_{s^*}^{l^{(k)}_{s^*} }
}
with $\{l^{(k)}\}$ an integral basis of the lattice
$L$ (cf. \latticeL); i.e.
\eqn\baseL{
L=\ZZ \, l^{(1)}\oplus \cdots \oplus \ZZ \, l^{(d)}  \quad ,
}
where $d=\tilde h^{2,1}(X^*)$. The integral basis is however
not unique, but we will find in section 4 that the variables
$x_k$, which are good coordinates of moduli space to
describe the large complex structure limit of $X^*$ and,
through the mirror map, the large radius limit of $X$,
are defined in terms of the basis of the Mori cone.
Since, for the moment, we do not need the detailed definition
of the Mori cone, we postpone its definition to section 5
where we show how it is obtained from the toric data.
In Appendix A we list this basis for $L$,
together with the resultant PF differential equations, for each model.
We notice that the appropriate basis does not always
consists of the shortest possible vectors.

For any $l^{(k)}\in L$, we can then rewrite \Dl~
acting on $\tilde\Pi(x)$ as
\eqn\generaldiff{
\left\{\prod_{l_j^{(k)}>0}\left(
\prod_{i=0}^{l_j^{(k)}-1}(\vartheta_j-i)\right)
-\prod_{i=1}^{|l_0^{(k)}|}
\Bigl(i-|l_0^{(k)}|-\vartheta_0 \Bigr)
\prod_{{l_j^{(k)}<0\atop j\neq0}}
\left(\prod_{i=0}^{|l_j^{(k)}|-1}
\Bigl(\vartheta_j+|l^{(k)}_j|-i)\Bigr)\right)x_k\right\}
\tilde\Pi(x)=0
}
where $\vartheta_j$ is $a_j\da{j}$ and is related to $\Theta_{x_k}$ by
\eqn\thetas{
\vartheta_j = \sum_{k=1}^d l^{(k)}_j \Theta_{x_k} \quad .
}
Depending on $l^{(k)}$, this operator will factorize,
leading thus to an operator of lower order.
For some of
our models, this leads directly to a complete set of
PF equations. For these cases the basis $\{l^{(k)}\}$
consists of the shortest possible vectors in $L$.
In Appendix A we have indicated the
differential operators which cannot be obtained directly for
some vector $l^{(k)}\in L$.

The completeness of the PF differential equations follows
{}from the application of the arguments presented in section 3.1.
Since the variable $x_k$ in our PF differential
equations are coordinates on the complex structure moduli space
in the vicinity of the large complex structure, each PF
differential equation can be brought to the form
\eqn\generalform{
\{ p_a(\Theta)+\sum_b f_{ab}(x)q_{ab}(\Theta)\} \Pi(x)=0
}
where $p_a,q_{ab}$ and $f_{ab}$ are polynomials with property
$f_{ab}(0)=0$ and $p_a$ is homogeneous.
The homogeneity of $p_a(\Theta)$ follows from
the characterization of the large complex structure
by the requirement that the indices of the PF differential equations
should be maximally degenerate and the gauge choice which
gives a power series solution that starts with a
constant.
The relation of the PF differential equations to the elements of
the local ring ${\cal R}^G$ described in section 3.1. also
holds in the large complex structure limit. Therefore
the criterion we should verify is that the ring
\eqn\PFring{
\IC[\Theta_1, \cdots,\Theta_{\tilde h^{2,1}}]/(p_a(\Theta))
}
is isomorphic to the local ring ${\cal R}^G$.
We can verify this for all models listed in Appendix A.

\vskip.5cm

\newsec{Logarithmic Solutions, Mirror Map and Yukawa Couplings}

In the previous section we have derived the Picard-Fuchs
differential equations starting from the $\Delta^*$-hypergeometric
system. Now we can argue
the general form of the solutions using results for the
generalized hypergeometric system.
After finding the point of maximally unipotent monodromy, we
define the mirror map.
Once we have the Picard-Fuchs
differential equations, we can determine the Yukawa
couplings on the complex moduli space of $X^*$.
We will see that these Yukawa couplings are expressed
in concise form using the discriminant of the surface.

\subsec{Solutions of the Picard-Fuchs Differential
Equations and Mirror Map}

When deriving the (Picard-Fuchs) differential equations,
we have defined the expansion variables as
\eqn\xk{
x_k:=(-1)^{l^{(k)}_0} a^{l^{(k)}} \quad (k=1, \cdots,\tilde h^{2,1})
}
with an integral basis $\{ l^{(k)} \}$ of the lattice
$L$ \latticeL~for $\Delta^*$. We find,
by solving the recursion relations for the coefficients
$c(n,\rho)$,
a power series solution around $x_k=0$
with the general form
\eqn\generalsol{
\eqalign{
&w(x;\rho)=\sum_n c(n,\rho)x^{n+\rho}\cr
&:=\!\!\!\!\sum_{n_1,\cdots,n_p\in\ZZ_{\geq 0}}\!\!\!\!
{\Gamma(1-\sum_k l^{(k)}_0(n_k+\rho_k))\over
\prod_{i>0}\Gamma(\sum_k l^{(k)}_i(n_k+\rho_k)+1)}
{\prod_{i>0}\Gamma(\sum_k l^{(k)}_i\rho_k+1)\over
\Gamma(1-\sum_k l^{(k)}_0\rho_k)}
x_1^{n_1+\rho_1}\cdots x_p^{n_p+\rho_p}\;\;,\cr
}}
where $p=\tilde h^{2,1}$ and the $\rho_j$ $(j=1,\cdots,p)$ are
the indices, i.e. the solutions of the indicial equations of
the differential equations. $c(n,\rho)$ is normalized
so that $c(0,\rho)=1$.
This is in fact of the form of
the general solution for the hypergeometric system
given in \GKZ~and thus applies for an arbitrary
choice for the integral basis $\{l^{(k)}\}$ of the lattice $L$.

Note that the power series solution can also be easily
obtained by explicitly performing the Cauchy integral \toricperiod~
in the limit $a_0\to\infty$ and choosing the
cycle $\gamma_i=\{(X_1,X_2,X_3,X_4)\in(\IC^*)^4|
|X_1|=\dots=|X_4|=\epsilon\}$.

For the mirror map we need to find the local solutions of
the PF equations with maximally unipotent monodromy
\ref\morrison2{D. Morrison, Amer. Math. Soc. 6 (1993) 223;
{\sl Compactifications of Moduli spaces inspired
by mirror symmetry} DUK-M-93-06, alg-geom/9304007}.
This means that when expanding in the appropriate variables
$x_k$, the solutions of the indicial equation will be
maximally degenerate (in fact all zero)
and there is a unique power series solution of the form
\generalsol~with all other solutions near $x_k=0$
containing logarithms.

We find that if we define
the expansion variables $x_k=(-1)^{l^{(k)}_0} a^{l^{(k)}} \;
(k=1,\cdots,\tilde h^{2,1})$ with $l^{(k)}$
being the basis for
the Mori cone in $L$,
we can take the large radius limit at $x_k=0$, \ie~by what was
said before, at this
point the monodromy becomes maximally unipotent with $\tilde h^{2,1}$
solutions linear in logarithms:
\eqn\linearlog{
w_k(x)=w_0(x)\,{\rm log} x_k + \tilde w_i(x,0) \quad
(k=1,\cdots,\tilde h^{2,1}) \;\; .
}
Here $w_0(x)=w(x,0)$ is the unique power series solution and
$\tilde w_i(x,0)$ are also power series. We will
normalize these solutions such that they do not contain
a constant term
(see also \BatyrevStraten).

Let us now turn to the explicit form of the logarithmic
solutions. Using standard arguments for their construction,
they are obtained by taking derivatives with respect to
the indices which are then set to zero. To get the solutions
containing higher powers of logarithms, one has to choose
certain linear combinations of derivatives with respect to the
$\rho_k$. This point is best illustrated by working out
an example, for which we choose the model $X_8(2,2,2,1,1)$.

It is easy to verify that the indicial equation
at $x_1={a_1 a_2 a_3 a_6\over a_0^4},
x_2={a_4 a_5\over a_6^2}\to 0$ has six ($={\rm dim}\,H^{2,1}(X^*)$)
solutions which are
all zero. (This is e.g. not the case if one expands around
$x_1,x_2\to\infty$.)
To find the logarithmic solutions at $x_1,x_2\to 0$
it suffices to note the relations
\eqn\df{
\eqalign{
& \cD_1 w(x;\rho)=\sum_{n_2\geq 0} c(0,n_2)\rho_1^2(-2n_2+\rho_1-2\rho_2)
x_1^{\rho_1}x_2^{n_2+\rho_2} \quad , \cr
& \cD_2 w(x;\rho)=-\sum_{n_1\geq 0}
c(n_1,0)\rho_2^2x_1^{n_1+\rho_1}x_2^{\rho_2} \quad ,
\cr}}
where the coefficients are
\eqn\cnm{
c(n_1,n_2;\rho)={\Gamma(4(n_1+\rho_1)+1)\Gamma(\rho_1+1)^3
\Gamma(\rho_2+1)^2\Gamma(\rho_1-2\rho_2+1) \over
\Gamma(n_1+\rho_1+1)^3\Gamma(n_2+\rho_2+1)^2
\Gamma(n_1-2n_2+\rho_1-2\rho_2+1) \Gamma(4\rho_1+1) }\; .
}
Due to the factor $\Gamma(n_1-2n_2+\rho_1-2\rho_2+1)$
in the denominator, we have that
$(i)$ $c(n_1,n_2)|_{\rho=0}=0 \; (n_1< 2n_2)$ and
$(ii)$ $(2\pd_{\rho_1}+\pd_{\rho_2})c(n_1,n_2)|_{\rho=0}=0 \; (n_1\leq
2n_2)$.
Usage of this and $[\cD_i,{\p\over\p\rho_k}]=0$ allows us to find
all five logarithmic solutions for this example:
\eqn\logsol{
\pd_{\rho_1}w(x;0)\;,\;
\pd_{\rho_2}w(x;0)\;;\;
\pd_{\rho_1}^2w(x;0)\;,\;
\pd_{\rho_1}\pd_{\rho_2}w(x;0)\;;\;
(\pd_{\rho_1}^3+{3\over2}\pd_{\rho_1}^2\pd_{\rho_2})w(x;0) \;.
}
The logarithmic solutions for the other models can be found
in Appendix A.

The mirror map, which relates the complex structure moduli
space on $X^*$ to the K\"ahler structure moduli space
on $X$, is described by the variables $t_k(x)$, which are
defined as
\eqn\mirrormap{
t_k(x)={\pd_{\rho_k} w(x;0) \over w(x;0) }
= {\rm log} x_k  +O(x)
\quad,}
In fact, in addition to the power series solution,
we can also give the general expression for the logarithmic
solutions that enter the mirror map. They are
\eqn\logsolutions{
\eqalign{
w_k(x)&=w_0(x)\,\log  x_k  \cr
&+\sum_{n \in\ZZ^p_{\geq 0}}\left\{|l_k^{(0)}|
\psi(\sum_j |l_j^{(0)}|n_j+1)
-\sum_{i>0}l_k^{(i)}\psi(\sum_j l_j^{(i)}n_j+1)\right\}
c(n) x^n \cr}
}
where $w_0(z)$ is the power series solution
$$
w_0(x)=\sum_{n\in\ZZ^p_{\geq0}}c(n)x^n
$$
with
$$
c(n)=
{(\sum_j|l_j^{(0)}|n_j)!\over
\prod_i(\sum_j l^{(i)}_j n_j)!}
$$

\vskip0.3cm

\subsec{Yukawa couplings}

Those Yukawa couplings which are functions of the
complex structure moduli, are
defined through the holomorphic three form $\Omega(x)$ as
(cf. e.g.
\ref\candelaossa{P. Candelas and X. De la Ossa,
                 Nucl. Phys. B355 (1991) 455})
\eqn\complexyukawa{
K_{x_ix_jx_k}(x)=\int\Omega(x)\wedge\pd_{x_i}\pd_{x_j}\pd_{x_k}\Omega(x)
\;\;.}
$\Omega(x)$ can be expanded in a basis of $H^3(X^*,\ZZ)$ as
\eqn\expandOmega{
\Omega(x)=\sum_{a=1}^{p+1}\(z^a(x)\alpha_a-{\cal G}_b(x)\beta^b \)
\quad,
}
where $p=h^{2,1}$ and $\alpha_a, \beta^b$ are a
symplectic basis of $H^3(X^*,\ZZ)$. $z^a$ and
${\cal G}_b$ are the period integrals with respect
to the cycles dual to $\alpha_a$ and $\beta^b$.
Then the Yukawa couplings can be expressed through
these periods as
\eqn\yukawa{
K_{x_ix_jx_k}=\sum_a\( z^a\pd_{x_i}\pd_{x_j}\pd_{x_k}{\cal G}_a
               -{\cal G}_a\pd_{x_i}\pd_{x_j}\pd_{x_k}z^a \)
\;\;.}
We now define
\eqn\W{
\eqalign{
W^{(k_1,\cdots,k_d)}
&=\sum_a\(z^a\pd_{x_1}^{k_1}\cdots\pd_{x_d}^{k_d}{\cal G}_a
               -{\cal G}_a\pd_{x_1}^{k_1}\cdots\pd_{x_d}^{k_d}z^a \)
\cr
&:= \sum_a(z^a\pd^{\bf k}{\cal G}_a -{\cal G}_a\pd^{\bf k}z^a ) \;\;.
}}
In this notation, $W^{({\bf k})}$ with $\sum k_i=3$
describes the various types of the Yukawa couplings
and $W^{({\bf k})}\equiv0$ for $\sum k_i=0,1,2$.

Now let us write the Picard-Fuchs differential
operators in the form
\eqn\picard{
\cD_l=\sum_{\bf k} f_l^{({\bf k})}\pd^{\bf k}
\quad ,
}
then we immediately obtain the relation
\eqn\picardrelation{
\sum_{\bf k}f_l^{({\bf k})}W^{({\bf k})}=0 \quad .
}
Further relations are obtained from operators
$\pd_{x_i}\cD_l$. If the PF differential equations are
complete in the sense of section 3.1., they are
sufficient for deriving linear relations among the Yukawa couplings
and their derivatives, which can be integrated to give
the Yukawa couplings up to an overall normalization.
In the derivation, we need to use the following relations
which are easily derived
\eqn\yukawarelations{
\eqalign{
W^{(4,0,0,0)}&=2\pd_{x_1}W^{(3,0,0,0)} \cr
W^{(3,1,0,0)}&=
{3\over2}\pd_{x_1}W^{(2,1,0,0)}+{1\over2}\pd_{x_2}W^{(3,0,0,0)} \cr
W^{(2,2,0,0)}&=\pd_{x_1}W^{(1,2,0,0)}+\pd_{x_2}W^{(2,1,0,0)} \cr
W^{(2,1,1,0)}&=\pd_{x_1}W^{(1,1,1,0)}+{1\over2}\pd_{x_2}W^{(2,0,1,0)}
+{1\over2}\pd_{x_3}W^{(2,1,0,0)} \cr
W^{(1,1,1,1)}&={1\over2}(\pd_{x_1}W^{(0,1,1,1)}
+\pd_{x_2}W^{(1,0,1,1)}
+\pd_{x_3}W^{(1,1,0,1)}
+\pd_{x_4}W^{(1,1,1,0)})
\;\; .}}
By symmetry the above relations exhaust all possibilities.

We have determined the Yukawa couplings for our models.
They are displayed in Appendix A. We should remark that they
are all of the form
$$
W={p(x)\over q(x)\,{\rm dis}_1(X^*)}
$$
where $p(x)$ and $q(x)$ are polynomials and
${\rm dis}_1(X^*)$ is a component of the discriminant
of the hypersurface,
the set of codimension one in moduli space where the
manifold becomes singular, i.e. where
$f_{\Delta^*}=X_1{\partial\over\partial X_1}f_{\Delta^*}=
\dots=X_4{\partial\over\partial X_4}f_{\Delta^*}=0$.
Other components of the discriminant surface can be read
from poles of the individual Yukawa couplings.

Note that for the models considered here the Laurant polynomials
remain transverse if we turn off the terms corresponding to the
divisors via the monomial divisor map, i.e. the corresponding points
in moduli space are regular.

\vskip.5cm

\newsec{Piecewise-Linear Functions and Asymptotic Form of the Mirror Map}

The mirror map is a local isometry
between two different kinds of
moduli spaces; the complex structure moduli space of $X^*$ and the
(complexified) K\"ahler moduli space of $X$.
We will be concerned with the real structure of the latter moduli
space in this section.
It has
the structure of a cone, the so-called K\"ahler cone. How this
cone structure appears in the definition of
the mirror map \mirrormap~can be seen
explicitly in our two  and three moduli models.
We should also remark that we are only discussing the toric
part of the K\"ahler cone.

\subsec{K\"ahler cone}

Let us consider a K\"ahler form $K$ on a Calabi-Yau manifold $X$.
The K\"ahler cone is defined by the requirements
\eqn\kahlerinequalities{
\int_X K\wedge K\wedge K >0 \;,\;
\int_S K\wedge K >0 \;,\;
\int_C K >0 \;,
}
with $S$ and $C$ homologically nontrivial
hypersurfaces and curves in $X$, respectively.
For toric varieties, Oda and Park
\ref\odapark{T. Oda and H.S. Park, Tohoku Math. J. 43 (1991) 375}
have shown how to determine the K\"ahler cone of $\IP_\Delta$
based on the toric data encoded in the polyhedron $\Delta$.
We will only sketch their construction and illustrate
it on the simplest example, the torus $X=X_3(1,1,1)$.

We start with the $n$-dimensional
polyhedron $\Delta$ and consider
its dual $\Delta^*$. We extend $\Delta^*$ to the
polyhedron $\bar \Delta^* \in \IR^{n+1}$ by considering
a convex hull of the origin and the set $(1,\Delta^*)$.
Then a simplicial decomposition of $\Delta^*$ induces a
corresponding simplicial decomposition $\Pi$ of
$\bar\Delta^*$. We denote the subset of the $k$-dimensional
simplices as $\Pi(k)$. We consider
piecewise linear functions, PL$(\Pi)$, on the union
$|\Pi|=\cup_{k=0}^{n+1} \Pi(k)$. A piecewise linear function
$u$ is defined by assigning real values $u_i$ to each
integral point $\nu_i^* \in \Delta^* \; (i=0, \cdots, s^*)$
which is not inside a codimension one face of $\Delta^*$
(we denote the set of such integral points as $\Xi$ with
$s^*=|\Xi|$ and its one dimensional extension by
$\bar\Xi=\{(1,\nu^*)|\nu^*\in\Xi\}$).
If the vertices of a simplex $\sigma \in \Pi(n+1)$
lying on $(1,\Delta^*)$ are given by $\bar\nu_{i_0}^*, \cdots,
\bar\nu_{i_n}^*$ then an arbitrary point $v \in \sigma$ can be
written as $v=c_{i_0}\bar\nu_{i_0}^* +\cdots +c_{i_n}\bar\nu_{i_n}^*
\; (c_{i_0}+ \cdots + c_{i_n} \leq 1, \, c_{i_k}\geq0)$ and the
piecewise linear function $u$ takes the value
\eqn\piecewiseu{
u(v)=c_{i_0}u_{i_0} + \cdots + c_{i_n}u_{i_n} \quad .
}
Equivalently, the piecewise linear function $u$
can be described by a collection of vectors $z_\sigma$
assigned to each simplex $\sigma\in \Pi(n+1)$ with
the property
\eqn\piecewise{
u(v)=\langle z_\sigma,v\rangle\quad{\rm for\;\;all }\;\; v\in\sigma ,
}
where $\langle*,*\rangle$ is the dual pairing.

A strictly convex piecewise linear function $u\in $CPL$(\Pi)$ is
a piecewise linear function with the property
\eqn\CPL{
\eqalign{
u(v)&=\langle z_\sigma,v\rangle\quad{\rm when}\;\;v\in\sigma\cr
u(v)&>\langle z_\sigma,v\rangle\quad {\rm when}\;\; v\notin\sigma\cr
}}
It is clear that if $u$ is a strictly convex piecewise linear
function then so is $\lambda u\;(\lambda\in\IR_{+})$.
Thus the set of the piecewise linear functions has
the structure of a cone.  In order to describe the cone
structure, we consider a vector space $W_1'$ whose basis vectors
are indexed by the set $\Xi$
\eqn\W{
W_1'=\sum_{\xi \in \Xi} \IR e_\xi \quad ,
}
with the basis $e_\xi$.
According to the construction of Oda and Park,
the convex piecewise linear functions CPL$(\Pi)$ constitute
a cone in the quotient space
\eqn\V{
V'= W_1'/
\{ \; \sum_{\xi\in \Xi} \langle x,\bar\xi \rangle e_\xi \;
\vert \;  x \in \IR^{n+1} \;\} \quad,
}
where $\bar\xi=(1,\xi)$. In our context this
cone can be identified with the K\"ahler cone of $\IP_{\Delta}$
(cf. also
\ref\quantumcohomology{
     V.V. Batyrev, {\it Quantum cohomology rings of toric manifolds},
     preprint (1993)}\batyrev\oda).
In the case of the torus $X_3(1,1,1)$, we have
$\bar\nu_0^*=(1,0,0) , \bar\nu_1^*=(1,1,0) , \bar\nu_2^*=(1,0,1)$
and $\bar\nu_3=(1,-1,-1)$ as the one dimensional
extension of the integral points of $\Delta^*$.
Simplicial decomposition of $\Delta^*$ and
$\bar\Delta^*$ are evident, and we have
$\Pi(3)=\{\sigma_1,\sigma_2,\sigma_3\}$ with
$\sigma_1=\langle 0,\bar\nu_0^*,\bar\nu_2^*,\bar\nu_3^*\rangle$,
$\sigma_2=\langle 0,\bar\nu_0^*,\bar\nu_1^*,\bar\nu_3^*\rangle$,
$\sigma_3=\langle 0,\bar\nu_0^*,\bar\nu_1^*,\bar\nu_2^*\rangle$.
Therefore a piecewise linear function $u$ is
described by either $(u_0,u_1,u_2,u_3)$
or $(z_{\sigma_1}, z_{\sigma_2}, z_{\sigma_3} )$
which are related through
\eqn\torusone{
\eqalign{
z_{\sigma_1}=(u_0,2u_0-u_2-u_3,u_2-u_0) \quad , \quad
&z_{\sigma_2}=(u_0,u_1-u_0, 2u_0-u_1-u_3) \quad, \cr
&z_{\sigma_3}=(u_0,u_1-u_0,u_2-u_0) \quad }
}
The condition of the strict convexity \CPL~ on $u$ becomes
the inequality
\eqn\torustwo{
u_1+u_2+u_3-3u_0 > 0 \quad .
}
This inequality produces a cone whose generic element $K_u$ is
\eqn\torusthree{
K_u=-\sum_{i=0}^{3}u_ie_{\nu_i^*}
\equiv {1\over3}((u_1+u_2+u_3)-3u_0)e_{\nu_0^*}  \quad,
}
where the second equivalence is modulo the relations
in \V~ which are
\eqn\torusfour{
e_{\nu_0^*} + e_{\nu_1^*} + e_{\nu_2^*} + e_{\nu_3^*} =0 \;\;,\;\;
e_{\nu_1^*} - e_{\nu_3^*} =0 \;\;,\;\;
e_{\nu_2^*} - e_{\nu_3^*} =0  \quad .
}
The inequality \torustwo~shows that $K_u$ is a generic
element of a cone, a half line in this case.
The identification of the base $e_{\nu_0^*}$ with a divisor
of $\IP_\Delta$, which is justified for a general toric variety,
results in the K\"ahler cone of this model.

\vskip0.3cm

Models with several moduli are treated similarly.
For example, in the case of $X_8(2,2,2,1,1)$ we obtain two
independent inequalities
\eqn\ineqI{
\eqalign{
-4u_0+u_1+u_2+u_3+u_6  & > 0 \cr
u_4+u_5 -2u_6        & > 0 \cr
}
}
{}from the condition \CPL. As a general element of the
divisor of $\IP_\Delta$, we have
\eqn\KI{
K_u \equiv {1\over8}(-8u_0+2u_1+2u_2+2u_3+u_4+u_5)e_{\nu_0^*}
    +{1\over2}(u_4+u_5-2u_6)e_{\nu_6^*} \quad.
}

This example already demonstrates the general situation.
If we write the inequalities in the form
$\langle u,l^{(k)}\rangle>0$, then the $l^{(k)}$ form
a particular integral basis for the lattice of relations $L$ of
the points $\bar\Xi$.  This basis generates a cone in
the lattice of relations, called Mori cone; it is dual
to the K\"ahler cone. The $l^{(k)}$ are exactly the basis of $L$
by which we have defined the variables $x_k$ (see eq.\xk) to
observe the maximally unipotent monodromy at $x_k=0$.
In terms of the $l^{(k)}$, $K_u$ can be written as
$K_u\equiv{1\over8}\langle u,2l^{(1)}+l^{(2)}\rangle e_{\nu_0^*}
+{1\over2}\langle u,l^{(2)}\rangle e_{\nu_6^*}$. We thus see
that the inequalities \ineqI~ give rise to a cone, the K\"ahler cone,
in the quotient space $V^\prime$.
{}From the general theory
of toric geometry it follows that we may identify the basis
$e_{\nu_0^*}$ and $e_{\nu_6^*}$ with the divisor $J$ associated to
the generating element of ${\rm Pic}(X)$
and the exceptional divisor $D$ on $\hat X$
coming from the resolution of the $\ZZ_2$ singular curve,
respectively.

The situation is now again easily described for the general case.
If we express each point corresponding to a divisor
as a linear relation of the vertices of $\bar\Delta^*$ in the form
$\sum\tilde l_j^{(i)}\bar\nu_j^*=0$ for the $i$-th divisor, such
that $\tilde l^{(i)}=\sum n_{ik}l^{(k)}$ with $\{l^{(k)}\}$ being
the basis of the Mori cone and the $n_{ik}$ positive
integers, then we have
\eqn\Ku{
K_u=\sum_i c_i\langle u,\tilde l^{(i)}\rangle e_{\nu_i^*}
=\sum_i c_i\Bigl(\sum_k n_{ik}\langle u,l^{(k)}\rangle\Bigr)
e_{\nu_i^*}
}
where the $c_i$ are rational numbers.

We should note that the identification
of the basis $e_{\nu_i^*}$ ($i=1,\dots,\tilde h^{1,1}(X)$)
with the divisors is justified only up to an as yet
unspecified constant, whose determination will be the subject
of the next subsection.

Let us finally give a simple example for a non-singular case,
the bi-cubic\break
$X_{(3,3)}(1,1,1|1,1,1)$. Starting from the
polyhedron $\Delta(1,1,1)\times\Delta(1,1,1)$, we
obtain the following independent inequalities
\eqn\ineqII{
u_1+u_2+u_3-3u_0 > 0 \quad,\quad
u_4+u_5+u_6-3u_0 > 0
}
and for the divisor of $\IP_\Delta$
\eqn\KII{
K_u \equiv -(u_1+u_2+u_3-3u_0)e_{\nu_3^*}
           -(u_4+u_5+u_6-3u_0)e_{\nu_6^*}
}

In appendix A we list the expressions of the generic divisor
$K_u$ for all models we are considering.

\subsec{Mirror map and instanton corrections}

In section 4.2. we have determined the Yukawa couplings
on the manifold $X^*$ up to a constant as a function of
the complex structure moduli utilizing the PF differential equations.
The results for the
models that we will consider in some detail have
been collected in Appendix A. We will now use these
results to determine the quantum
Yukawa couplings
on $X$ as a function of its
K\"ahler moduli. This will be achieved by close study of  the
mirror map $t_k(x)$\mirrormap.

As we have seen, the variable
$t_k$ is associated with Mori's basis for the lattice of relations $L$.
We now  need  to find
the variable $\tilde t_k$ which corresponds to the integral
cohomology basis $h_i \in H^{1,1}(X,\ZZ)$, to
reproduce the intersection numbers \intersectionII,\intersectionIII~
summarized in sect.2. In terms of the
$\tilde t_i$, we have an expansion of the K\"ahler form
\eqn\intkahler{
K(X)=\sum_{i=1}^{\tilde h^{(1,1)}} \tilde t_i h_i \quad .
}
After identifying
the integral basis we will be able to read off the degrees of the
rational curves with respect to the divisors $J, D, E$ introduced
in the sect.2.
We take the relation between the two sets of parameters
to be linear:
\eqn\defttilde{
t_i(x)= \sum_{j=1}^{\tilde h^{1,1}} m_{ij} \tilde t_j(x) \quad.
}

Those Yukawa couplings on $X$ which are functions of the
K\"ahler moduli, are described
by those on $X^*$ which are functions of the complex structure
moduli through the mirror map \cdgp\morrison.
To obtain them one first changes coordinates from the $x_i$
to
the $\tilde t_i$ coordinates and goes to a physical gauge by dividing by
$w_0(x(\tilde t))^2$ \cdgp\morrison.
Here $w(x)$ is the power series solution of the
Picard-Fuchs differential equations normalized by setting
$w_0(x)=1+O(x)$.
The transformation properties of the Yukawa couplings under a change
of coordinates follows from eq.\complexyukawa~ and the fact that
$\int \Omega\wedge\pd_i\Omega = \int \Omega\wedge \pd_i\pd_j\Omega =0$.
We then obtain the following expression for the Yukawa couplings on $X$
as a function of the K\"ahler moduli $\tilde t_i$:
\eqn\yukawaonX{
K_{\tilde t_i\tilde t_j\tilde t_k}(\tilde t)={1\over w_0(x(\tilde t))^2}
\sum_{l,m,n}{\p x_l\over\p \tilde t_i}
{\p x_m\over\p \tilde t_j}{\p x_n\over\p \tilde t_k}
K_{x_l x_m x_n}(x(\tilde t))
}
Introducing variables $q_i=e^{\tilde t_i}$, we expect
the quantum Yukawa coupling in the form of a series
\foot{Recall that for (2,2) string models there are no
further corrections from curves of finite genus.}
\eqn\correctedyukawa{
\eqalign{
K_{\tilde t_i\tilde t_j\tilde t_k}
&=\int_X h_i \wedge h_j \wedge h_k+\sum_{C}
\int_C h_i\int_C h_j
\int_C h_k{e^{\int_C K(X)}
\over 1-e^{\int_C K(X)}}\cr
&=K^0_{ijk}+\sum_{n_i}{N(\{n_l\})n_i n_j n_k\over
1-\prod_l q_l^{n_l}}\prod_l q_l^{n_l}}
}
where we have defined $n_i=\int_C h_i$,
which is an integer
since $h_i\in H^{1,1}(\hat X,\ZZ)$.
The sum in the first line is over all instantons $C$
of the $\sigma$-model based on $X$ and the denominator
in the integrand takes care of multiple covers of them.
$N(\{n_i\})$ is also an integer
which is the instanton number with degrees $\{n_i\}$.
By considering specific examples below, however, we will see
that it is not necessarily a positive integer.
For more than one
K\"ahler modulus the $n_i$ do not have to be positive,
especially for the manifolds obtained from singular varieties
by resolution. The integral $\int_C K(X)$ however does have to
be positive for $K(x)$ to lie within the K\"ahler cone.
These requirements on the series expansion of \yukawaonX~result in
several constraints on the $m_{ij}$ in
\defttilde~and
the integration
constant for the Yukawa couplings $K_{x_ix_jx_k}$ on $X^*$.

In our calculations, the constraints from the topological
triple coupling (the leading
term of \correctedyukawa) allow several possible values
for the $m_{ij}$.
The additional constraints which stem from the form due to
the multiple covering
turns out to be satisfied by almost all solutions
which satisfy the first constraint.
In order to fix the parameters $m_{ij}$
we need to take a closer look at the mirror
map \mirrormap~in the large radius limit.

In the previous subsection we have described the K\"ahler cone by
using its isomorphism with the class of strictly convex
piecewise linear functions. These functions were
defined by their values $u_i$ on the integral points of
$\Delta^*(\vec w)$ not lying inside codimension-one faces.
The condition of strict convexity resulted in inequalities
$\langle u,l^{(k)}\rangle > 0$, with the
$l^{(k)}$ a basis of the Mori cone.
In terms of the $l^{(k)}$, a general element of the
K\"ahler class of $\IP_\Delta$ can be written in the form \Ku.

On the other hand, from the definition of the $x_k$ through the basis
of the Mori cone,
we have in the large radius limit $x_k \rightarrow 0$,
\eqn\asympt{
t_k \sim {\rm log} x_k \sim \sum_i({\rm log}a_i) l^{(k)}_i  \quad.
}
The similarity of the condition for the
large radius limit $-t_k>>0$ to the inequality for
the K\"ahler cone $\langle u,l^{(k)}\rangle > 0$ then suggests
to identify $u_i$ with ${\rm log}a_i$ as an
asymptotic form of the mirror map.
If we impose this asymptotic relation $u_i={\rm log}a_i$ when
$x_k\rightarrow 0$, we can translate the expression
\Ku~for the K\"ahler class to
$\sum_i\sum_j c_i n_{ij}t_j e_{\nu_i^*}
=\sum_i\sum_{j,k}c_i n_{ij}m_{jk}\tilde t_k \,e_{\nu_i^*}$.
For each model we can find an integer solution $m_{ij}$  with the
property
\eqn\normalizationKu{
K_u\equiv \sum_i\sum_{j,k=1}^{\tilde h^{1,1}}
c_i n_{ij}m_{jk}\tilde t_k \,e_{\nu_i^*}
= \sum_i \tilde c_i e_{\nu_i^*} \tilde t_{k_i} \quad .
}
Here $\tilde c_i$'s are giving the normalization
factor to the integral basis.
In this way, we fix the solution $m_{ij}$ which reproduces
the topological triple couplings together with the normalization
of the basis $e_{\nu_i^*}$
under the Ansatz of the asymptotic form of the mirror map.
This suggest that we associate $\tilde c_i e_{\nu_i^*}$
with the element $h_i\in H^{1,1}(\hat X,\ZZ)$ and get
the K\"ahler cone as the part of moduli space in which
the $\tilde t_i$ may lie such that \kahlerinequalities~is satisfied.

The asymptotic form of the mirror map was also considered by Batyrev
\quantumcohomology~
in his definition of the quantum cohomology ring, (these
asymptotic relations also appeared in
ref.\topologychange).
Our analysis described above is consistent with these references.

We will apply this recipe in the next section to some examples.

\newsec{Predictions and Discussions }

In this section we will present the instanton expansions
and calculate the topological invariants $N(\{n_i\})$
for various two and three moduli cases.
If at a given degree $\{n_i\}$
the manifold has only isolated, nonsingular
instantons, $N$  simply counts their number.
However for non-isolated, singular instantons the situation
becomes less clear and further detailed studies are needed.

Let us turn to our examples and fix the mirror map by
applying the formalism described in the previous section.
For the singular hypersurface  $X_8(2,2,2,1,1)$ $K_u$ is given
in eq.\KI~and we have $l^{(1)}=(-4,1,1,1,0,0,1)$ and
$l^{(2)}=(0,0,0,0,1,1,-2)$ for the generators of
Mori's cone. Using eqs. \KI~and \normalizationKu~we get for the
variables $m_{ij}$ in the Ansatz
\defttilde~ $m=\pmatrix{4\,\tilde c_1&-\tilde c_2\cr 0&2\,
\tilde c_2\cr}$.
We now compare the intersection numbers given in
section 2.4 as $K^0= 8\,J^3-8\,JD^2-16\,D^3$
with the $O(q^0)$ terms in the expansion of the Yukawa couplings
\yukawaonX
\eqn\Kttt{
\eqalign{
K_{\tilde t_1 \tilde t_1 \tilde t_1}
&=8+{\rm O}(q)={4\over c^3}\,{{{m_{11}^2}\,
\left(2\,m_{11}+3\,m_{21}\right)}}+{\rm O}(q)\cr
K_{\tilde t_1 \tilde t_1 \tilde t_2}
&=0+{\rm O}(q)=
{4\over c^3}\,{{m_{11}\,\left(2\,m_{11}\,m_{12}+2\,m_{12}\,m_{21}+
m_{11}\,m_{22}\right)}}+{\rm O}(q)\cr
K_{\tilde t_1 \tilde t_2 \tilde t_2}
&= - 8 + {\rm O}(q)=
{4\over c^3}\,{{m_{12}\,\left( 2\,m_{11}\,m_{12} + m_{12}\,m_{21} +
2\,m_{11}\,m_{22} \right) }} +  {\rm O}(q)\cr
K_{\tilde t_2 \tilde t_2 \tilde t_2}
&=-16+{\rm O}(q)=
{4\over c^3}\,{{m_{12}^2\,\left(2\,m_{12}+3\,m_{22}\right)}} +
\rm O(q);\cr
}}
here we have taken an integration constant
$8/c^3$ into account; it arises when integrating the
first order differential equations satisfied by the
Yukawa couplings.

One constraint on the $m_{ij}$ is that they have to be
integers since the exponents of the $q_i$ are
the degrees of the rational curves with respect to the
various $h_i\in H^{1,1}(\hat X,\ZZ)$.
As the only solution which leads to integer $m$
we can identify
$m=\pmatrix{c&\phantom{-2}c\cr0&-2 c\cr}$ with $c\in \ZZ$.
With this Ansatz
we obtain $K_{\tilde t_1 \tilde t_1 \tilde t_1 }
= -16 - 32\,  {1\over  q_2^{2 c}} + \cdots $.
The second term gives rise to a fractional topological
invariant $N(0,-2 c) = 32/(2 c)^3$ if $|c|>1$.
The two choices of the sign just correspond to an overall
sign of the two K\"ahler moduli. Our sign convention will
always be such that $n_J\geq 0$.
Requiring integral topological invariants we therefore
conclude that $m=\pmatrix{1&\phantom{-}1\cr0&-2\cr}$ and
$\tilde c=({1\over4},-1)$.
Thus we may associate $J$ and $D$
to ${1\over4}e_{\nu_0^*}$ and $-e_{\nu_6^*}$, respectively.
If we combine this with the general description for the K\"ahler
cone given before, we can determine the
K\"ahler cone $\sigma(K)$ as
\eqn\Kahlercone{
\sigma(K)=\{\tilde t_1 h_J + \tilde t_2 h_D \; |\;
\tilde t_1 + \tilde t_2 > 0 \;\;,\;\; \tilde t_2 < 0 \; \} \;\;.
}
It describes possible directions for the
large radius limit where the instanton corrections are suppressed.
The topological invariants $N(\{n_i\})$
can now be read off the expansion of
$K_{\tilde t_i \tilde t_j \tilde t_k}(q_1,q_2)$.
{}From the relation between the basis $t_i$ and $\tilde t_i$ in
terms of the integer matrix $m$ we find that the degrees
are of the form $(n_J,n_D)=(p,p-2q)$, with $p,q=0,1,2,\dots$
We have listed the topological invariants up to order $p+q\leq 10$
and find non-zero numbers only at degrees $(n_J,n_D)$ within
the wedge $n_J \geq |n_D|,\, n_J+n_D$ even, and in addition at $(0,-2)$.
Whereas $n_J=\int_C h_J\geq 0$,
$n_D=\int_C h_D$ also takes negative integer values.
We observe the symmetry $N(n_J,n_D)=N(n_J,-n_D)$ for
$n_J>0$, and have thus listed only the former.
All topological invariants are non-negative integers for this model.

The other models can be discussed similarly. In Appendix
A we give the K\"ahler cones and in Appendix B the topological
invariants $N(\{n_i\})$ at low degrees.

The model $X_{12}(6,2,2,1,1)$ is very similar to the
model discussed above. It also has a singular $\ZZ_2$
curve.
Here the degrees are of the form $(n_J,n_D)=(p,p-3q)$ and
we have listed them again up to order $p+q=10$.

There are two more models
with the singular set being a $\ZZ_2$ curve, namely
$X_{12}(4,3,2,2,1)$ and $X_{14}(7,2,2,2,1)$.
We get from the Yukawa couplings
the topological invariants with degrees
$(n_J,n_D)=(p,3p-2q)$ and
$(n_J,n_D)=(p,7p-2q),\,p,q=0,1,2,\dots$, respectively.
In contrast to the first two models some of the invariants now
are negative integers.

Let us note some observations which relate these four models
to the one-moduli complete intersections discussed in
\libtei~ and \ktthree. If for fixed $n_J>0$ we compute
$\sum_{n_D}N(n_J,n_D)$, we find for the four models discussed
above the same numbers as for the one modulus models
$X_{(4|2)}(1,1,1,1,1,1)$, $X_{(6|2)}(1,1,1,1,1,3)$,
$X_{(6|4)}(1,1,1,2,2,3)$ and $X_{8}(4,1,1,1,1)$,
respectively \ktone\ktthree\libtei.

In contrast to these three models,
$X_{18}(9,6,1,1,1)$ has a $\ZZ_3$ point singularity.
The topological invariants appear in the instanton expansion
of the Yukawa couplings at degrees
$(n_J,n_E)=(p,p-3q)$ with $p,q=0,1,2,\dots$. We have listed
them for $p+q\leq 6$. We find non-zero values for all
degrees within the cone generated by $(1,1)$ and $(0,-1)$.
We again find that some of the topological invariants are negative.

As examples for hypersurfaces in $\IP^4(\vec w)$ with three
moduli we have picked from table 2 three models, representing
the three different types of singularities which occur.
The hypersurface  $X_{24}(12,8,2,1,1)$ has a singular $\ZZ_2$
curve with an exceptional $\ZZ_4$
point. The exceptional divisors correspond to the ruled surface $C\times \IP^1$
and the Hirzebruch surface $\Sigma_2$. Here the degrees of rational
curves are $(n_J,n_D,n_E)=(n,2m-p,n-2)$. We have displayed the
topological invariants for $(n+m+p)\leq 6$. Again, some of the invariants
are negative. For the case $X_{12}(3,3,3,2,1)$ we have a singular
$\ZZ_3$ curve. The two exceptional divisors are the
irreducible components in $C\times (\IP^1\wedge \IP^1)$.
Nonvanishing contributions to the instanton sum occur at degrees
$(n_J,n_{D_1},n_{D_2})=(n,m-2 p,2 n-2m + p)$. As before
the topological invariants take both signs and
are tabulated for $(n+m+p)\leq 6$ in Appendix B.
In the model $X_{12}(6,3,1,1,1)$
we have a two-fold degenerate
$\ZZ_3$ fixed point, which results in two exceptional divisors
$E_1$ and $E_2$, each isomorphic to $\IP^2$.
The interesting point is that they correspond
in the Landau-Ginzburg description to one invariant and one
twisted state.
The Picard-Fuchs equations derived as in sect. 3 contain only two
parameters $x$, $y$. A consistent instanton sum emerges,
if we interpret the corresponding
parameters $\tilde t_1$ and $\tilde t_2$ after the mirror map,
as  associated to $J$ and the symmetric combination $E_1+E_2$.
In doing so, the $m_{ij}$ have to be adjusted s.t. they fit
the intersections $K^0=18 J^2+ 18(E_1+E_2)^3$,
which results in $m=\pmatrix{1&{\phantom{-}}1\cr 0 &-3}$.


For the model $X_{(3|3)}(1,1,1|1,1,1)$ the topological invariants
are all non-negative and positive for $n_{J_1},n_{J_2}\geq0$,
and satisfy $N(n_{J_1},n_{J_2})=N(n_{J_2},n_{J_1})$,
as expected. We have listed them for $n_{J_1}+n_{J_2}\leq 10$.
Some of these numbers can in fact be compared
with results in \BatyrevStraten~ where the same model
was studied on a one-dimensional submanifold of the
K\"ahler structure moduli space which corresponds to requiring
symmetry under exchange of the two $\IP^2$ factors which
leaves only one parameter in \bicubic. This corresponds to
$h_J=h_{J_1}+h_{J_2}$ and the numbers $N(n_J)$ given
in \BatyrevStraten~are related to the numbers listed
in the appendix by $N(n_J)=\sum_{n_{J_1}+n_{J_2}=
n_J}N(n_{J_1},n_{J_2})$. Especially the number of rational
curves of degrees $(1,0)$ and $(0,1)$ agrees with the
explicit calculation in \BatyrevStraten. We also want
to point out the periodicity of the topological invariants at
degrees $(0,n)$.

The following observation about the numbers $N(0,n)$ for the
model $X_{(3|3)}(1,1,1|1,1,1)$ has been related to us by
Victor Batyrev. He points out that there are no rational curves
on this manifold for $n>3$. Yet we do find non-zero instanton
numbers. The mathematical explanation of this fact is connected
with covers of degenerated rational curves.

We have furthermore listed the first few
topological invariants for the models
$X_{(2|4)}(1,1|1,1,1,1)$ and $X_{(2|2|3)}(1,1|1,1|1,1,1)$.
One observes an equality of the invariants $N(k,0) (k\geq 0)$ for
$X_{(2|4)}(1,1|1,1,1,1)$ with those $N(k,k)$ for the model
$X_{8}(2,2,2,1,1)$.

Let us conclude with some remarks.
We have extended the analysis that was initiated in \cdgp~
to models with more than one modulus. It turned out
that one encounters several new features as compared to
the one-modulus models. For instance, the fact that
some of the topological invariants $N(\{n_i\})$ turn out to be
negative integers was a priori unexpected since the
experience with the one-modulus model showed that
they are simply the number of rational curves at a
given degree. This simple interpretation does however
have to be extended in the case where one has non-isolated
or singular curves and our results show that the topological invariants
are then no longer necessarily positive.

To push the analysis further to models with many moduli
seems to be a difficult task. Even though straightforward in
principle, it becomes exceedingly tedious to set up the
Picard-Fuchs equations and especially to obtain the
Yukawa couplings.

We have restricted ourselves in this paper to a computation of
the Yukawa couplings in the large radius limit.
The couplings that were computed are however not
normalized appropriately to yield the physical couplings.
To achieve this, one needs to know the K\"ahler potential.
It can be obtained from the knowledge of all the periods,
i.e. all the solutions of the Picard-Fuchs equations,
as was first done explicitly for a one-modulus model in
\cdgp.
It is however largely determined by the Yukawa couplings,
since they are third derivatives with respect to the
moduli of the prepotential from which the K\"ahler
potential can be derived. This leaves only terms polynomial
of order two in the moduli undetermined. The only relevant
term is however the quadratic one which is known to be
proportional to the Euler number of the Calabi-Yau manifold.

Let us finally point out again the relevance of mirror
symmetry in the analysis presented here.
Even though it is still a mystery from the mathematical
point of view, we have given further compelling evidence
by giving an explicit construction of the mirrors of all
Calabi-Yau manifolds which are hypersurfaces in weighted
projective space. The successful framework which is general
enough to discuss mirror symmetry for these spaces is
that of toric geometry.

\vskip1cm
\noindent
{\bf Acknowledgement:} Two of us (A.K. and S.T.)
would like to thank Victor Batyrev for patiently explaining
some details of his work and Yu. I. Manin for a useful discussion.
We also thank H.P. Nilles for
discussions and for providing computer facilities.
A.K. acknowledges several conversations with D. Dais
on topological triple couplings and thanks C. Vafa
for an invitation to the Lyman Laboratory where this
work was initiated.
We are also grateful to P. Candelas
for discussions and informing us of similar work done by
him and his co-workers.
While finishing writing up these
results we also learned from him that he has explicitly checked
for all Calabi-Yau
hypersurfaces in $\IP^4(\vec w)$ that the corresponding
polyhedron is reflexive. One of us (S.H) would like to
thank T.Eguchi, B.Greene, A.Matsuo, T.Nakanishi, H.Ooguri
and C.Vafa for valuable discussions and continuous encouragements.
S.H. and S.-T.Y are supported by DOE grant \#DEFG02-88ER25065
and A.K and S.T. by Deutsche Forschungsgemeinschaft and
the EC under contract SC1-CT92-0789.
\vskip1.5cm
\noindent
{\bf Note added:} The work of Candelas et al. has now appeared
as preprints
\ref\bcdfhjq{P. Berglund, P. Candelas, X. de la Ossa,
             A. Font, T. H\"ubsch, D. Jancic and F. Quevedo,
             {\it Periods for Calabi-Yau and Landau-Ginzburg Vacua},
             preprint CERN-TH.6865/93} and
\ref\pdfkm{P. Candelas, X. de la Ossa, A. Font, S. Katz
           and D. Morrison, {\it Mirror Symmetry for
           Two-Parameter Models}, preprint CERN-TH.6884/93}.

\vfill\eject


\def\ub{\bar x}
\def\vb{\bar y}
\def\wb{\bar z}
\def\tu{\Theta_x}
\def\tv{\Theta_y}
\def\tw{\Theta_z}
\def\u{ x }
\def\v{ y }
\def\w{ z }
\def\t{ \Theta }

\appendix{A}{ Picard-Fuchs Differential Equations, Discriminant
Surface and Yukawa Couplings }
In this appendix we give the basis  $l^{(k)}$ for Mori's cone in
the lattice $L$ of linear relations \latticeL~
and the Picard-Fuchs differential operators $\cD_{k}$,
acting on $\tilde \Pi$. Differential operators,
which are not directly obtained by factorizing eq.
\generaldiff~for some $l^{(k)}$ are marked with a star.  For convenience
we abbreviate the variables
$x_k=(-1)^{l^{(k)}_0}a^{l^{(k)}}$, $k=1,\ldots
{\tilde h}^{2,1}$, as $\u,\v$ etc.

We also give the logarithmic
solutions around the point of maximal unipotent monodromy ,
as linear combinations of derivatives of the
power series solution $w_0$ with respect to the
indices $\rho_k$, evaluated at $\rho_k=0$.

Next we provide the discriminant and the Yukawa couplings.
To simplify the formulas for the discriminant hypersurface
$\dis$ and the Yukawa couplings
$W^{(k_1,\ldots,k_{{\tilde h}^{2,1}})}$ we use rescaled variables
$\ub,\vb$ etc.
Furthermore, to save space, we list $\tK=\dis\, W$
and write, for example,
$\tK^{(2,1)}=\dis\,K_{\ub\ub\vb}$.
Also, the PF equations determine the Yukawa couplings
in each model only up to a common overall constant, which
we have suppressed below.

We finally give the matrix $m$ (eq.\defttilde)
and the K\"ahler cone.

\subsec{Hypersurfaces in $\IP^4(\vec w)$ }

\leftline{ $\underline{X_{8}(2,2,2,1,1)}$ }

\eqn\whIlattice{
l^{(1)} =(-4,1,1,1,0,0,1) \;\;,\;\; l^{(2)} =(0,0,0,0,1,1,-2) }
$$\eqalign{
&\cD_1=\tu^2\,(\tu-2\,\tv)-4\,\u\,(4\,\tu+3)\,(4\,\tu+2)\,(4\,\tu+1)
\;  \cr
&\cD_2=\tv^2-\v\,(2\,\tv-\tu+1)\,(2\,\tv-\tu)  \cr
}
$$
$$
w_0\quad ; \quad
\pd_{\rho_1}w_0,\;
\pd_{\rho_2}w_0\quad;\quad
\pd_{\rho_1}^2w_0,\;
\pd_{\rho_1}\pd_{\rho_2}w_0\quad ;\quad
(\pd_{\rho_1}^3+{3\over2}\pd_{\rho_1}^2\pd_{\rho_2})w_0
$$
\eqn\whIg{
\ub=2^8\u,\quad \vb=4\v}
$$
\dis=(1-\ub)^2-\ub^2\vb
$$
$$
\tK^{(3,0)} = {1\over\ub^3},\;\;
\tK^{(2,1)} = {2(1-\ub)\over\ub^2\,\vb},\;\;
\tK^{(1,2)} = {4(2\,\ub-1)\over\ub\,\vb\,(1-\vb)},\;\;
\tK^{(0,3)} = {8(1-\ub+\vb-3\,\ub\,\vb)\over\vb^2\,(1-\vb)^2}
$$
\vskip0.3cm
\eqn\whIKu{
K_u={1\over8}\langle u,2l^{(1)}+l^{(2)}\rangle e_{\nu_0^*}
   +{1\over2}\langle u,l^{(2)}\rangle e_{\nu_6^*} }

\eqn\whImij{
m=\pmatrix{1&\phantom{-}1\cr 0&-2\cr}
}
\eqn\kcI{
\sigma(K)=\{\;\tilde t_J h_J+\tilde t_D h_D\;|\;
            \tilde t_J + \tilde t_D > 0,\; \tilde t_D < 0 \;\}
}
\vfill\eject

\leftline{ $\underline{X_{12}(6,2,2,1,1)}$ }

\eqn\whIIlattice{
l^{(1)} =(-6,3,1,1,0,0,1) \;\;,\;\;
l^{(2)}  =(0,0,0,0,1,1,-2)
}
$$\eqalign{
&\cD_1=\tu^2\,(\tu-2\,\tv)-8\,\u\,(6\,\tu+5)\,(6\,\tu+3)\,(6\,\tu+1)
\;  \cr
&\cD_2=\tv^2-\v\,(2\,\tv-\tu+1)\,(2\,\tv-\tu)  \cr
}
$$
$$
w_0 \quad ; \quad
\pd_{\rho_1}w_0,\;
\pd_{\rho_2}w_0\quad;\quad
\pd_{\rho_1}^2w_0,\;
\pd_{\rho_1}\pd_{\rho_2}w_0\quad ;\quad
(\pd_{\rho_1}^3+{3\over2}\pd_{\rho_1}^2\pd_{\rho_2})w_0
$$
\eqn\whIIg{
\ub=2^6 3^3\,\u,\quad \vb=4\, \v}
$$
\dis = (1-\ub)^2-\ub^2\,\vb
$$
$$
\tK^{(3,0)} = {1\over {4\,{\ub^3}}},\;\;
\tK^{(2,1)} = {{1 - \ub}\over {2\,{\ub^2}\,\vb}},\;\;
\tK^{(1,2)} = {{ 2\,\ub-1}\over
    {\ub\,\vb\,\left(1-  \vb\right)}} ,\;\;
\tK^{(0,3)} = {2(1 - \ub + \,\vb - 3\,\ub\,\vb)\over
    {{\vb^2}\,{{\left( 1 - \vb \right)}^2}}}
$$
\eqn\whIIKu{
K_u={1\over12}\langle u,2l^{(1)}+l^{(2)}\rangle e_{\nu_0^*}
   +{1\over2}\langle u,l^{(2)}\rangle e_{\nu_6^*} }

\eqn\whIImij{
m=\pmatrix{1&\phantom{-}1\cr 0&-2\cr}
}
\eqn\kcII{
\sigma(K)=\{\;\tilde t_J h_J + \tilde t_D h_D\;|\;
             \tilde t_J + \tilde t_D > 0,\; \tilde t_D < 0 \;\}
}

\vskip0.3cm

\leftline{ $\underline{X_{12}(4,3,2,2,1)}$ }

\eqn\whIIIlattice{
l^{(1)}=(-6,2,0,1,1,-1,3) \quad l^{(2)}=(0,0,1,0,0,1,-2)
}

$$
\eqalign{
& \cD^\star_1= \tu^2\,(3\,\tu-2\,\tv)-36\,\u\,(6\,\tu+5)\, (6\,\tu+1)\,
               (\tv-\tu+2\v(1+6\tu-2\tv))\cr
& \cD_2=(\tv-\tu)\,\tv-\v\,(3\,\tu-2\tv-1)\,(3\,\tu-2\tv) \cr
}
$$
Here $\cD^\star_1$ is obtained by extending the
hypergeometric system as described in sect. 3.
$$
w_0,\quad ; \;
\p_{\rho_1}w_0,\;
\p_{\rho_2}w_0\;;\;
\p_{\rho_1}^2w_0,\;
(\p_{\rho_1}\p_{\rho_2}+2\p_{\rho_2}^2)w_0\;;\;
(\p_{\rho_1}^3+{3\over2}\p_{\rho_1}^2\p_{\rho_2}
+{9\over2}\p_{\rho_1}\p_{\rho_2}^2+{9\over2}\p_{\rho_2}^3)w_0
$$
\eqn\whIIIg{
\u=2^3 3^3\,\ub,\quad \v=2^2 3\,\vb}
$$
\dis =  1 + 2\,\ub - 6\,\ub\,\vb - 9\,{\ub^2}\,\vb +
6\,{\ub^2}\,{\vb^2} - {\ub^2}\,{\vb^3}$$
$$\eqalign{
\tK^{(3,0)} &=  {{1 + 3\,\ub - \ub\,\vb}\over {{\ub^3}}},\,\,\,
\tK^{(2,1)}  =  {{3\,\left( 1 + 2\,\ub - 2\,\ub\,\vb \right) }
\over {2\,{\ub^2}\,\vb}}
\cr
\tK^{(1,2)} & ={{9\,\left(2+4\,\ub-\vb-5\,\ub\,\vb + 3\,\ub\,{\vb^2}
\right) }\over {4\,\ub\,\left( 3 - \vb \right) \,{\vb^2}}}
\cr
\tK^{(0,3)} &= {{27\,\left(4+8\,\ub-3\,\vb-12\,\ub\,\vb+{\vb^2}+
8\,\ub\,{\vb^2} - 4\,\ub\,{\vb^3} \right) }\over
{8\,{{\left( 3 - \vb \right) }^2}\,{\vb^3}}}}
$$

\eqn\whIIIKu{
K_u={1\over12}\langle u,2l^{(1)}+3l^{(2)}\rangle e_{\nu_0^*}
   +{1\over2}\langle u,l^{(2)}\rangle e_{\nu_6^*} }

\eqn\whIIImij{
m=\pmatrix{1&\phantom{-}3\cr 0&-2\cr}
}
\eqn\kcIII{
\sigma(K)=\{\; \tilde t_J h_J + \tilde t_D h_D \;|\;
             \tilde t_J + 3\tilde t_D > 0,\; \tilde t_D < 0 \;\}
}
\vskip0.3cm

\leftline{ $\underline{X_{14}(7,2,2,2,1)}$ }

\eqn\whIVlattice{
  l^{(1)}=(-7,0,1,1,1,-3,7) \;\;,\;\;
  l^{(2)}=(0,1,0,0,0,1,-2)
}
$$
\eqalign{
\cD^\star_1=&\tu^2(7\,\tu-2\tv) -
7\u \left(\v\,( 28\, \tu- 4\,\tv+18)+\tv-3\,\tu -2 \right)
\cr
&\times\left(\v\,( 28\, \tu- 4\,\tv+10)+\tv-3\,\tu -1\right)
\left(\v\,( 28\, \tu- 4\,\tv+2)+\tv-3\,\tu \right)\cr
\cD_2=&(\tv-3\,\tu)\tv-\v(7 \tu-2\tv-1)(7\tu-2\tv)\cr
}
$$
Here $\cD^\star_1$ is obtain by extending the
hypergeometric system as described in sect. 3.
$$
w_0 \; ; \;
\pd_{\rho_1} w_0,\;
\pd_{\rho_2}w_0\; ;\;
\pd_{\rho_1}^2 w_0,\,
(\pd_{\rho_2}^2+{2\over 3} \pd_{\rho_1}\pd_{\rho_2}) w_0
\; ;\;
({2\over 63}\pd_{\rho_1}^3+{1\over 3}\pd_{\rho_1}^2\pd_{\rho_2}+
\pd_{\rho_1}\pd_{\rho_2}^2+\pd_{\rho_2}^3) w_0
$$
\eqn\whIVg{
\ub=\u,\quad \vb=7\,\v}
$$\eqalign{
\dis=&\,1+27\,\ub-63\,\ub\,\vb+56\,\ub\,{\vb^2}-112\,\ub\,{\vb^3}-
(7-4\vb)^4 \ub^2\vb^3}
$$
$$\eqalign{
\tK^{(3,0)} &=
{{2 + 63\,\ub - 155\,\ub\,\vb + 152\,\ub\,{\vb^2} - 48\,\ub\,{\vb^3}}
\over{{\ub^3}}},
\cr
\tK^{(2,1)} &=
 {{7\,\left(1+27\,\ub - 66\,\ub\,\vb+64\,\ub\,{\vb^2}-
        32\,\ub\,{\vb^3}\right)}\over{{\ub^2}\,\vb}},
\cr
\tK^{(1,2)} &={{49\,\left(3+81\,\ub-2\,\vb-243\,\ub\,\vb+
        301\,\ub\,{\vb^2}-200\,\ub\,{\vb^3}+80\,\ub\,{\vb^4}\right)}
\over{\ub\,{\vb^2}\,\left(4\,\vb-7\right)}}
\cr
\tK^{(0,3)} &=
{{343\,\left(9 + 243\,\ub - 11\,\vb - 864\,\ub\,\vb + 4\,{\vb^2} +
1305\,\ub\,{\vb^2} - 1092\,\ub\,{\vb^3} + 560\,\ub\,{\vb^4} - 192\,\ub\,
{\vb^5} \right) }\over {{\vb^3}\,{{\left(  4\,\vb -7\right) }^2}}}}
$$
\eqn\whIVKu{
K_u={1\over14}\langle u,2l^{(1)}+7l^{(2)}\rangle e_{\nu_0^*}
   +{1\over2}\langle u,l^{(2)}\rangle e_{\nu_6^*}}

\eqn\whIVmij{
m=\pmatrix{1&\phantom{-}7\cr 0&-2\cr}
}
\eqn\kcIV{
\sigma(K)=\{\;\tilde t_J h_J + \tilde t_D h_D \;|\;
              \tilde t_J + 7\tilde t_D > 0,\; \tilde t_D < 0 \;\}
}

\vfill\eject

\leftline{ $\underline{X_{18}(9,6,1,1,1)}$ }

\eqn\whVlattice{
l^{(1)}=(-6,3,2,0,0,0,1)   \;\;,\;\;
 l^{(2)}=(0,0,0,1,1,1,-3)
}
$$
\eqalign{
&\cD_1=\tu\, (\tu-3\, \tv)-12\,\u\,(6\,\tu+5)\,(6\,\tu+1) \cr
&\cD_2=\tv^3-\v\,(\tu-3\,\tv-2)\,(\tu-3\,\tv-1)\,(\tu-3\,\tv) \cr
}
$$
$$
w_0\quad ;\quad
\pd_{\rho_1} w_0,\;
\pd_{\rho_2} w_0 \quad;\quad
\pd_{\rho_2}^2 w_0,\;
(\pd_{\rho_1}^2+{2\over 3} \pd_{\rho_1}\pd_{\rho_2})w_0 \quad;\quad
(3\pd_{\rho_1}^3+3\,\pd_{\rho_1}^2\pd_{\rho_2}+\pd_{\rho_1}
\pd_{\rho_2}^2)w_0
$$
\eqn\whIg{
\ub= 2^4 3^3\u,\quad \vb=3^3\v
}
$$
\dis = (1-\ub)^3-\ub^3 \vb
$$
$$
\tK^{(3,0)} = {1\over\ub^3},\;\;
\tK^{(2,1)} = {3(1-\ub)\over\ub^2\,\vb},\;\;
\tK^{(1,2)} = {9(1-\ub)^2\over\ub\,\vb^2},\;\;
\tK^{(0,3)} = {27(1-3\,\ub+3\,\ub^2)\over\vb^2\,(1+\vb)}
$$
\eqn\whVKu{
K_u={1\over18}\langle u,3l^{(1)}+l^{(2)}\rangle e_{\nu_0^*}
   +{1\over3}\langle u,l^{(2)}\rangle e_{\nu_6^*}}
\eqn\whVmij{
m=\pmatrix{1&\phantom{-}1\cr 0&-3\cr}
}
\eqn\kcV{
\sigma(K)=\{\;\tilde t_J h_J + \tilde t_E h_E\;|\;
             \tilde t_J + \tilde t_E > 0,\; \tilde t_E < 0 \;\}
}

\vskip0.3cm

\leftline{ $\underline{X_{12}(6,3,1,1,1)}$ }

\eqn\whVIlattice{
l^{(1)}=(-4, 2, 1, 0, 0, 0, 1 ),  \,\,
l^{(2)}=( 0, 0, 0, 1, 1, 1, -3 ), \,\,
}
$$
\eqalign{
&\cD_1=\tu\,(\tu-3\tv)-4\u (4\tu+3)(4\tu+1)\cr
&\cD_2=\tv^3 + \v\, (3\,\tv-\tu+2)(3\,\tv-\tu+1)
(3\,\tv-\tu)
\cr}
$$
$$
w_0\quad;\quad
\p_{\rho_1}w_0,\;\p_{\rho_2}w_0\quad;\quad
(\p_{\rho_1}^2+{2\over3}\p_{\rho_1}\p_{\rho_2})w_0,\;
\p_{\rho_2}^2 w_0\quad;\quad
(3\p_{\rho_1}^3+3\p_{\rho_1}^2\p_{\rho_2}
+\p_{\rho_1}\p_{\rho_2}^2)w_0
$$
Note that this is a model with $h^{2,1}=3$,
but only two moduli can be represented as monomial
deformations.
\eqn\whVIg{
\ub= 2^6 \u,\quad \vb=\v
}
$$
\dis = 1 - 3\ub + 3\ub^2 - \ub^3 -27\ub^3\vb
$$
$$
\tK^{(3,0)} = {18\over\ub^3},\;\;
\tK^{(2,1)} = {6(1-\ub)\over\ub^2\,\vb},\;\;
\tK^{(1,2)} = {2(1-\ub)^2\over\ub\,\vb^2},\;\;
\tK^{(0,3)} = {18(1-3\,\ub+3\,\ub^2)\over\vb^2\,(1+27\vb)}
$$
\eqn\whVIKu{
K_u={1\over12}\langle u,2l^{(1)}+l^{(2)}\rangle e_{\nu_0^*}
   +{1\over2}\langle u,l^{(2)}\rangle e_{\nu_6^*}}

\eqn\whVImij{
m=\pmatrix{1&\phantom{-}1\cr 0&-3\cr}
}
\eqn\kcV{
\sigma(K)=\{\; \tilde t_J h_J + \tilde t_E h_E \;|\;
              \tilde t_J + \tilde t_E > 0,\; \tilde t_E<0 \;\}
}

\vfill\eject


\leftline{ $\underline{X_{12}(3,3,3,2,1)}$ }

\eqn\whVIIIlattice{\eqalign{
l^{(1)}&=(-4,1,1,1,0,-1,0,2),  \,\,
l^{(2)} =(0,0,0,0,0,1,1,-2),\cr
l^{(3)}&=(0,0,0,0,1,0,-2,1)
}}
$$
\eqalign{
\cD_2 =& (\tu - \tv)\,(2\, \tw - \tu)-
\v \,(2\, \tv - 2\, \tu- \tw+1) \,(2\, \tv - 2\, \tu- \tw)
\cr
\cD_3 =& \tw\,(2\, \tu -2\, \tv + \tw) -
\w \, (2 \tw - \tv +1) \,(2 \tw - \tv )}
$$
The remaining second order and the two third order differential
operators are rather complicated, so we have not included them here.
The leading terms are $\lim_{x,y,z\rightarrow0}\cD^\star_1=
5\,\tu\tw+2\,\tv^2 +2\,\tw^2 -2\,\tu\,\tv - 5\,\tv\,\tw$ and
$\lim_{x,y,z\rightarrow 0}\cD^\star_4=\tu^2 (2\tv-2\tu-\tw)$,
$\lim_{x,y,z\rightarrow 0}\cD^\star_5=\tu^2\, (2 \tw- \tv)$
\eqn\whVIIg{\ub={2^8}\,\u,\quad\vb=2\v,\quad \wb=\w }
$$
\eqalign{
\dis=& 1 + \ub - 6\,\ub\,\vb - 4\,{\ub^2}\,\vb + 12\,{\ub^2}\,{\vb^2}
   + 4\,{\ub^3}\,{\vb^2} - 8\,{\ub^3}\,{\vb^3}\cr
   & - 18\,{\ub^2}\,{\vb^2}\,\wb - 16\,{\ub^3}\,{\vb^2}\,\wb +
   36\,{\ub^3}\,{\vb^3}\,\wb - 27\,{\ub^3}\,{\vb^4}\,{\wb^2}}
$$
The expressions for the Yukawa couplings, even in the variables
$\ub,\vb,\wb$, are by far too lengthy to be reproduced here.

\eqn\whVIIIKu{
K_u={1\over12}\langle u,3l^{(1)}+4l^{(2)}+2l^{(3)}\rangle e_{\nu_0^*}
   +{1\over3}\langle u,l^{(2)}+2l^{(3)}\rangle e_{\nu_6^*}
   +{1\over3}\langle u,2l^{(2)}+l^{(3)}\rangle e_{\nu_7^*}}

\eqn\whVIIImij{
m=\pmatrix{1&\phantom{-}0&\phantom{-}2\cr
           0&\phantom{-}1&-2\cr
           0&-2&\phantom{-}1\cr}
}
\eqn\kcVIII{
\sigma(K)=\{\tilde t_J h_J+\tilde t_{D_1}h_{D_1}
              +\tilde t_{D_2}\tilde h_{D_2}|
              \tilde t_J+2\tilde t_{D_2}>0,\,
              \tilde t_{D_1}-2\tilde t_{D_2}>0,\,
              \tilde t_{D_2}-2\tilde t_{D_1}>0\}
}

\vskip0.3cm

\leftline{ $\underline{X_{24}(12,8,2,1,1)}$ }

\eqn\whVIIlattice{\eqalign{
l^{(1)}&=(-6,3,2,0,0,0,0,1),  \,\,
l^{(2)}=(0, 0, 0, 0, 1, 1, 0, -2),\cr
l^{(3)}&=(0, 0, 0, 1, 0, 0, -2, 1)}
}
$$
\eqalign{
&\cD_1=\tu\,(\tu-2\,\tw)-12\,\u\,(6\,\tu+5)\,(6\,\tu+1)\cr
&\cD_2=\tv^2-\v\, (2\,\tv-\tw+1)\,(2\,\tv-\tw)\cr
&\cD_3=\tw\,(\tw-\tv)- \w (2\, \tw-\tu +1)\,(2\, \tw-\tu )
}
$$
$$
\eqalign{
w_0\quad;\quad&
\p_{\rho_1}w_0,\;\p_{\rho_2}w_0,\;\p_{\rho_3}w_0\quad;\quad
(\p_{\rho_1}^2+\p_{\rho_1}\p_{\rho_3})w_0,\;
\p_{\rho_1}\p_{\rho_2}w_0,\;
(\p_{\rho_3}^2+2\p_{\rho_2}\p_{\rho_3})w_0\quad;\cr
&(\p_{\rho_1}^3+{3\over2}\p_{\rho_1}^2\p_{\rho_2}
+{3\over2}\p_{\rho_1}^2\p_{\rho_3}+{3\over4}\p_{\rho_1}
\p_{\rho_3}^2+{3\over2}\p_{\rho_1}\p_{\rho_2}\p_{\rho_3})w_0}
$$
\eqn\whVIIg{
\ub=2^4 3^3\, \u,\quad
\vb=2^2 \v,   \quad
\wb=2^2 \w }
$$
\dis=(1-\ub)^4-2\wb(1-\ub)^2+\ub^4\wb^2(1-\vb)
$$
$$
\eqalign{
\tK^{(3,0,0)}&={1-\ub\over\ub^3},\;\;
\tK^{(2,1,0)}=
{1-2\,\ub+\ub^2-\,\ub^2\,\wb\over4\,\ub^2\,\vb},\;\;
\tK^{(2,0,1)}=-{(1-\ub)^2\over\ub^2\,\wb},\cr
\tK^{(1,2,0)}&={(1-\ub)\,
(1-2\,\ub+\ub^2-2\,\ub^2\,\wb)\over16\ub\vb(\vb-1)},\;
\tK^{(1,1,1)}={(1-\ub)\,(1-2\,\ub+\ub^2-\ub^2\,\wb)
\over4\,\ub\,\vb\,\wb},\cr
\tK^{(1,0,2)}&={(1-\ub)^3\over\ub\,\wb^2},\;\;
\tK^{(0,2,1)}={2(2\,\ub-1)\,\wb\,
(1- 2\,\ub+2\,\ub^2-2\,\ub^2\,\wb)\over
16\,\vb\,(1-2\,\wb+\wb^2-\vb\,\wb^2)},\cr
\tK^{(0,3,0)}&=
{(1-2\,\ub)\,\wb\,
\left((1-\ub)^2\,(1-\wb-\vb\wb)-\ub^2(\wb+\vb\,\wb-\,\wb^2-
3\vb\wb^2)\right)\over
64\,(\vb-1)\,\vb^2\,
(1-2\,\wb+\wb^2-\vb\,\wb^2)},\cr
\tK^{(0,1,2)}&={(2\,\ub-1)\,
\left((1-\ub)^2\,(1-\wb+\vb\wb)-\ub^2(\wb-\vb\,\wb\
-\wb^2+\vb\wb^2)\right)\over
4\,\vb\,\wb\,(1-2\,\wb+\wb^2-\vb\,\wb^2)},\cr
\tK^{(0,0,3)}&=
{(2\,\ub-1)\,
\left(2(1-\ub)^2+\wb(\vb-1)(1-2\ub+2\ub^2)\right)\over
\wb^2\,(1-2\,\wb+\wb^2-\vb\,\wb^2)}}
$$
\eqn\whVIIKu{
K_u={1\over24}\langle u,4l^{(1)}+l^{(2)}+2l^{(3)}\rangle e_{\nu_0^*}
   +{1\over4}\langle u,2l^{(2)}+l^{(3)}\rangle e_{\nu_6^*}
   +{1\over2}\langle u,l^{(2)}\rangle e_{\nu_7^*}}

\eqn\whVIImij{
m=\pmatrix{1&\phantom{-}0&\phantom{-}1\cr
           0&\phantom{-}2&\phantom{-}0\cr
           0&-1&-2\cr}
}
\eqn\kcVII{
\sigma(K)=\{\;\tilde t_J h_J+\tilde t_Dh_D+\tilde t_Eh_E\;|\;
\tilde t_J+\tilde t_E>0,\,\tilde t_D>0,\,\tilde t_D+2\tilde t_E<0\}
}

\vskip0.3cm

\subsec{Hypersurfaces in Products of Projective Spaces}

\leftline{ $\underline{X_{(3|3)}(111|111)}$ }

\eqn\phIlattice{
l^{(1)}=(-3,1,1,1,0,0,0)  \quad
l^{(2)}=(-3,0,0,0,1,1,1)
}
By factorizing $\cD_1+\cD_2\equiv(\tu+\tv)\cD^\star_2$ one obtains:
$$
\eqalign{
&\cD_1=\tu^3 -
(3\,\tu+3\,\tv)\,(3\,\tu+3\,\tv-1)\,(3\,\tu+3\,\tv-2)\,\u\cr
&\cD^\star_2=(\tu^2-\tu\tv+\tv^2)-3\,
(3\,\tu+3\,\tv-1)(3\,\tu+3\,\tv-2)\,(\u+\v)\cr}
$$
$$
w_0\quad ;\quad
\pd_{\rho_1} w_0,\;
\pd_{\rho_2} w_0 \quad;\quad
(\pd_{\rho_2}^2 +\pd_{\rho_1}\pd_{\rho_2})w_0,\;
(\pd_{\rho_1}^2 +\pd_{\rho_1}\pd_{\rho_2})w_0 \quad;\quad
(\pd_{\rho_1}^2\pd_{\rho_2}+\pd_{\rho_1}\pd_{\rho_2}^2)w_0
$$

\eqn\phI{
\ub=3^3\,\u,\quad\vb=3^3\,\v}
$$\dis
=1- (1-\ub)^3+(1-\vb)^3 + 3\, \ub\, \vb\,
(\ub+\vb+7)
$$
$$
\tK^{(3,0)}={-2-\ub-\vb\over27\,{\ub}^2},\;\;
\tK^{(2,1)}={{\ub\,(2\,\ub+\vb-1)-(1-\vb)^2}\over
81\, {\ub}^2\, \vb}$$
For symmetry reasons $\tK^{(0,3)}$, $\tK^{(1,2)}$ are
given by the above expressions but with $\ub$ and $\vb$
exchanged.
\eqn\phIKu{
K_u=-\langle u,l^{(1)}\rangle e_{\nu_3^*}
   -\langle u,l^{(2)}\rangle e_{\nu_6^*}}

\eqn\phImij{
m_{ij}=\delta_{ij}
}
\eqn\kcphI{
\sigma(K)=\{\;\tilde t_{J_1}h_{J_1}+\tilde t_{J_2}h_{J_2}\;|\;
                \tilde t_{J_1},\tilde t_{J_2} > 0 \;\}
}

\vskip0.3cm

\leftline{ $\underline{X_{(2|4)}(11|1111)}$ }

\eqn\phIIlattice{
l^{(1)}=(-2,1,1,0,0,0,0) \quad
l^{(2)}=(-4,0,0,1,1,1,1) \quad
}
By  factorizing $\tv^2\, \cD_1-4\cD_2\equiv(\tu+2\,\tv)\cD^\star_2$
one obtains:
$$
\eqalign{
&\cD_1=\tu^2-(4\, \tv+2\, \tu)(4\, \tv+2\, \tu-1)\, \u
\cr
&\cD^\star_2= \tu \tv^2-2\tv^3
- 2\,(4\, \tv+2\, \tu-1)\,( \tv^2 \u - 4\, (4\, \tv+2\, \tu-2)\,
(4\, \tv+2\,\tu-3)\, \v) \cr}
$$
$$
w_0\quad ;\quad
\pd_{\rho_1} w_0,\;
\pd_{\rho_2} w_0 \quad;\quad
\pd_{\rho_2}^2 w_0,\;
\pd_{\rho_1}\pd_{\rho_2} w_0 \quad;\quad
(\pd_{\rho_2}^2 \pd_{\rho_1}+{1\over 6} \pd_{\rho_2}^3)w_0
$$
\eqn\phII{
\ub=2^2\,\u,\quad\vb=2^8\,\v}
$$
\dis=(1 -2\,(2 \ub+\vb)+6\,{\ub}^2+{\vb}^2-12\,
\ub \, \vb- 4 {\ub}^3 -2\, {\ub}^2 \, \vb+{\ub}^4)
$$
$$
\eqalign{
\tK^{(3,0)}&={\vb-6\,\ub-{\ub}^2-1\over 4\,{\ub}^2},\;\;
\tK^{(2,1)}= {{ 2 \ub - \vb + {\ub}^2 -3  }
\over 8 \,  \ub\, \vb}\cr
\tK^{(1,2)}&=
{(1+\ub)\, (2\, \ub + \vb - {\ub}^2-1)
\over 16\, \ub\, {\vb}^2},\;\;
\tK^{(0,3)} = {{ 3 \ub - 3\, \vb - 3\, {\ub}^2 -\ub \,
\vb + {\ub}^3 -1  }\over
32 \, { \vb}^3 }}
$$

\eqn\phIIKu{
K_u=-\langle u,l^{(1)}\rangle e_{\nu_2^*}
    -\langle u,l^{(2)}\rangle e_{\nu_6^*}}

\eqn\phIImij{
m_{ij}=\delta_{ij}
}
\eqn\kcphII{
\sigma(K)=\{\;\tilde t_{J_1}h_{J_1}+\tilde t_{J_2}h_{J_2}\;|\;
             \tilde t_{J_1},\tilde t_{J_2} > 0 \;\}
}

\vskip0.3cm

\leftline{ $\underline{X_{(2|2|3)}(11|11|111)}$ }

\eqn\phIIIlattice{
\eqalign{
l^{(1)}&=(-2,1,1,0,0,0,0,0 ),\,\,
l^{(2)}=(-2,0,0,1,1,0,0,0 ),\cr
l^{(3)}&=(-3,0,0,0,0,1,1,1 )}}
By  factorizing $16\, (\tv \cD_1+\tu\cD_2)
- 27\, \cD_3-12 \, \tw \, (\cD_1+\cD_2)\equiv
(2\tu+2\tv+3\tw)\cD^\star_3$ one obtains:
$$
\eqalign{
\cD_1=&\tu^2-\u\, (2\,\tu+2\, \tv+3\, \tw+1)\, (2\,\tu+2\, \tv+3\, \tw)
\cr
\cD_2=&\tv^2-\v\, (2\,\tu+2\, \tv+3\, \tw+1)\, (2\,\tu+2\, \tv+3\, \tw)
\cr
\cD^\star_3=&3\,\tw\,(3\,\tw-2\,\tu-2\,\tv)+8\tu\,\tv-(3\,\tw+2\,\tu+2
\tv-1)\, \cr
&\bigl(3^3 \w (3\, \tw+2\,\tu+2 \, \tv+1)-
4\,\u\,(3\,\tw-4\,\tv)-4\, \v\, (3\,\tw-4\,\tu)\bigr)}
$$

$$\eqalign{w_0\,:
\pd_{\rho_1} w_0,\,
\pd_{\rho_2} w_0,\,
\pd_{\rho_3} w_0\; ;\;
(3\pd_{\rho_1} \pd_{\rho_3}\!\! & +\!\pd_{\rho_3}^3)w_0,\,
(\pd_{\rho_1} \pd_{\rho_3}\!\!-\!\pd_{\rho_2} \pd_{\rho_3})w_0,\,
(\pd_{\rho_2} \pd_{\rho_3}\!\! +\!{4\over 3}\pd_{\rho_1}
\pd_{\rho_2})w_0\; \cr
(\pd_{\rho_1} \pd_{\rho_3}^2\!\!+\!
 \pd_{\rho_2} &\pd_{\rho_3}^2\!\!+\!
 3\pd_{\rho_1} \pd_{\rho_2} \pd_{\rho_3})w_0}
$$

\eqn\phIII{
\ub=2^2\,\u,\quad\vb=2^2\,\v,\quad\wb=3^3\,\w}

$$\eqalign{\dis =&
1-6\,\ub+15\,{\ub^2}-20\,{\ub^3}+15\,{\ub^4}-6\,{\ub^5}+{\ub^6}-
   6\,\vb+18\,\ub\,\vb-12\,{\ub^2}\,\vb
\cr &
   -12\,{\ub^3}\,\vb+18\,{\ub^4}\,\vb-
   6\,{\ub^5}\,\vb+15\,{\vb^2}-12\,\ub\,{\vb^2}-6\,{\ub^2}\,{\vb^2}
\cr & -
   12\,{\ub^3}\,{\vb^2}+15\,{\ub^4}\,{\vb^2}-
   20\,{\vb^3}-12\,\ub\,{\vb^3}-
   12\,{\ub^2}\,{\vb^3}-20\,{\ub^3}\,{\vb^3}
\cr &+
   15\,{\vb^4} + 18\,\ub\,{\vb^4} +
   15\,{\ub^2}\,{\vb^4} - 6\,{\vb^5} -
   6\,\ub\,{\vb^5} + {\vb^6} - 4\,\wb + 24\,{\ub^2}\,\wb
\cr & -
   32\,{\ub^3}\,\wb + 12\,{\ub^4}\,\wb -
   144\,\ub\,\vb\,\wb + 96\,{\ub^2}\,\vb\,\wb +
   48\,{\ub^3}\,\vb\,\wb + 24\,{\vb^2}\,\wb
\cr & +
   96\,\ub\,{\vb^2}\,\wb - 120\,{\ub^2}\,{\vb^2}\,\wb -
   32\,{\vb^3}\,\wb + 48\,\ub\,{\vb^3}\,\wb +
   12\,{\vb^4}\,\wb + 6\,{\wb^2} + 18\,\ub\,{\wb^2}
\cr & +
   42\,{\ub^2}\,{\wb^2} - 2\,{\ub^3}\,{\wb^2} +
   18\,\vb\,{\wb^2} - 36\,\ub\,\vb\,{\wb^2} -
   30\,{\ub^2}\,\vb\,{\wb^2} + 42\,{\vb^2}\,{\wb^2}
\cr & -
   30\,\ub\,{\vb^2}\,{\wb^2} -
   2\,{\vb^3}\,{\wb^2} - 4\,{\wb^3} -
   12\,\ub\,{\wb^3} - 12\,\vb\,{\wb^3} + {\wb^4}
}$$

The expressions for the Yukawa couplings, even in the variables
$\ub,\vb,\wb$, are too lengthy to be reproduced here.

\eqn\phIIIKu{
K_u=-\langle u,l^{(1)}\rangle e_{\nu_2^*}
    -\langle u,l^{(2)}\rangle e_{\nu_4^*}
    -\langle u,l^{(3)}\rangle e_{\nu_7^*} }

\eqn\phIIImij{
m_{ij}=\delta_{ij}
}
\eqn\kcphIII{
\sigma(K)=\{\;\tilde t_{J_1}h_{J_1}+\tilde t_{J_2}h_{J_2}
+\tilde t_{J_3}h_{J_3}\;|\;
\tilde t_{J_1},\tilde t_{J_2},\tilde t_{J_3} > 0 \;\}
}

\vskip0.3cm

\leftline{ $\underline{X_{(2|2|2|2)}(11|11|11|11)}$ }

\eqn\phIVlattice{
\eqalign{
l^{(1)}&=(-2,1,1,0,0,0,0,0,0 ),\,\,
l^{(2)} =(-2,0,0,1,1,0,0,0,0 ),\cr
l^{(3)}&=(-2,0,0,0,0,1,1,0,0 ),\,\,
l^{(4)}=(-2,0,0,0,0,0,0,1,1 )}}
By factorizing $(\cD_1-\cD_2)(\t_3-\t_4)+(\cD_3-\cD_4)(\t_1-\t_2)=
(\t_1+\t_2+\t_3+\t_4)\cD^\star_5$ we define $\cD^\star_5$
and similarly, by exchanging in the above equation the
indices $2\leftrightarrow 3$, $\cD^\star_6$,
s.t. the system reads
$$
\eqalign{
\cD_i=&\t_i^2-x_i\, (2\,\t_1+2\, \t_2+2\, \t_3+ 2\t_4 + 1)\,
(2\,\t_1+2\, \t_2+2\, \t_3 + 2\, \t_4),\,\,\, {\rm for}\,\, i=1,2,3,4
\cr
\cD^\star_5=& (\t_1-\t_2)(\t_3-\t_4) +
2\,(2\,\t_1+2\, \t_2+2\, \t_3+ 2\t_4-1)\,
\cr & \cdot
\bigl(x_1\, (\t_4-\t_3)+
x_2\,(\t_3-\t_4)+x_3\, (\t_2-\t_1)+x_4\,(\t_1-\t_2)\bigr)
\cr
\cD^\star_6=&  (\t_1-\t_3)(\t_2-\t_4)+
2\,(2\,\t_1+2\, \t_3+2\, \t_2+ 2\t_4-1)\,
\cr & \cdot
\bigl( x_1\, (\t_4-\t_2)+
x_2\,(\t_2-\t_4)+x_3\, (\t_3-\t_1)+x_4\,(\t_1-\t_3)\bigr)}
$$
$$
\eqalign{
&\!\!\!\!\!\! w_0\, ; \cr
\pd_{\rho_1} w_0,\,
\pd_{\rho_2} w_0,\,
&\pd_{\rho_3} w_0,\
\pd_{\rho_4} w_0\, ;\cr
(\pd_{\rho_1} \pd_{\rho_2}\!\! -\! \pd_{\rho_3} \pd_{\rho_4})w_0,\,
(\pd_{\rho_1} \pd_{\rho_3}\!\! - \! \pd_{\rho_2} \pd_{\rho_4})w_0,\,
&(\pd_{\rho_1} \pd_{\rho_4}\!\!- \!\pd_{\rho_2} \pd_{\rho_3})w_0,\,
(\pd_{\rho_1} \pd_{\rho_2} \!\!+
\!\pd_{\rho_2} \pd_{\rho_3}\!\! +
\! \pd_{\rho_1}
\pd_{\rho_4} ) w_0;\cr
(\pd_{\rho_1} \pd_{\rho_2}\pd_{\rho_3}\!\!+\!
\pd_{\rho_1} \pd_{\rho_2}\pd_{\rho_4}\!+\,
&\pd_{\rho_1} \pd_{\rho_3}\pd_{\rho_4}\!\!+\!
\pd_{\rho_2} \pd_{\rho_3}\pd_{\rho_4})w_0.
}
$$

\eqn\phIVKu{
K_u=-\langle u,l^{(1)}\rangle e_{\nu_2^*}
    -\langle u,l^{(2)}\rangle e_{\nu_4^*}
    -\langle u,l^{(3)}\rangle e_{\nu_6^*}
    -\langle u,l^{(4)}\rangle e_{\nu_8^*}}

\eqn\phIVmij{
m_{ij}=\delta_{ij}
}
\eqn\kcphIV{
\sigma(K)=\{\;\tilde t_{J_1}h_{J_1}+\tilde t_{J_2}h_{J_2}
             +\tilde t_{J_3}h_{J_3}+\tilde t_{J_4}h_{J_4} \;|\;
   \tilde t_{J_1},\tilde t_{J_2},\tilde t_{J_3},\tilde t_{J_4} > 0 \;\}
}

\vfill\eject

\appendix{B}{Topolocical Invariants $N(\{n_i\})$}
\vskip.5cm
Here we append the tables for the first few topological
invariants $N(\{n_i\})$ for the discussed cases.
In the first column of the tables we list the degree.
The first entry is always the degree
with respect to $h_J$, the
others with respect to $h_D$ or $h_E$.
In the second column we list the non-zero invariants
within the indicated range of degrees.

\subsec{Hypersurfaces in $\IP^4(\vec w)$ }

\leftline{\bf$\underline{X_{8}(2,2,2,1,1)}$ }
\vskip.5cm

{}From the relation between the basis $t_i$ and $\tilde t_i$
in term of the matrix $m$ listed in appendix A, we
find that the degrees are of the form
$(n,m)=(p,p-2q)$ with $p,q=0,1,2,\dots$.
We find non-zero invariants
only for integers $(n,m)$ within
the wedge $n\geq |m|,\,n+m\,$even, and in addition
at $(n,m)=(0,-2)$. We also observe the
symmetry $N(n,m)=N(n,-m)$ for $n>0$ and only list the former.

Below we list the topological invariants for $p+q\leq 10$.

$$
\vbox{\offinterlineskip
\halign{ &\vrule# & \strut\quad\hfil#\quad\cr
\noalign{\hrule}
\noalign{\hrule}
height1pt&\omit&   &\omit&\cr
&(0,-2)&&  4 &\cr
height1pt&\omit&   &\omit&\cr
\noalign{\hrule}
\noalign{\hrule}}
\vskip 1cm
\hrule
\halign{ &\vrule# & \strut\quad\hfil#\quad\cr
\noalign{\hrule}
height1pt&\omit&   &\omit&\cr
&(1,1)&&  640 &\cr
height1pt&\omit&   &\omit&\cr
\noalign{\hrule}
\noalign{\hrule}
height1pt&\omit&   &\omit&\cr
&(2,0)&&  72224 &\cr
&(2,2)&&  10032 &\cr
height1pt&\omit&   &\omit&\cr
\noalign{\hrule}
\noalign{\hrule}
height1pt&\omit&   &\omit&\cr
&(3,1)&&  7539200 &\cr
&(3,3)&&  288384 &\cr
height1pt&\omit&   &\omit&\cr
\noalign{\hrule}
\noalign{\hrule}
height1pt&\omit&   &\omit&\cr
&(4,0)&&  2346819520 &\cr
&(4,2)&&  757561520 &\cr
&(4,4)&&  10979984 &\cr
height1pt&\omit&   &\omit&\cr
\noalign{\hrule}
\noalign{\hrule}
height1pt&\omit&   &\omit&\cr
&(5,1)&&  520834042880 &\cr
&(5,3)&&  74132328704 &\cr
&(5,5)&&  495269504 &\cr
height1pt&\omit&   &\omit&\cr
\noalign{\hrule}}
\hrule}
\quad
\vbox{\offinterlineskip
\halign{ &\vrule# & \strut\quad\hfil#\quad\cr
\noalign{\hrule}
\noalign{\hrule}
height1pt&\omit&   &\omit&\cr
&(6,0)&&  212132862927264 &\cr
&(6,2)&&  95728361673920 &\cr
&(6,4)&&  7117563990784 &\cr
&(6,6)&&  24945542832 &\cr
height1pt&\omit&   &\omit&\cr
\noalign{\hrule}
\noalign{\hrule}
height1pt&\omit&   &\omit&\cr
&(7,1)&&  64241083351008256 &\cr
&(7,3)&&  15566217930449920 &\cr
&(7,5)&&  673634867584000 &\cr
&(7,7)&&  1357991852672 &\cr
height2pt&\omit&   &\omit&\cr
\noalign{\hrule}
\noalign{\hrule}
height1pt&\omit&   &\omit&\cr
&(8,4)&&  2320662847106724608 &\cr
&(8,6)&&  63044114100112216 &\cr
&(8,8)&&  78313183960464 &\cr
height2pt&\omit&   &\omit&\cr
\noalign{\hrule}
\noalign{\hrule}
height1pt&\omit&   &\omit&\cr
&(9,7)&&  5847130694264207232 &\cr
&(9,9)&&  4721475965186688 &\cr
height2pt&\omit&   &\omit&\cr
\noalign{\hrule}
\noalign{\hrule}
height1pt&\omit&   &\omit&\cr
&(10,10)&&  294890295345814704 &\cr
\noalign{\hrule}
}
\hrule}
$$
\vfill\eject

\leftline{\bf  $\underline{X_{12}(6,2,2,1,1)}$ }
\vskip.5cm

{}From the relation between the basis $t_i$ and $\tilde t_i$
in term of the matrix $m$ we find, as in the perevious model,
that the degrees are of the form
$(n,m)=(p,p-2q)$ with $p,q=0,1,2,\dots$ and non-zero
topological invariants at the same points as indicated there.
Also, for $n>0$ the symmetry
$N(n,m)=N(n,-m)$ is again present.
We list the non-zero topological invariants
again for $p+q\leq 10$ and $n\geq 0$
$$
\vbox{\offinterlineskip
\halign{ &\vrule# & \strut\quad\hfil#\quad\cr
\noalign{\hrule}
\noalign{\hrule}
height1pt&\omit&   &\omit&\cr
&(0,-2)&&  2 &\cr
height1pt&\omit&   &\omit&\cr
\noalign{\hrule}
\noalign{\hrule}}
\vskip 1cm
\hrule
\halign{ &\vrule# & \strut\quad\hfil#\quad\cr
\noalign{\hrule}
height1pt&\omit&   &\omit&\cr
&(1,1)&&  2496 &\cr
height1pt&\omit&   &\omit&\cr
\noalign{\hrule}
\noalign{\hrule}
height1pt&\omit&   &\omit&\cr
&(2,0)&&  1941264 &\cr
&(2,2)&&  223752 &\cr
height1pt&\omit&   &\omit&\cr
\noalign{\hrule}
\noalign{\hrule}
height1pt&\omit&   &\omit&\cr
&(3,1)&&  1327392512 &\cr
&(3,3)&&  38637504 &\cr
height1pt&\omit&   &\omit&\cr
\noalign{\hrule}
\noalign{\hrule}
height1pt&\omit&   &\omit&\cr
&(4,0)&&  2859010142112 &\cr
&(4,2)&&  861202986072 &\cr
&(4,4)&&  9100224984 &\cr
height1pt&\omit&   &\omit&\cr
\noalign{\hrule}
\noalign{\hrule}
height1pt&\omit&   &\omit&\cr
&(5,1)&&  4247105405354496 &\cr
&(5,3)&&  540194037151104 &\cr
&(5,5)&&  2557481027520 &\cr
height1pt&\omit&   &\omit&\cr
\noalign{\hrule}}
\hrule}
\quad
\vbox{\offinterlineskip
\halign{ &\vrule# & \strut\quad\hfil#\quad\cr
\noalign{\hrule}
\noalign{\hrule}
height1pt&\omit&   &\omit&\cr
&(6,0)&&  11889148171148384976 &\cr
&(6,2)&&  5143228729806654496 &\cr
&(6,4)&&  331025557765003648 &\cr
&(6,6)&&  805628041231176 &\cr
height1pt&\omit&   &\omit&\cr
\noalign{\hrule}
\noalign{\hrule}
height1pt&\omit&   &\omit&\cr
&(7,1)&&  24234353788301851080192 &\cr
&(7,3)&&  5458385566105678112256 &\cr
&(7,5)&&  199399229066445715968 &\cr
&(7,7)&&  274856132550917568 &\cr
height2pt&\omit&   &\omit&\cr
\noalign{\hrule}
\noalign{\hrule}
height1pt&\omit&   &\omit&\cr
&(8,4)&&  5277289545342729071440512 &\cr
&(8,6)&&  118539665598574460315052 &\cr
&(8,8)&&  99463554195314072664 &\cr
height2pt&\omit&   &\omit&\cr
\noalign{\hrule}
\noalign{\hrule}
height1pt&\omit&   &\omit&\cr
&(9,7)&&  69737063786422755330975040 &\cr
&(9,9)&&  37661114774628567806400 &\cr
height2pt&\omit&   &\omit&\cr
\noalign{\hrule}
\noalign{\hrule}
height1pt&\omit&   &\omit&\cr
&(10,10)&&  14781417466703131474388040 &\cr
\noalign{\hrule}
}
\hrule}
$$
\vfill\eject

\leftline{\bf$\underline{X_{12}(4,3,2,2,1)}$ }
\vskip.5cm

The degrees are of the form
$(n,m)=(p,3p-2q)$ with $p,q=0,1,2,\dots$.
Here we find non-zero topological invariants
only for integers $(n,m)$ within
the cone generated by $(1,\pm3)$ and, as in the previous
two cases at $(n,m)=(0,-2)$. Again, there is the
symmetry $N(n,m)=N(n,-m)$ for $n>0$.
We give the topological invariants for $p+q\leq 8$.

$$
\vbox{\offinterlineskip
\halign{ &\vrule# & \strut\quad\hfil#\quad\cr
\noalign{\hrule}
\noalign{\hrule}
height1pt&\omit&   &\omit&\cr
&(0,-2)&& 6 &\cr
height1pt&\omit&   &\omit&\cr
\noalign{\hrule}
\noalign{\hrule}}
\vskip 1cm
\hrule
\halign{ &\vrule# & \strut\quad\hfil#\quad\cr
\noalign{\hrule}
height1pt&\omit&   &\omit&\cr
&(1,1)&& 7524 &\cr
&(1,3)&& 252 &\cr
height1pt&\omit&   &\omit&\cr
\noalign{\hrule}
\noalign{\hrule}
height1pt&\omit&   &\omit&\cr
&(2,0)&& 16761816 &\cr
&(2,2)&& 5549652 &\cr
&(2,4)&& 30780 &\cr
&(2,6)&& -9252 &\cr
height1pt&\omit&   &\omit&\cr
\noalign{\hrule}
\noalign{\hrule}
height1pt&\omit&   &\omit&\cr
&(3,1)&& 56089743576 &\cr
&(3,3)&& 10810105020 &\cr
&(3,5)&& 45622680 &\cr
&(3,7)&& -4042560 &\cr
&(3,9)&& 848628 &\cr
height1pt&\omit&   &\omit&\cr
\noalign{\hrule}
\noalign{\hrule}
height1pt&\omit&   &\omit&\cr
&(4,0)&& 427990123181952 &\cr
&(4,2)&& 230227010969940 &\cr
&(4,4)&& 31014597012048 &\cr
&(4,6)&& 107939555010 &\cr
&(4,8)&& -6771588480 &\cr
&(4,10)&& 691458930 &\cr
&(4,12)&& -114265008 &\cr
height1pt&\omit&   &\omit&\cr
\noalign{\hrule}}
\hrule}
\quad
\vbox{\offinterlineskip
\halign{ &\vrule# & \strut\quad\hfil#\quad\cr
\noalign{\hrule}
\noalign{\hrule}
height1pt&\omit&   &\omit&\cr
&(5,5)&& 110242870186236480 &\cr
&(5,7)&& 348378053579208 &\cr
&(5,9)&& -16730951255208 &\cr
&(5,11)&& 1299988453932 &\cr
&(5,13)&& -138387180672 &\cr
&(5,15)&& 18958064400 &\cr
height1pt&\omit&   &\omit&\cr
\noalign{\hrule}
\noalign{\hrule}
height1pt&\omit&   &\omit&\cr
&(6,10)&& -53592759845826120 &\cr
&(6,12)&& 3355331493727332 &\cr
&(6,14)&& -288990002251968 &\cr
&(6,16)&& 30631007909100 &\cr
&(6,18)&& -3589587111852 &\cr
height1pt&\omit&   &\omit&\cr
\noalign{\hrule}
\noalign{\hrule}
height1pt&\omit&   &\omit&\cr
&(7,15)&& -778844028150225792 &\cr
&(7,17)&& 70367764763518200 &\cr
&(7,19)&& -7266706161056640 &\cr
&(7,21)&& 744530011302420 &\cr
height2pt&\omit&   &\omit&\cr
\noalign{\hrule}
\noalign{\hrule}
height1pt&\omit&   &\omit&\cr
&(8,20)&& -18212970597635246400 &\cr
&(8,22)&& 1813077653699325510 &\cr
&(8,24)&& -165076694998001856 &\cr
height2pt&\omit&   &\omit&\cr
\noalign{\hrule}
\noalign{\hrule}
height1pt&\omit&   &\omit&\cr
&(9,25)&& -470012260531104088320 &\cr
&(9,27)&& 38512679141944848024 &\cr
height2pt&\omit&   &\omit&\cr
\noalign{\hrule}
\noalign{\hrule}
&(10,30)&& -9353163584375938364400 &\cr
height1pt&\omit&   &\omit&\cr
\noalign{\hrule}
}
\hrule}
$$
\vfill\eject

\leftline{\bf$\underline{X_{14}(7,2,2,2,1)}$ }
\vskip.5cm

The degrees are of the form
$(n,m)=(p,7p-2q)$ with $p,q=0,1,2,\dots$.
Here we find non-zero topological invariants
only for integers $(n,m)$ within
the cone generated by $(1,\pm 7)$ and, as in the previous
two cases at $(n,m)=(0,-2)$. Again, there is the
symmetry $N(n,m)=N(n,-m)$ for $n>0$.
We give the topological invariants for $p+q\leq 10$.

$$
\vbox{\offinterlineskip
\halign{ &\vrule# & \strut\quad\hfil#\quad\cr
\noalign{\hrule}
\noalign{\hrule}
height1pt&\omit&   &\omit&\cr
& (0,-2) && 28 & \cr
height1pt&\omit&   &\omit&\cr
\noalign{\hrule}
\noalign{\hrule}}
\vskip 1cm
\hrule
\halign{ &\vrule# & \strut\quad\hfil#\quad\cr
\noalign{\hrule}
height1pt&\omit&   &\omit&\cr
&(1,1)&& 14427 &\cr
&(1,3)&&   378 &\cr
&(1,5)&&   -56 & \cr
&(1,7)&&     3 & \cr
height1pt&\omit&   &\omit&\cr
\noalign{\hrule}
\noalign{\hrule}
height1pt&\omit&   &\omit&\cr
&(2,0)&& 68588248 & \cr
&(2,2)&& 29683962 &\cr
&(2,4)&&   500724 &\cr
&(2,6)&&   -69804 &\cr
&(2,8)&&     9828 &\cr
&(2,10)&&   -1512 &\cr
&(2,12)&&     140 &\cr
&(2,14)&&      -6 &\cr
height1pt&\omit&   &\omit&\cr
\noalign{\hrule}
\noalign{\hrule}
height1pt&\omit&   &\omit&\cr
&(3,7)&&  -258721916 &\cr
&(3,9)&&    27877878 &\cr
&(3,11)&&   -5083092 &\cr
&(3,13)&&     837900 &\cr
&(3,15)&&    -122472 &\cr
&(3,17)&&      13426 &\cr
&(3,19)&&      -896  &\cr
&(3,21)&&         27 &\cr
height1pt&\omit&   &\omit&\cr
\noalign{\hrule}}
\hrule}
\quad
\vbox{\offinterlineskip
\halign{ &\vrule# & \strut\quad\hfil#\quad\cr
\noalign{\hrule}
\noalign{\hrule}
height1pt&\omit&   &\omit&\cr
&(4,16)&&  -652580600 &\cr
&(4,18)&&   109228644 &\cr
&(4,20)&&   -15811488 &\cr
&(4,22)&&     1841868 &\cr
&(4,24)&&     -154280 &\cr
&(4,26)&&        8008 &\cr
&(4,28)&&        -192 &\cr
height1pt&\omit&   &\omit&\cr
\noalign{\hrule}
\noalign{\hrule}
height1pt&\omit&   &\omit&\cr
&(5,25)&& -2613976470 &\cr
&(5,27)&&   315166313 &\cr
&(5,29)&&   -29721888 &\cr
&(5,31)&&     2006914 &\cr
&(5,33)&&      -85064 &\cr
&(5,35)&&        1695 &\cr
height1pt&\omit&    &\omit&\cr
\noalign{\hrule}
\noalign{\hrule}
height1pt&\omit&   &\omit&\cr
&(6,34)&& -6314199584 &\cr
&(6,36)&&   496850760 &\cr
&(6,38)&&   -28067200 &\cr
&(6,40)&&     1004360 &\cr
&(6,42)&&    -17064 &\cr
height1pt&\omit&   &\omit&\cr
\noalign{\hrule}
\noalign{\hrule}
height1pt&\omit&   &\omit&\cr
&(7,43)&&  -8479946160  &\cr
&(7,45)&&    411525674  &\cr
&(7,47)&&    -12736640  &\cr
&(7,49)&&       188454  &\cr
height2pt&\omit&   &\omit&\cr
\noalign{\hrule}
\noalign{\hrule}
height1pt&\omit&   &\omit&\cr
&(8,52)&& -6238001000  &\cr
&(8,54)&&   170052708  &\cr
&(8,56)&&    -2228160  &\cr
height2pt&\omit&   &\omit&\cr
\noalign{\hrule}
\noalign{\hrule}
&(9,61)&& -2360463560  &\cr
&(9,63)&&    27748899  &\cr
height1pt&\omit&   &\omit&\cr
height2pt&\omit&   &\omit&\cr
\noalign{\hrule}
\noalign{\hrule}
&(10,70)&& -360012150  &\cr
height1pt&\omit&   &\omit&\cr
\noalign{\hrule}
}
\hrule}
$$

\vfill\eject

\leftline{ $\underline{X_{18}(9,6,1,1,1)}$ }

{}From the relation between the basis $t_i$ and $\tilde t_i$
in term of the matrix $m$ we now find
$(n,m)=(p,p-3q)$ with $p,q=0,1,2,\dots$.
We find non-zero topological invariants
on all of these points. Below are our results
for $p+q\leq 6$.

$$
\vbox{\offinterlineskip
\hrule
\halign{ &\vrule# & \strut\quad\hfil#\quad\cr
\noalign{\hrule}
height1pt&\omit&   &\omit&\cr
&(1,1)&&  540 &\cr
&(2,2)&&  540 &\cr
&$\vdots$&& $\vdots$ &\cr
&(6,6)&&  540 &\cr
height1pt&\omit&   &\omit&\cr
\noalign{\hrule}
\noalign{\hrule}
height1pt&\omit&   &\omit&\cr
&(0,-3)&&  3 &\cr
&(1,-2)&&  -1080 &\cr
&(2,-1)&& 143370 &\cr
&(3,0)&& 204071184 &\cr
&(4,1)&& 21772947555 &\cr
&(5,2)&& 1076518252152 &\cr
height1pt&\omit&   &\omit&\cr
\noalign{\hrule}}
\hrule}
\quad
\vbox{\offinterlineskip
\halign{ &\vrule# & \strut\quad\hfil#\quad\cr
\noalign{\hrule}
\noalign{\hrule}
height1pt&\omit&   &\omit&\cr
&(0,-6)&&  -6 &\cr
&(1,-5)&& 2700 &\cr
&(2,-4)&&  -574560 &\cr
&(3,-3)&&  74810520 &\cr
&(4,-2)&& -49933059660 &\cr
height1pt&\omit&   &\omit&\cr
\noalign{\hrule}
\noalign{\hrule}
height1pt&\omit&   &\omit&\cr
&(0,-9)&&  27 &\cr
&(1,-8)&& -17280 &\cr
&(2,-7)&& 5051970 &\cr
&(3,-6)&&  -913383000 &\cr
height1pt&\omit&   &\omit&\cr
\noalign{\hrule}}
\hrule}
\quad
\vbox{\offinterlineskip
\halign{ &\vrule# & \strut\quad\hfil#\quad\cr
\noalign{\hrule}
\noalign{\hrule}
height1pt&\omit&   &\omit&\cr
&(0,-12)&&  -192 &\cr
&(1,-11)&& 154440 &\cr
&(2,-10)&& -57879900 &\cr
height1pt&\omit&   &\omit&\cr
\noalign{\hrule}
\noalign{\hrule}
height1pt&\omit&   &\omit&\cr
&(0,-15)&& 1695 &\cr
&(1,-14)&& -1640520 &\cr
height1pt&\omit&   &\omit&\cr
\noalign{\hrule}
\noalign{\hrule}
height1pt&\omit&   &\omit&\cr
&(0,-18)&& -17064       &\cr
\noalign{\hrule}
}
\hrule}
$$
\vskip 15 mm
\leftline{ $\underline{X_{12}(6,3,1,1,1)}$ }

{}From the relation between the basis $t_i$ and $\tilde t_i$
in term of the matrix $m$ we now find
$(n,m)=(p,p-3q)$ with $p,q=0,1,2,\dots$.
We find non-zero topological invariants
on all of these points. Below are our results
for $p+q\leq 6$.

$$
\vbox{\offinterlineskip
\hrule
\halign{ &\vrule# & \strut\quad\hfil#\quad\cr
\noalign{\hrule}
height1pt&\omit&   &\omit&\cr
&(1,1)&&  216 &\cr
&(2,2)&&  324 &\cr
&(3,3)&&  216 &\cr
&(4,4)&&  324 &\cr
&(5,5)&&  216 &\cr
&(6,6) && 324 &\cr
height1pt&\omit&   &\omit&\cr
\noalign{\hrule}
\noalign{\hrule}
height1pt&\omit&   &\omit&\cr
&(0,-3)&& 6 &\cr
&(1,-2)&& -432 &\cr
&(2,-1)&& 10260 &\cr
&(3,0)&&   1233312 &\cr
&(4,1)&&   26837190 &\cr
&(5,2)&&  368683056  &\cr
height1pt&\omit&   &\omit&\cr
\noalign{\hrule}}
\hrule}\quad
\vbox{\offinterlineskip
\halign{ &\vrule# & \strut\quad\hfil#\quad\cr
\noalign{\hrule}
\noalign{\hrule}
height1pt&\omit&   &\omit&\cr
&(0,-6)&&  -12 &\cr
&(1,-5)&&  1080 &\cr
&(2,-4)&& -41688 &\cr
&(3,-3)&&   810864 &\cr
&(4,-2)&&   -61138584 &\cr
height1pt&\omit&   &\omit&\cr
\noalign{\hrule}
\noalign{\hrule}
height1pt&\omit&   &\omit&\cr
&(0,-9)&& 54 &\cr
&(1,-8)&&  -6912 &\cr
&(2,-7)&& 378756 &\cr
&(3,-6)&& -11514096 &\cr
height1pt&\omit&   &\omit&\cr
\noalign{\hrule}}
\hrule}\quad
\vbox{\offinterlineskip
\halign{ &\vrule# & \strut\quad\hfil#\quad\cr
\noalign{\hrule}
\noalign{\hrule}
height1pt&\omit&   &\omit&\cr
&(0,-12)&& -384 &\cr
&(1,-11)&&   61776 &\cr
&(2,-10)&& -4411260 &\cr
height1pt&\omit&   &\omit&\cr
\noalign
{\hrule}
\noalign{\hrule}
height1pt&\omit&   &\omit&\cr
&(0,-15)&& 3390 &\cr
&(1,-14)&&  -656208 &\cr
height1pt&\omit&   &\omit&\cr
\noalign{\hrule}
\noalign{\hrule}
height1pt&\omit&   &\omit&\cr
&(0,-18)&& -34128       &\cr
\noalign{\hrule}
}
\hrule}
$$
\vfill\eject

\leftline{$\underline{X_{12}(3,3,3,2,1)}$}
\vskip.5 cm

The degrees are $(n,m-2p,2n-2m+p)$, $n,m,p=0,1,2,\dots$.
For $n+m+p\leq6$, the non-zero invariants are
$$
\vbox{\offinterlineskip
\hrule
\halign{ &\vrule# & \strut\quad\hfil#\quad\cr
\noalign{\hrule}
height1pt&\omit&   &\omit&\cr
&(0,-1,-1)&& 2 & \cr
height1pt&\omit&   &\omit&\cr
\noalign{\hrule}
\noalign{\hrule}
height1pt&\omit&   &\omit&\cr
&(0,1,-2)&& 2 & \cr
height1pt&\omit&   &\omit&\cr
\noalign{\hrule}
\noalign{\hrule}
height1pt&\omit&   &\omit&\cr
&(0,-2,1)&& 2 & \cr
height1pt&\omit&   &\omit&\cr
\noalign{\hrule}
\noalign{\hrule}
height1pt&\omit&   &\omit&\cr
&(1,-2,0)&& -28 & \cr
height1pt&\omit&   &\omit&\cr
\noalign{\hrule}
\noalign{\hrule}
height1pt&\omit&   &\omit&\cr
&(1,0,-1)&& -296 & \cr
height1pt&\omit&   &\omit&\cr
\noalign{\hrule}
\noalign{\hrule}
height1pt&\omit&   &\omit&\cr
&(2,0,1)&& 32272 & \cr
height1pt&\omit&   &\omit&\cr
\noalign{\hrule}}
\hrule}
\quad
\vbox{\offinterlineskip
\halign{ &\vrule# & \strut\quad\hfil#\quad\cr
\noalign{\hrule}
\noalign{\hrule}
height1pt&\omit&   &\omit&\cr
&(1,0,2)&& -28 & \cr
&(2,0,4)&& -129 & \cr
&(3,0,6)&& -1620 & \cr
&(4,0,8)&&  -29216 & \cr
&(5,0,10)&& -651920 & \cr
height1pt&\omit&   &\omit&\cr
\noalign{\hrule}
\noalign{\hrule}
height1pt&\omit&   &\omit&\cr
&(1,2,-2)&& -28 & \cr
height1pt&\omit&   &\omit&\cr
\noalign{\hrule}
\noalign{\hrule}
height1pt&\omit&   &\omit&\cr
&(1,1,0)&&  -296 & \cr
&(2,2,0)&&  4646 & \cr
height1pt&\omit&   &\omit&\cr
\noalign{\hrule}}
\hrule}
\quad
\vbox{\offinterlineskip
\halign{ &\vrule# & \strut\quad\hfil#\quad\cr
\noalign{\hrule}
\noalign{\hrule}
height1pt&\omit&   &\omit&\cr
&(1,-1,1)&& -296 & \cr
\noalign{\hrule}
\noalign{\hrule}
height1pt&\omit&   &\omit&\cr
&(2,-1,3)&& 276 & \cr
\noalign{\hrule}
\noalign{\hrule}
height1pt&\omit&   &\omit&\cr
&(3-1,5)&& 4544 & \cr
\noalign{\hrule}
\noalign{\hrule}
height1pt&\omit&   &\omit&\cr
&(2,1,2)&& 276 & \cr
height1pt&\omit&   &\omit&\cr
\noalign{\hrule}
\noalign{\hrule}
height1pt&\omit&   &\omit&\cr
&(3,1,4)&& 4544 & \cr
height1pt&\omit&   &\omit&\cr
\noalign{\hrule}
\noalign{\hrule}
height1pt&\omit&   &\omit&\cr
&(4,1,6)&& 100134 & \cr
height1pt&\omit&   &\omit&\cr
\noalign{\hrule}
\noalign{\hrule}
height1pt&\omit&   &\omit&\cr
&(2,3,-2)&& 276 & \cr
height1pt&\omit&   &\omit&\cr
\noalign{\hrule}
\noalign{\hrule}
height1pt&\omit&   &\omit&\cr
&(3,2,2)&& -7720 & \cr
height1pt&\omit&   &\omit&\cr
\noalign{\hrule}}
\hrule}
$$

\vskip 15 mm
\leftline{$\underline{X_{24}(12,8,2,1,1)}$}
\vskip.5 cm

The non-zero topological invariants, whose degree
is of the general form
$(n,2m-p,n-2p)$ where $n,m,p=0,1,2,\dots$.In the range
$n+m+p\leq6$ we find them to be
$$
\vbox{\offinterlineskip
\hrule
\halign{ &\vrule# & \strut\quad\hfil#\quad\cr
\noalign{\hrule}
height1pt&\omit&   &\omit&\cr
&(0,-3,-10)&& -10 & \cr
height1pt&\omit&   &\omit&\cr
\noalign{\hrule}
\noalign{\hrule}
height1pt&\omit&   &\omit&\cr
&(0,-2,-8)&& -8 & \cr
height1pt&\omit&   &\omit&\cr
\noalign{\hrule}
\noalign{\hrule}
height1pt&\omit&   &\omit&\cr
&(0,-1,-6)&& -6 & \cr
height1pt&\omit&   &\omit&\cr
\noalign{\hrule}
\noalign{\hrule}
height1pt&\omit&   &\omit&\cr
&(0,-1,-2)&& -2 & \cr
height1pt&\omit&   &\omit&\cr
\noalign{\hrule}
\noalign{\hrule}
height1pt&\omit&   &\omit&\cr
&(0,0,-4)&& -4 & \cr
&(0,0,-8)&& -32 & \cr
\noalign{\hrule}
\noalign{\hrule}
height1pt&\omit&   &\omit&\cr
&(0,1,-6)&& -6 & \cr
height1pt&\omit&   &\omit&\cr
\noalign{\hrule}
\noalign{\hrule}
height1pt&\omit&   &\omit&\cr
&(0,1,-2)&& -2 & \cr
height1pt&\omit&   &\omit&\cr
\noalign{\hrule}
\noalign{\hrule}
height1pt&\omit&   &\omit&\cr
&(1,-2,-7)&& 3360 & \cr
height1pt&\omit&   &\omit&\cr
\noalign{\hrule}
\noalign{\hrule}
height1pt&\omit&   &\omit&\cr
&(1,-1,-5)&& 2400 & \cr
height1pt&\omit&   &\omit&\cr
\noalign{\hrule}}
\hrule}
\quad
\vbox{\offinterlineskip
\halign{ &\vrule# & \strut\quad\hfil#\quad\cr
\noalign{\hrule}
\noalign{\hrule}
height1pt&\omit&   &\omit&\cr
&(1,-1,-1)&& 480 & \cr
&(2,-2,-2)&& 480 & \cr
&(3,-3,-3)&& 480 & \cr
height1pt&\omit&   &\omit&\cr
\noalign{\hrule}
\noalign{\hrule}
height1pt&\omit&   &\omit&\cr
&(1,0,1)&& 480 & \cr
& $\vdots$ && $\vdots$ & \cr
&(6,0,6)&& 480 & \cr
height1pt&\omit&   &\omit&\cr
\noalign{\hrule}
\noalign{\hrule}
height1pt&\omit&   &\omit&\cr
&(1,0,-3)&& 1440 & \cr
height1pt&\omit&   &\omit&\cr
\noalign{\hrule}
\noalign{\hrule}
height1pt&\omit&   &\omit&\cr
&(1,1,-5)&& 2400 & \cr
height1pt&\omit&   &\omit&\cr
\noalign{\hrule}
\noalign{\hrule}
height1pt&\omit&   &\omit&\cr
&(1,1,-1)&& 480 & \cr
&(2,2,-2)&& 480 & \cr
height1pt&\omit&   &\omit&\cr
\noalign{\hrule}
\noalign{\hrule}
height1pt&\omit&   &\omit&\cr
&(2,-1,-4)&& -452160 & \cr
height1pt&\omit&   &\omit&\cr
\noalign{\hrule}}
\hrule}
\quad
\vbox{\offinterlineskip
\halign{ &\vrule# & \strut\quad\hfil#\quad\cr
\noalign{\hrule}
\noalign{\hrule}
height1pt&\omit&   &\omit&\cr
&(2,-1,0)&& 282888 & \cr
&(4,-2,0)&& 8606976768 & \cr
height1pt&\omit&   &\omit&\cr
\noalign{\hrule}
\noalign{\hrule}
height1pt&\omit&   &\omit&\cr
&(2,1,0)&& 282888 & \cr
height1pt&\omit&   &\omit&\cr
\noalign{\hrule}
\noalign{\hrule}
height1pt&\omit&   &\omit&\cr
&(3,-2,-1)&& 17058560 & \cr
height1pt&\omit&   &\omit&\cr
\noalign{\hrule}
\noalign{\hrule}
height1pt&\omit&   &\omit&\cr
&(3,-1,1)&& 17058560 & \cr
height1pt&\omit&   &\omit&\cr
\noalign{\hrule}
\noalign{\hrule}
height1pt&\omit&   &\omit&\cr
&(3,0,-1)&& 51516800 & \cr
height1pt&\omit&   &\omit&\cr
\noalign{\hrule}
\noalign{\hrule}
height1pt&\omit&   &\omit&\cr
&(3,1,1)&& 17058560 & \cr
height1pt&\omit&   &\omit&\cr
\noalign{\hrule}
\noalign{\hrule}
height1pt&\omit&   &\omit&\cr
&(4,-1,2)&& 477516780 & \cr
height1pt&\omit&   &\omit&\cr
\noalign{\hrule}
\noalign{\hrule}
height1pt&\omit&   &\omit&\cr
&(4,1,2)&& 477516780 & \cr
height1pt&\omit&   &\omit&\cr
\noalign{\hrule}
\noalign{\hrule}
height1pt&\omit&   &\omit&\cr
&(5,-1,3)&& 8606976768 & \cr
\noalign{\hrule}}
\hrule}
$$

\vfil
\eject

\subsec{Hypersurfaces in products of ordinary projective spaces }

\leftline{$\underline{X_{(3|3)}(1,1,1|1,1,1)}$}
\vskip.5cm
Due to the symmetry under exchange of $J_1$ and $J_2$, we
list only the curves of bi-degree $(n_{J_1},n_{J_2})$ with
$n_{J_1}\leq n_{J_2}$.
The table is for $n_{J_1}+n_{J_2}\leq 10$.

$$
\vbox{\offinterlineskip
\hrule
\halign{ &\vrule# & \strut\quad\hfil#\quad\cr
\noalign{\hrule}
height1pt&\omit&   &\omit&\cr
&(0,1)&& 189 &\cr
&(0,2)&& 189 &\cr
&(0,3)&& 162 &\cr
&(0,4)&& 189 &\cr
&(0,5)&& 189 &\cr
&(0,6)&& 162 &\cr
&(0,7)&& 189 &\cr
&(0,8)&& 189 &\cr
&(0,9)&& 162 &\cr
&(0,10)&& 189 &\cr
height1pt&\omit&   &\omit&\cr
\noalign{\hrule}
\noalign{\hrule}
height1pt&\omit&   &\omit&\cr
&(1,1)&&  8262 &\cr
&(2,2)&&  13108392 &\cr
&(3,3)&&  55962304650 &\cr
&(4,4)&&  366981860765484 &\cr
&(5,5)&&  3057363233014221000 &\cr
height1pt&\omit&   &\omit&\cr
\noalign{\hrule}
\noalign{\hrule}
height1pt&\omit&   &\omit&\cr
&(1,2)&&  142884 &\cr
&(2,4)&&  12289326723 &\cr
&(3,6)&&  2978764837454880 &\cr
height1pt&\omit&   &\omit&\cr
\noalign{\hrule}
\noalign{\hrule}
height1pt&\omit&   &\omit&\cr
&(1,3)&&  1492290 &\cr
&(2,6)&&  2673274744818 &\cr
height1pt&\omit&   &\omit&\cr
\noalign{\hrule}}
\hrule}
\quad
\vbox{\offinterlineskip
\halign{ &\vrule# & \strut\quad\hfil#\quad\cr
\noalign{\hrule}
\noalign{\hrule}
height1pt&\omit&   &\omit&\cr
&(1,4)&&  11375073 &\cr
&(2,8)&&  256360002145128 &\cr
height1pt&\omit&   &\omit&\cr
\noalign{\hrule}
\noalign{\hrule}
height1pt&\omit&   &\omit&\cr
&(1,5)&&69962130 &\cr
height1pt&\omit&   &\omit&\cr
height1pt&\omit&   &\omit&\cr
&(1,6)&&  368240958 &\cr
height1pt&\omit&   &\omit&\cr
\noalign{\hrule}
\noalign{\hrule}
height1pt&\omit&   &\omit&\cr
&(1,7)&&  1718160174 &\cr
height1pt&\omit&   &\omit&\cr
\noalign{\hrule}
\noalign{\hrule}
height1pt&\omit&   &\omit&\cr
&(1,8)&&  7278346935 &\cr
height1pt&\omit&   &\omit&\cr
\noalign{\hrule}
\noalign{\hrule}
height1pt&\omit&   &\omit&\cr
&(1,9)&&  28465369704 &\cr
height1pt&\omit&   &\omit&\cr
\noalign{\hrule}
\noalign{\hrule}
height1pt&\omit&   &\omit&\cr
&(1,2)&&  142884 &\cr
&(2,4)&&  12289326723 &\cr
&(3,6)&&  2978764837454880 &\cr
height1pt&\omit&   &\omit&\cr
\noalign{\hrule}
\noalign{\hrule}
height1pt&\omit&   &\omit&\cr
&(2,3)&&  516953097 &\cr
&(4,6)&&  1182543546601766871 &\cr
height1pt&\omit&   &\omit&\cr
\noalign{\hrule}
\noalign{\hrule}
height1pt&\omit&   &\omit&\cr
&(2,5)&&  206210244204 &\cr
height1pt&\omit&   &\omit&\cr
\noalign{\hrule}
\noalign{\hrule}
height1pt&\omit&   &\omit&\cr
&(2,7)&&  28368086706594 &\cr
height1pt&\omit&   &\omit&\cr
\noalign{\hrule}
\noalign{\hrule}
height1pt&\omit&   &\omit&\cr
&(3,4)&&  3154647509010 &\cr
height1pt&\omit&   &\omit&\cr
\noalign{\hrule}
\noalign{\hrule}
height1pt&\omit&   &\omit&\cr
&(3,5)&&  114200061474474 &\cr
height1pt&\omit&   &\omit&\cr
\noalign{\hrule}
\noalign{\hrule}
height1pt&\omit&   &\omit&\cr
&(3,7)&&  60186196491885072 &\cr
height1pt&\omit&   &\omit&\cr
\noalign{\hrule}
\noalign{\hrule}
height1pt&\omit&   &\omit&\cr
&(4,5)&&  25255131122299086 &\cr
height1pt&\omit&   &\omit&\cr
\noalign{\hrule}}
\hrule}
$$

\vfill\eject

\leftline{$\underline{X_{(2|4)}(1,1|1,1,1,1)}$}
\vskip.5cm

We list the non-zero topological invariants at degrees
$(n_{J_1},n_{J_2})$ with $n_{J_1},n_{J_2}\geq 0$
and $n_{J_1}+n_{J_2}\leq 10$.

$$
\vbox{\offinterlineskip
\hrule
\halign{ &\vrule# & \strut\quad\hfil#\quad\cr
\noalign{\hrule}
height1pt&\omit&   &\omit&\cr
&(1,0)&& 64  &\cr
height1pt&\omit&   &\omit&\cr
\noalign{\hrule}
\noalign{\hrule}
height1pt&\omit&   &\omit&\cr
&(0,1)&& 640  &\cr
&(0,2)&& 10032  &\cr
&(0,3)&& 288384  &\cr
&(0,4)&& 10979984  &\cr
&(0,5)&& 495269504  &\cr
&(0,6)&& 24945542832  &\cr
&(0,7)&& 1357991852672  &\cr
&(0,8)&& 78313183960464  &\cr
&(0,9)&& 4721475965186688  &\cr
&(0,10)&& 294890295345814704  &\cr
height1pt&\omit&   &\omit&\cr
\noalign{\hrule}
\noalign{\hrule}
height1pt&\omit&   &\omit&\cr
&(1,1)&&  6912 &\cr
&(2,2)&&  8271360 &\cr
&(3,3)&&  26556152064 &\cr
&(4,4)&&  130700405114112 &\cr
&(5,5)&&  816759204484794624 &\cr
height1pt&\omit&   &\omit&\cr
\noalign{\hrule}
\noalign{\hrule}
height1pt&\omit&   &\omit&\cr
&(2,1)&&  14400 &\cr
&(4,2)&&  48098560 &\cr
&(6,3)&&  445404149568 &\cr
\noalign{\hrule}
height1pt&\omit&   &\omit&\cr
&(3,1)&&  6912 &\cr
&(6,2)&&  8271360 &\cr
height1pt&\omit&   &\omit&\cr
\noalign{\hrule}
\noalign{\hrule}
height1pt&\omit&   &\omit&\cr
&(4,1)&&  640 &\cr
&(8,2)&&  10032 &\cr
height1pt&\omit&   &\omit&\cr
\noalign{\hrule}
\noalign{\hrule}
height1pt&\omit&   &\omit&\cr
&(1,2)&&  742784 &\cr
&(2,4)&&  532817161216 &\cr
&(3,6)&&  1084895026038311424 &\cr
height1pt&\omit&   &\omit&\cr
\noalign{\hrule}}
\hrule}
\quad
\vbox{\offinterlineskip
\halign{ &\vrule# & \strut\quad\hfil#\quad\cr
\noalign{\hrule}
\noalign{\hrule}
height1pt&\omit&   &\omit&\cr
&(3,2)&&  31344000 &\cr
&(6,4)&&  2485623412554752 &\cr
height1pt&\omit&   &\omit&\cr
\noalign{\hrule}
\noalign{\hrule}
height1pt&\omit&   &\omit&\cr
&(5,2)&&  31344000 &\cr
height1pt&\omit&   &\omit&\cr
\noalign{\hrule}
\noalign{\hrule}
height1pt&\omit&   &\omit&\cr
&(7,2)&&  742784 &\cr
height1pt&\omit&   &\omit&\cr
\noalign{\hrule}
\noalign{\hrule}
height1pt&\omit&   &\omit&\cr
&(1,3)&&  75933184 &\cr
&(2,6)&&  15714262788770816 &\cr
height1pt&\omit&   &\omit&\cr
\noalign{\hrule}
\noalign{\hrule}
height1pt&\omit&   &\omit&\cr
&(2,3)&&  2445747712 &\cr
&(4,6)&&  33831527906249235456 &\cr
height1pt&\omit&   &\omit&\cr
\noalign{\hrule}
\noalign{\hrule}
height1pt&\omit&   &\omit&\cr
&(4,3)&&  130867460608 &\cr
height1pt&\omit&   &\omit&\cr
\noalign{\hrule}
\noalign{\hrule}
height1pt&\omit&   &\omit&\cr
&(5,3)&&  329212616704 &\cr
height1pt&\omit&   &\omit&\cr
\noalign{\hrule}
\noalign{\hrule}
height1pt&\omit&   &\omit&\cr
&(7,3)&&  329212616704 &\cr
height1pt&\omit&   &\omit&\cr
\noalign{\hrule}
\noalign{\hrule}
height1pt&\omit&   &\omit&\cr
&(1,4)&&  7518494784 &\cr
&(2,8)&&  325754044147209418752 &\cr
height1pt&\omit&   &\omit&\cr
\noalign{\hrule}
\noalign{\hrule}
height1pt&\omit&   &\omit&\cr
&(3,4)&&  12305418469184 &\cr
height1pt&\omit&   &\omit&\cr
\noalign{\hrule}
\noalign{\hrule}
height1pt&\omit&   &\omit&\cr
&(5,4)&&  746592735013952 &\cr
height1pt&\omit&   &\omit&\cr
\noalign{\hrule}
\noalign{\hrule}
height1pt&\omit&   &\omit&\cr
&(1,5)&&  728114777344 &\cr
height1pt&\omit&   &\omit&\cr
\noalign{\hrule}
\noalign{\hrule}
height1pt&\omit&   &\omit&\cr
&(2,5)&&  97089446866176 &\cr
height1pt&\omit&   &\omit&\cr
\noalign{\hrule}
\noalign{\hrule}
height1pt&\omit&   &\omit&\cr
&(3,5)&&  4074651399444224 &\cr
height1pt&\omit&   &\omit&\cr
\noalign{\hrule}
\noalign{\hrule}
height1pt&\omit&   &\omit&\cr
&(4,5)&& 78142574531195136 &\cr
height1pt&\omit&   &\omit&\cr
\noalign{\hrule}
\noalign{\hrule}
height1pt&\omit&   &\omit&\cr
&(1,6)&&  69368161314176 &\cr
height1pt&\omit&   &\omit&\cr
\noalign{\hrule}
\noalign{\hrule}
height1pt&\omit&   &\omit&\cr
&(1,7)&&  6526028959787520 &\cr
height1pt&\omit&   &\omit&\cr
\noalign{\hrule}
\noalign{\hrule}
height1pt&\omit&   &\omit&\cr
&(2,7)&&  2336268973133447168 &\cr
height1pt&\omit&   &\omit&\cr
\noalign{\hrule}
\noalign{\hrule}
height1pt&\omit&   &\omit&\cr
&(3,7)&&  247572316458452288000 &\cr
height1pt&\omit&   &\omit&\cr
\noalign{\hrule}
\noalign{\hrule}
height1pt&\omit&   &\omit&\cr
&(1,8)&&  607840242136069376 &\cr
height1pt&\omit&   &\omit&\cr
\noalign{\hrule}
\noalign{\hrule}
height1pt&\omit&   &\omit&\cr
&(1,9)&&  56154770246801057024 &\cr
height1pt&\omit&   &\omit&\cr
\noalign{\hrule}}
\hrule}
$$
\vfil
\eject

\leftline{$\underline{X_{(2|2|3)} (1,1|1,1|1,1,1)}$}
\vskip.5cm

We have the obvious symmetry
$N(n_{J_1},n_{J_2},n_{J_3})=N(n_{J_2},n_{J_1},n_{J_3})$
and we will list the non-zero
invariants only for $n_{J_1}\leq n_{J_2}$
and for $n_{J_1}+n_{J_2}+n_{J_3}\leq 6$

$$
\vbox{\offinterlineskip
\halign{ &\vrule# & \strut\quad\hfil#\quad\cr
\noalign{\hrule}
\noalign{\hrule}
height1pt&\omit&   &\omit&\cr
&(0,0,1)&&    168  &\cr
&(0,0,2)&&    168  &\cr
&(0,0,3)&&    144  &\cr
&(0,0,4)&&    168  &\cr
&(0,0,5)&&    168  &\cr
&(0,0,6)&&    144   &\cr
height1pt&\omit&   &\omit&\cr
\noalign{\hrule}
\noalign{\hrule}
height1pt&\omit&   &\omit&\cr
&(0,1,0)&&   54  &\cr
height1pt&\omit&   &\omit&\cr
\noalign{\hrule}
\noalign{\hrule}
height1pt&\omit&   &\omit&\cr
&(0,1,1)&&    1080    &\cr
&(0,2,2)&&    55080   &\cr
&(0,3,3)&&    5686200 &\cr
height1pt&\omit&   &\omit&\cr
\noalign{\hrule}
\noalign{\hrule}
height1pt&\omit&   &\omit&\cr
&(0,1,2)&&    9504 &\cr
&(0,2,4)&&    12531888 &\cr
height1pt&\omit&   &\omit&\cr
\noalign{\hrule}
\noalign{\hrule}
height1pt&\omit&   &\omit&\cr
&(0,1,3)&&    55080  &\cr
height1pt&\omit&   &\omit&\cr
\noalign{\hrule}
\noalign{\hrule}
height1pt&\omit&   &\omit&\cr
&(0,1,4)&&    258876   &\cr
height1pt&\omit&   &\omit&\cr
\noalign{\hrule}
\noalign{\hrule}
height1pt&\omit&   &\omit&\cr
&(0,1,5)&&    1045440   &\cr
height1pt&\omit&   &\omit&\cr
\noalign{\hrule}
\noalign{\hrule}
height1pt&\omit&   &\omit&\cr
&(0,2,1)&&    1080 &\cr
&(0,4,2)&&    55080 &\cr
height1pt&\omit&   &\omit&\cr
\noalign{\hrule}
\noalign{\hrule}
height1pt&\omit&   &\omit&\cr
&(0,2,3)&&    1045440 &\cr
height1pt&\omit&   &\omit&\cr
\noalign{\hrule}}
\hrule}
\quad
\vbox{\offinterlineskip
\halign{ &\vrule# & \strut\quad\hfil#\quad\cr
\noalign{\hrule}
\noalign{\hrule}
height1pt&\omit&   &\omit&\cr
height1pt&\omit&   &\omit&\cr
&(0,3,1)&&    168 &\cr
height1pt&\omit&   &\omit&\cr
\noalign{\hrule}
\noalign{\hrule}
height1pt&\omit&   &\omit&\cr
&(0,3,2)&&    94248   &\cr
height1pt&\omit&   &\omit&\cr
\noalign{\hrule}
\noalign{\hrule}
height1pt&\omit&   &\omit&\cr
&(1,1,1)&&    22968 &\cr
&(2,2,2)&&    212527800 &\cr
height1pt&\omit&   &\omit&\cr
\noalign{\hrule}
\noalign{\hrule}
height1pt&\omit&   &\omit&\cr
&(1,1,2)&&    801720 &\cr
height1pt&\omit&   &\omit&\cr
\noalign{\hrule}
\noalign{\hrule}
height1pt&\omit&   &\omit&\cr
&(1,1,3)&&    14272344 &\cr
height1pt&\omit&   &\omit&\cr
\noalign{\hrule}
\noalign{\hrule}
height1pt&\omit&   &\omit&\cr
&(1,2,0)&&    54 &\cr
height1pt&\omit&  &\omit&\cr
\noalign{\hrule}
\noalign{\hrule}
height1pt&\omit&   &\omit&\cr
&(1,2,1)&&    84240 &\cr
height1pt&\omit&   &\omit&\cr
\noalign{\hrule}
\noalign{\hrule}
height1pt&\omit&   &\omit&\cr
&(1,2,2)&&    9589752 &\cr
height1pt&\omit&   &\omit&\cr
\noalign{\hrule}
\noalign{\hrule}
height1pt&\omit&   &\omit&\cr
&(1,2,3)&&    422121240  &\cr
height1pt&\omit&   &\omit&\cr
\noalign{\hrule}
\noalign{\hrule}
height1pt&\omit&   &\omit&\cr
&(1,3,1)&&    84240 &\cr
height1pt&\omit&   &\omit&\cr
\noalign{\hrule}
\noalign{\hrule}
height1pt&\omit&   &\omit&\cr
&(1,3,2)&&    37017000   &\cr
height1pt&\omit&   &\omit&\cr
\noalign{\hrule}
\noalign{\hrule}
height1pt&\omit&   &\omit&\cr
&(1,4,1)&&    22968   &\cr
\noalign{\hrule}
\noalign{\hrule}
height1pt&\omit&   &\omit&\cr
&(2,2,1)&&    823968 &\cr
height1pt&\omit&   &\omit&\cr
\noalign{\hrule}
\noalign{\hrule}
height1pt&\omit&   &\omit&\cr
&(2,3,0)&&    54 &\cr
height1pt&\omit&   &\omit&\cr
\noalign{\hrule}
\noalign{\hrule}
height1pt&\omit&   &\omit&\cr
&(2,3,1)&&    2286360   &\cr
height1pt&\omit&   &\omit&\cr
\noalign{\hrule}}
\hrule}
$$
\listrefs
\bye